\documentclass[11pt]{article}
\usepackage[]{amsmath,amssymb}
\usepackage{amsfonts}
\usepackage{graphics,epsfig}
\usepackage{epic}

\let\sect=\section
\def\section{\newpage\sect}

\def\text#1{\mbox{\rm #1\ }}

\def\ie{{\rm i.e.,\/}\ }
\def\etc{{\rm etc.\/}\ }

\def \otimesdot {\stackrel{\cdot}{\otimes}}
\newcommand{\ZZ}{\mathbb{Z}}

\newcommand{\CC}{\mathbb{C}}

\newcommand{\ch}[1]{\chi_{#1}}
\newcommand{\och}[1]{\overline{\chi_{#1}}}
\newcommand{\ud}[1]{\underline{#1}}

\newcommand{\ep}{\epsilon}
\newcommand{\xa}[1]{|\chi_{#1}|^2}
\newcommand{\xaa}[2]{|\chi_{#1} + \chi_{#2}|^2}
\newcommand{\xaaa}[3]{|\chi_{#1} + \chi_{#2} + \chi_{#3}|^2}
\newcommand{\xaaaa}[4]{|\chi_{#1} + \chi_{#2}+ \chi_{#3}+ \chi_{#4}|^2}
\newcommand{\xaaaaa}[5]{|\chi_{#1} + \chi_{#2}+ \chi_{#3}+ \chi_{#4}+
\chi_{#5}|^2}
\newcommand{\xaaaaaa}[6]{|\chi_{#1} + \chi_{#2}+ \chi_{#3}+ \chi_{#4}+
                             \chi_{#5} + \chi_{#6}|^2}
\newcommand{\xaaaaaaa}[7]{|\chi_{#1} + \chi_{#2}+ \chi_{#3}+ \chi_{#4}+
                             \chi_{#5} + \chi_{#6}+ \chi_{#7}|^2}
\newcommand{\xaaaaaaaa}[8]{|\chi_{#1} + \chi_{#2}+ \chi_{#3}+ \chi_{#4}+
                               \chi_{#5} + \chi_{#6}+ \chi_{#7}+ \chi_{#8}|^2}
\newcommand{\xx}[2]{\chi_{#1}\overline{\chi_{#2}}}

\title{
Twisted partition functions for ADE boundary conformal field theories
       and Ocneanu algebras of quantum  symmetries
         \vspace{0.8cm}}

\author{R. Coquereaux${}^{1}$  \thanks{~Email:
Robert.Coquereaux@cpt.univ-mrs.fr}$\;$,
              G. Schieber${}^{1,}$  ${}^2$ \thanks{~Email:
schieber@if.ufrj.br} \\
\\
${}^1$ {\it Centre de Physique Th\'eorique - CNRS} \\
             {\it Campus de Luminy - Case 907}           \\
             {\it F-13288 Marseille - France}            \\
\\
${}^2$ {\it Instituto de F\'{\i}sica, Universidade Federal do Rio de
Janeiro} \\
             {\it Ilha do Fund\~ao, Caixa Postal 68528}\\
             {\it 21945-970, Rio de Janeiro, Brasil}\\
}

\date{}
\begin{document}
\thispagestyle{empty}
\begin{titlepage}
\maketitle
\abstract{For every ADE Dynkin diagram, we give a realization, in
terms of usual fusion algebras (graph algebras), of the
algebra of quantum symmetries described by the associated Ocneanu
graph. We give explicitly, in each case, the list of the
corresponding twisted partition functions.}

\vspace{1.cm}

\begin{center}
We dedicate this article to the memory of our friend  Prof. Juan A. Mignaco
deceased, 6 June 2001.
\end{center}

\vspace{2. cm}

\noindent Keywords: conformal field theories, ADE, modular invariance,
quantum symmetries, Hopf algebras, quantum groups.
\vspace{1.0cm}


\vspace{2. cm}

\noindent {\tt  hep-th/0107001}\\
\noindent CPT-2001/P.4216 \\

\vspace*{0.3 cm}

\end{titlepage}



\section{Introduction}

For each $ADE$ Dynkin diagram $G$, we consider the corresponding
Ocneanu graph $Oc(G)$, as given by $\cite{Ocneanu:paths}$, and build
explicitly an algebra (the algebra of quantum symmetries
of the given Dynkin diagram)
whose multiplication table is encoded by this Ocneanu graph. Using
this algebra structure, we obtain explicitly -- and easily -- the
expression of all
the twisted partition functions that one may associate with
the given Dynkin diagram (one for each vertex of its corresponding
Ocneanu graph).

Our first purpose is not to deduce the graph $Oc(G)$ from $G$,
since that was already done in $\cite{Ocneanu:paths}$ (actually the details
have never been made available in printed form\ldots), but to give a
simple  presentation of the corresponding algebra of quantum
symmetries.
In each case, this algebra will be given in terms of a quotient of
the tensor square of the graph algebra (fusion algebra) associated
with some particular ADE Dynkin diagram. These algebras $Oc(G)$ are commutative
in almost all cases (for $D_{2n}$ they involve $2\times 2$ matrices).

Our other purpose is to use this structure to obtain explicitly the
corresponding ``toric matrices'' (terminology taken form
\cite{Coque:qtetra}) and the corresponding ``twisted partition
functions'' (terminology taken form \cite{PetZub:bcft} and
\cite{PetZub:Oc}).

The torus structure of all ADE models has  been worked out by
A. Ocneanu himself several years ago (unpublished).
Explicit expressions for the eight toric matrices of
dimension $6\times 6$ of the $D_{4}$ model can be found in \cite{PetZub:bcft}
(where they are interpreted
physically in terms of the 3-state Potts model) and for the twelve toric matrices
of dimension $12\times 12$ of the  $E_{6}$ model in \cite{Coque:qtetra}
(where one can also find a general method of calculation for these 
quantities). One of our purposes is to give explicit results, in particular for
all exceptional cases, by following the method explained in this last 
reference \cite{Coque:qtetra} and summarized in section 2.
Starting from Conformal Field Theory (CFT), another general method for obtaining the
structure of these twisted partition functions has been described in the
subsequent article \cite{PetZub:Oc} which contains closed formulae;
we do not use this formalism. Actually, 
the constructions performed in the sequel avoid, deliberately, the use of CFT concepts.

Again, we insist upon the fact that we take for granted the data given
by the Ocneanu graphs themselves; otherwise, we should either have to
diagonalize the convolution product of the quantum Racah-Wigner bi-algebra
associated with the given ADE diagram, or to solve the problem of finding
what are the irreducible elements for the set of  ``connections''
that one can define on a pair of graphs (system of generalized
Boltzman weights, see also \cite{Roche:Oc}). This was done by Ocneanu himself.
The present paper can be read independently of \cite{Coque:qtetra} since 
all the necessary information is gathered in section 2.

As already stressed, our own 
presentation, which follows \cite{Coque:qtetra}, uses neither the 
language nor the techniques  of conformal field theory, but 
the results themselves can be interpreted in terms of CFT. For 
instance the toric matrices lead to quantities that can be
interpreted in terms of partition functions for boundary counformal 
field theories in presence of defect lines. The reader interested in 
those CFT aspects should look at the article (\cite{PetZub:Oc}) which contains
many results of independent interest and is
probably the most complete published work on this subject, in
relation with conformal field theories. 

For every ADE example, the particular toric matrix associated with the 
``unit vertex''  
of the corresponding Ocneanu graph is the usual modular invariant for the 
associated  
ADE model (in the classification of \cite{CapItzZub}), \ie the
corresponding sesquilinear form gives the usual modular invariant partition
function. The other partition functions (the non trivially ``twisted''
ones), those associated with the other points of the Ocneanu graph,
are not modular invariant.

It is unfortunately almost impossible to provide a unified (or
uniform) treatment for all ADE diagrams; indeed, all of them are
``special'', in one way or another. The $A_{n}$ are a bit too
``simple'' (many interesting constructions just coincide in that
case), the $D_{2n}$ are the only ones to give rise to a non abelian
algebra of quantum symmetries, the $E_{7}$ does not define a positive
integral graph algebra and the $D_{2n+1}$ do not define any integral
graph algebra at all; ``only'' $E_{6}$ and $E_{8}$ lead, somehow, to
a similar
treatment.

The structure of the present paper is as follows:
after a first section devoted to a general overview of the theory, we
examine separately all types of $ADE$ Dynkin diagrams. In each case,
\ie in each section, after having presented the graph algebra associated
with the chosen diagram (when it exists), we describe
explicitly the structure of an associative algebra that we can
associate with its corresponding Ocneanu graph, express it
in terms of (usual) graph algebras and deduce, from this
algebra structure, the corresponding toric matrices.
In order not to clutter the paper with sparse matrices of big size,
we list only the
sesquilinear forms -- \ie the twisted partition functions --
associated with these toric matrices. For pedagogical reasons we prefer to
perform this analysis in the following order: $A_n$, $E_{6}$, $E_{8}$,
       $D_{2n}$, $D_{2n+1}$, $E_{7}$.

\section{Summary of the algebraic constructions}
\subsection{Foreword}
To every pair of $ADE$ Dynkin diagrams with the same Coxeter number,
one may associate (Ocneanu) an algebra of quantum symmetries. Its
elements (also called ``connections'' on the given pair of graphs) can
be added and multiplied in a way analogous to what is done for
representation of groups; in particular, this algebra has a unit, and one may
consider
a set of ``irreducible'' quantum symmetries, that, by definition, build up a
basis of linear generators for this algebra (an analogue of the notion
of irreducible representations). Using multiplication, we may also
single out, in each case, two (algebraic) generators, usually called
``chiral left'' and ``chiral right'' generators,  playing the role of
fundamental representations for groups: all other irreducible elements can be
obtained as linear combinations of products of these two
generators. The Ocneanu graph precisely encodes this algebraic
structure: its number of vertices is equal to the number
of irreducible elements and edges encode multiplication by the two
generators.

When the two chosen Dynkin diagrams coincide, we can find
another interpretation for the Ocneanu graph (and algebra)
of quantum symmetries. This is actually the case of interest, for us, in
the present paper. Here is a sketch of the theory.
One first considers elementary paths (\ie genuine
paths) on the chosen Dynkin diagram $G$; one then build the
Hilbert space $Path(G)$ of all paths, by taking linear
combinations of elementary paths and declaring that elementary paths
are orthogonal. This vector space provides a path model for the Jones
algebra associated with $G$. The next step is to consider the vector subspace
$EssPath(G)$ of essential paths: by definition, they span the
intersection of kernels of all Jones projectors; in the classical
situation, the essential paths starting from the origin
would correspond to projectors on the symmetric
representations of
$SU(2)$ (or of finite subgroups of $SU(2)$). We refer to the paper
\cite{Coque:Karpacz} (to be contrasted with \cite{Coque:qtetra}) where a
geometrical study of the classical binary polyhedral groups \cite{Klein}
(symmetries of Platonic
bodies) is performed, using McKay correspondence \cite{McKay}, by
studying paths and essential paths on the
affine Dynkin diagrams of type $E_{6}^{(1)}$, $E_{7}^{(1)}$ and
$E_{8}^{(1)}$.
Essential paths start somewhere ($a$), end somewhere ($b$) and have a
certain length ($n$).
The finite dimensional vector space $EssPath(G)$ is therefore graded by
the length $n$ of the paths: $EssPath(G) =
\bigoplus_{n} EssPath^{n}(G)$. Notice that essential paths are usually
linear combinations of elementary paths. We may then build the graded
algebra ${\cal A} \doteq \bigoplus_{n} End(EssPath^{n}(G))$ where
each summand is the space of endomorphisms of $EssPath^{n}(G)$;
they can be explicitly written as  square matrices.
The algebra ${\cal A}$ is not only an algebra (for the obvious
composition $\circ$ of endomorphisms) but also a {\sl bi}-algebra:
using concatenation of elementary paths together with the existence 
of a scalar product on $Path(G)$,  one can define
a convolution product $\ast$ on ${\cal A}$.
Details concerning this construction, also due to Ocneanu, and about
its interpretation in
the case of affine $ADE$ diagrams (\ie in the case of $SU(2)$ itself
and the usual polyhedral groups) will be found in \cite{CoGaTr:triangles}.
${\cal A}$ is therefore a kind of finite dimensional and quantum
analogue of the
Racah-Wigner bi-algebra. Being semi-simple for both algebra
structures $\circ$ and $\ast$, we may decompose ${\cal A}$ as
a sum of square matrices (blocks) in two different ways. For the first
structure ($\circ$), which is obvious from its very definition,
the corresponding projectors are labeled by $n$ (the length of
essential paths). For the second structure ($\ast$), the blocks
are labeled by an index, that we shall call $x$; the Ocneanu algebra
is then precisely the algebra spanned by those $x$, \ie by the
corresponding projectors: this is an analogue of the table of multiplication of
characters (convolution product) for a  finite group.
The underlying vector space of ${\cal A}$
possesses two adapted basis, one is expressed in terms of the ``double
triangles of Ocneanu'' (that we prefer to draw as a ``fermionic''
diffusion graph
with a connecting vertical ``photon'' line labeled by $n$), the other in terms
of diffusion graphs with horizontal ``very thick lines'' labeled by $x$, the
vertices of the Ocneanu graph. The change of basis between the two
adapted basis can be thought of as a duality relation; it is a kind
of generalized Fourier
transform involving quantum Racah symbols at a particular root of unity
depending on the chosen
Dynkin diagram. It will also be conceptually important to
consider the length $n$ as labelling a particular vertex of a $A_{N}$
graph (the first vertex to the left being labeled $0$).

Several constructions used in our paper can certainly be understood in terms of
planar algebras \cite{Jones:planar} (see also \cite{Jones:book}), nets of subfactors
\cite{Evans-1}, \cite{Evans-2},  or in
terms of braided categories
\cite{Kirilov:ADE}, but we shall not discuss this here.
We do not plan, in the present paper, to give any interpretation of
these constructions in terms of standard Hopf algebra constructions:
this has not been worked out, yet.

\subsection{Structure of the following sections}

\subsubsection{The diagram (ADE) and its adjacency matrix}

We give the diagram $G$ itself, choose a particular labelling for vertices
and give the adjacency matrix ${\cal G}$ in a specified basis.
We consider vertices $\sigma_{v}$ of $G$
as would-be irreducible representations for a quantum analogue of a
group algebra ${\cal
H}_{G}$ that we do not need to define.
We also write down the norm $\beta$ of $G$ (the biggest
eigenvalue of ${\cal G}$) and the Perron-Frobenius eigenvector $D$ (\ie
the normalized eigenvector corresponding  to $\beta$, with its
smallest component, associated with the vertex $\sigma_0$,
normalized to the integer
$1$).
The components of $D$ give (by definition)
the quantum dimensions  of the
irreducible representations $\sigma_{v}$.
In all cases, $\beta$ is equal to the $q$-number\footnote{We define
$[n]_{q}=\frac{q^n - q^{-n}}{q - q^{-1}}$, where $q=exp(\frac{i\pi}{\kappa})$.}
$[2]_{q} = q + \frac{1}{q} = 2 \cos(\frac{\pi}{\kappa})$.
This value $\kappa$ is, by  definition, the Coxeter number of the graph.
In all cases, the $q$-dimension of $\sigma_{0}$ (the marked vertex) is
$[1]_{q}=1$.
More information can be gathered, for instance, from the book \cite{FMS:book}.

We should remember the values of Coxeter numbers for the various ADE
Dynkin diagrams:
$$\begin{array}{|c|ccccc|}
\hline
{} & A_n & D_{n+2} & E_6 & E_7 & E_8 \\
\hline
\text{Coxeter number} & n+1 & 2(n+1) & 12 & 18 & 30 \\
\hline
\end{array}
$$

\subsubsection{The graph algebra of the Dynkin diagram and the quantum table of
characters}

The next step is to associate with the diagram $G$, when possible, a
commutative algebra playing the role of an algebra of characters.
This algebra is linearly generated, as a vector space, by the vertices
$\sigma_v$ of $G$. As an  associative algebra, it admits a unit $\sigma_0$ and
one generator $\sigma_{1}$, with quantum dimension $[2]_{q}$. The
relations of this associative algebra
are defined by the graph $G$ itself, considered as encoding
multiplication by the generator $\sigma_{1}$: the irreducible representations
appearing in the decomposition of $\sigma_1 \sigma$, with $\sigma$, a
vertex of $G$, are the neighbours of
$\sigma$ on the diagram $G$.
We impose, furthermore, that the structure constants of this algebra
should be positive integers, as it is the case for irreducible
representations of
groups or, more generally, of Hopf algebras.
It is (almost well) known, since  \cite{Pasquier:alg} that, for
ADE diagrams, the solution to the above problem does not exist for
       $E_{7}$ and $D_{odd}$. For all other $ADE$ diagrams, there exists a
unique solution. This algebra is called the the graph algebra
associated with $G$, or the fusion algebra associated
with $G$ and sometimes  (\cite{Zuber:Bariloche}),
the dual Pasquier algebra of $G$.
Such a commutative algebra is also  a ``positive integral hypergroup'',
or simply an hypergroup, when no confusion arises (see \cite{hypergroups} and
references therein).
We shall denote this algebra by the same symbol as the graph itself,
and hope that no confusion with the simple Lie group bearing the same
name will arise.
Practically, we have to build a multiplication table,
the first two rows and columns being already known (multiplication by
the unit $\sigma_{0}$ and
by the generator $\sigma_{1}$).
The table is  built in a very straightforward way, by imposing associativity.
For instance, in the case of the graph $A_{n}$, $n>4$, let us calculate,
\begin{eqnarray*}
\sigma_{2} \sigma_{2}  & = & (\sigma_{1}\sigma_{1} - \sigma_{0})\sigma_{2}
= \sigma_{1}\sigma_{1} \sigma_{2}- \sigma_{0}\sigma_{2}  \\ {} & = &
\sigma_{1}(\sigma_{1}+\sigma_{3}) - \sigma_{2} =
\sigma_{0}+\sigma_{2} + \sigma_{2} + \sigma_{4} - \sigma_{2} =
\sigma_{0}+\sigma_{2}  + \sigma_{4}
\end{eqnarray*}

In every case (except $E_{7}$ and $D_{odd}$)
we shall give the multiplication table of the graph algebra. When
writing down this table, and in order to save space, we shall drop the
symbols $\sigma$ and refer to the different vertices only by their subscript.

The  graph matrix algebra of the $ADE$ diagram $G$, with $r$ vertices,
       is a matrix algebra linearly generated by $r$ matrices of
size $r\times r$  providing a
faithful realization of the  graph algebra spanned by the $\sigma_{a}$'s. Its
construction is straightforward: to $\sigma_{0}$ one associates the
unit matrix (call it $G_0$) and to the generator  $\sigma_{1}$ we associate a
matrix $G_1$ equal to the adjacency matrix ${\cal G}$ of the diagram;
to the other vertices $\sigma_a$, expressed in terms of
$\sigma_0$ and $\sigma_1$ we associate the corresponding matrices $G_a$
given in terms of $G_0$ and $G_1$.
Since these last two matrices are already explicitly known, in order
to save space,
we shall just give the
polynomial expressions giving the $G_j$ in terms of these two.

In the particular case of $A_N$ graphs, the fusion matrices $G_{i}$ will
be also called $N_i$.

The $r$ matrices $G_a$ commute with one another (they are all polynomials in one and 
the same $G_1$) and can be simultaneously
diagonalized thanks to a matrix $S_{G}$.
If the $\sigma_a$'s were irreducible representations of a finite group,
this matrix $S_{G}$ would be the table of characters for this finite
group, \ie the
result of the pairing between conjugacy classes and irreducible characters
(notice that we do not need
to build explicitly the conjugacy classes!). In the present situation,
$S_{G}$ is the quantum analogue of a table of characters.
In the case of $A_{N}$ graphs, the same matrix, simply denoted by
$S$, represents one of the
generators of the
modular group $SL(2,\ZZ)$ (Verlinde - Hurwitz representation).

\subsubsection{Essential matrices and paths}
The general definition of essential paths on a graph was defined by
A. Ocneanu (\cite{Ocneanu:paths}), but we do not need this precise
definition here because we just need
to count these particular paths. We shall nevertheless recall this
definition in the appendix. It is enough to know that
the general notion of essential paths generalizes the notion of symmetric
(or $q-symmetric$) representations (at least for those paths starting
from the origin). Some general comments and
particular cases (diagrams $E_6$ and
$E_6^{(1)}$) can be read in \cite{Coque:Karpacz} and \cite{Coque:qtetra}.
The number $E_a[p,b]$ of essential paths of length $p$ starting at some
vertex $a$ and ending on the vertex $b$ is given by $b$-th component of the
row vector $E_a(p)$ defined as follows:
\begin{itemize}
\item $E_a(0)$ is the  (line) vector characterizing the chosen initial
vertex
\item $E_a(1) \doteq E_a(0).{\cal G}$
\item $E_a (p) \doteq  E_a(p-1).{\cal G} - E_a (p-2)$
\end{itemize}

The expression of $E_a(0)$ depends on the chosen ordering of vertices; it is
convenient
anyway to set $E_0(0)=(1,0,0,\ldots)$ for the unit $\sigma_{0}$ of $G$, and
$E_1(0)=(0,1,0,\ldots)$ for the generator
$\sigma_1$.
For a graph with $r$ vertices, starting from $E_a(0)$, we {\sl
would} obtain in this way $r$ rectangular
matrices $E_a$ with infinitely many rows (labeled by $p=0,1,2,\ldots$)
       and $r$ columns (labeled by $b$).

The reader can check that, for Dynkin $ADE$ diagrams,
the numbers $E_a(p)$ are positive integers {\sl
provided} $0 \leq p \leq \kappa-2$ ($\kappa$ being the Coxeter number
of the graph),
but this ceases to be true as soon as $p>\kappa-2$. For instance, in
the case of the $E_6$ graph,
where $\kappa = 12$, we get $E_0(11) = (0,0,0,0,0,0)$,  $E_0(12) =
(0,0,0,0,-1,0)$.
This reflects the fact (Ocneanu \cite{Ocneanu:paths}) that essential
paths on these graphs,with a length bigger
than $\kappa-2$, do not exist.
We call ``essential matrices''  the $r$ rectangular $(\kappa-1) \times
r$ matrices
obtained by keeping only the first $\kappa-1$ rows of the $E_a(.)'s.$
For every $ADE$ diagram, these finite dimensional rectangular
matrices will still be
denoted\footnote{
not to be confused with the symbol used for the exceptional Dynkin diagrams
themselves!} by $E_a$.
The components of the rectangular matrix $E_a$ are denoted by  $E_a[p,b]$.
Matrix elements of these matrices can be displayed as vertices
with three edges labeled by $a,b,p$ (or, dually, as triangles).
Warning: the smallest value for $p$, the length of essential paths,
is $0$, and not $1$.

In order to save space, we shall not give explicitly all these
matrices $E_a$, although
they are absolutely crucial for obtaining the next results;
however their calculation,
using the above recurrence formulae, is totally straightforward, once
the matrix
$G_1 \equiv {\cal G}$ is known.
The pattern of non-zero entries of an essential matrix $E_a$,
associated with some
graph $G$,
gives a figure expressing ``visually'' the structure of the space of
essential paths
starting
from $a$. These essential matrices were introduced in \cite{Coque:qtetra} as a
convenient
tool, but the geometrical patterns themselves were first obtained  by
A.Ocneanu (the essential paths starting from all possible vertices
are displayed, for all $ADE$ graphs, in the appendix of
\cite{Ocneanu:paths}).  In the coming sections, we shall only display the
essential matrix $E_{0}$ that encodes essential paths starting from the
origin. Let us finally mention that a list of
the rectangular matrices $E_{0}$'s (not the $E_{a}$'s), interpreted 
in the context of RSOS models, can be found in
\cite{PearceZhou:RSOS}.

When the diagram is not an $ADE$ but an {\sl affine} $ADE$, the
essential matrices are no longer of finite size: they have infinitely
many rows; they can be interpreted in terms of the classical
induction/restriction theory for representations of $SU(2)$ and
its finite subgroups (binary polyhedral groups). See the lecture notes
\cite{Coque:Karpacz} for a study of the corresponding classical geometries,
along the above lines.

For Dynkin diagrams of type $A_N$, we have the relation
$N_i =  N_{i-1}.{\cal G} - N_{i-2}$,
and from the above definition of essential matrices, we see that
there is no difference, in this case, between the
fusion graph matrices $G_{i} = N_i$ and the essential matrices $E_i$.

Before ending this section, we should point out the fact that since
rows of the essential
matrices associated with a particular Dynkin diagram $G$ have labels
$p$ running
from $0$ to $\kappa-1$, they are therefore also indexed by the
vertices $\tau_p$
of the Dynkin diagram $A_{\kappa - 1}$. In this way, we can interpret these
essential
matrices as a kind of quantum analogue of the theory of induction/restriction:
irreducible representations of $A_{\kappa - 1}$ can be ``reduced'' to
irreducible
representations
of $G$ (essential matrices can be read ``horizontally'' in this way) and
irreducible representations of $G$ can ``induce'' irreducible
representations of
$A_{\kappa - 1}$
       (essential matrices can be read ``vertically'' in this way).
Rather than displaying the essential matrices, or the corresponding
spaces of paths,
we shall only give, for each vertex of the graph $G$, the list of induced
representations of $A_{\kappa - 1}$.
This information can be deduced immediately from the essential matrix
$E_0$. In other
words,
we consider, for  each vertex $\sigma_v$ of $G$  an associated quantum vector
bundle
and decompose the space of its sections into irreducible representations
of $A_{\kappa
- 1}$.

\subsubsection{Dimensions of blocks for the  Racah-Wigner-Ocneanu
bi-algebras}
The Racah-Wigner-Ocneanu  bi-algebra ${\cal A}$ is a direct sum of
blocks in two different ways (see section 2.1). Its dimension
is obtained either by
summing the squares $d_n^2$ where $d_n$ is the number of essential
paths of length
$n$, or by summing the squares $d_{x}^{2}$, where the $d_{x}$ are
the sizes of the Ocneanu blocks.
The integers $d_{n}$ are obtained by summing all matrix elements of
the row $n+1$
over all essential matrices $E_{a}$ (all vertices $a$ of a given diagram).
This first calculation is relatively easy.

The integers $d_{x}$ giving the number of ``vertices'' labeled by
$(a,b,x)$ can be obtained from the multiplication table of ${\cal H}_{Oc(G)}$.
If the label $x$ of an Ocneanu block
is of the type $a \otimesdot b$, or a linear combination of such elements
(the notation $\otimesdot$ is
introduced later in the text), and
when ${\cal H}_{Oc(G)}$ is commutative and contains two (left and right)
subalgebras  isomorphic with the graph algebra of the Dynkin diagram $G$,
the integers $d_{x} = d_{a \otimesdot b}$ can be obtained simply by
summing all matrix elements ${(\Sigma_{x})^{c}}_{d}$ of the matrices
$\Sigma_{x} \doteq G_{a}G_{b}$, where $G_{a}$ and $G_{b}$ are fusion matrices
of the Dynkin diagram $G$. This holds in particular for $A_{N}$ and
for the exceptional cases $E_{6}$ and $E_{8}$.
The other cases -- in particular the case of $E_{7}$ -- are slightly more
involved. We refer to the corresponding sections.

The knowledge of integers $d_{n}$, $d_{x}$ was implicit in the work
of Ocneanu, already presented to several audiences years ago (for
instance \cite{Ocneanu:Marseille}). The values of $d_{n}$ and $d_{x}$
were first published, for the $E_{6}$ case, in \cite{Coque:qtetra} (the
treatment  of the $E_{8}$ case being the same). General results, for all
cases, were published in \cite{PetZub:Oc}.
We take advantage of the explicit realization that we
find for the bialgebra ${\cal H}_{Oc(G)}$ to recover easily all the
results, including the more difficult $E_{7}$ case (see the
corresponding section).

We give the integers $d_{n}$, $d_{x}$ and the sums
$\sum d_n$, $\sum d_x$ and $\sum d_n^2 = \sum d_x^2$.
The equality of squares is a direct consequence of the bi-algebra
structure. In most cases (not $D_{even}$) one finds also that $\sum d_n =
\sum d_x$; this can be understood as coming from a change of basis in
the vector
space $EssPath(G)$. The equality of sums can actually be also
achieved for $D_{even}$ by performing the summation only on particular
classes of elements (see the discussion made in \cite{PetZub:Oc}.)

\subsubsection{The Ocneanu graph corresponding to a Dynkin diagram and
its algebra}

The Ocneanu graph $Oc(G)$ associated with a Dynkin diagram $G$ was already
discussed in the present introduction.
As already stated, we take it directly from reference \cite{Ocneanu:paths}.

One of our purposes is to give an explicit presentation for the
corresponding algebras, that will be called
${\cal H}_{Oc(G)}$. In most cases it will be obtained from the tensor square
of some graph algebra, by
taking the tensor product over a particular subalgebra (not over the complex
numbers).
The multiplication is the natural
one, namely: $(a_1 \otimes b_1)\times (a_2 \otimes b_2) = a_1 a_2
\otimes b_1 b_2$, and we shall identify $a u \otimes b$ and $a \otimes
ub$, whenever $u$ belongs to
the particular subalgebra over which the tensor product is taken 
(we use the notation $\otimesdot$).
In other words we take the quotient of the tensor square of the
appropriate graph algebra by the two-sided ideal generated by
elements $0\otimes u - u \otimes 0$, where $0$ is the unit of the
graph algebra of $G$. In the cases of $D_{odd}$ and $E_{7}$, the above
construction has to be ``twisted'': some elements $a u \otimes b$
have to be identified with $a \otimes \rho(u) b$, but $\rho$ is not
the identity map.

In most cases, the graph algebra to be used
in the above construction is the graph algebra of  $G$ itself.
In the case of
the diagram $E_7$, however,
one has to use the graph algebra of $D_{10}$.
For the diagram $D_{2n+1}$ one has to use the graph algebra of
$A_{4n-1}$. For the diagram $D_{2n}$, elements of ${\cal H}_{Oc(D_{2n})}$
also involve  $2\times 2$ matrices.

In general, the elements $u$ that are used to define the appropriate
two-sided ideal belong to a subalgebra $U$ that
admits a complementary subspace
$P$ which is invariant by left and right $U$-multiplications (a 
general feature since the algebra $U$ is semi-simple).
This property implies that elements
of ${\cal H}_{Oc(G)}$ can be decomposed into linear combinations of
only four types of elements belonging to $0\otimesdot U$, $0\otimesdot
P$, $P\otimesdot 0$ and $P \otimesdot P$.

Following Ocneanu terminology, we call ``chiral left subalgebra'' or ``chiral
right subalgebra'' the subalgebras spanned by left or right generators ($\sigma_1 \otimes \sigma_0$ or $\sigma_0
\otimes \sigma_1$)
       and ``ambichiral'' the intersection of the chiral parts. Left and
       right subalgebra are respectively described on Ocneanu graphs by
fat continuous lines,
       and fat dashed lines. The thin lines (continuous or dashed) represent
       right or left cosets.

       \smallskip

       Warning:  A given algebra  ${\cal H}_{Oc(G)}$ is, in a
sense, already defined
       by its graph $Oc(G)$ since the later describes multiplication by
       the two chiral generators. What we do in this paper is to
propose, for all Dynkin diagram
       $G$, an explicit  realization of these algebras ${\cal
       H}_{Oc(G)}$, in terms of usual graph algebras. In turn,
       this realization allows us in a simple way to determine all the
toric matrices
       associated with a given diagram (see below).
         We stress the fact that the quantum graphs $Oc(G)$ are taken from
       \cite{Ocneanu:paths}, however the proposed realizations for the
       algebras ${\cal H}_{Oc(G)}$ are ours.

       \smallskip
       
The number of vertices of
$Oc(G)$ depends very much of the choice of $G$ itself (for instance
$Oc(E_6)$ contains $12$
points, $Oc(E_7)$ contains $17$ points, $Oc(E_8)$ contains $32$
points).

\subsubsection{Modular invariant partition functions and twisted
partition functions}

To every vertex $x$ of the Ocneanu graph $Oc(G)$ of the Dynkin diagram $G$, one
associates a particular ``toric matrix'' $ W_{x}$. 
These matrices are related to the study of paths on the Ocneanu
graphs: the matrix element $ ({W_{x}})_{i,j}$ of  $ W_{x}$ gives the 
number of independent  paths leaving the vertex $x$ of 
$Oc(G)$ and reaching the origin $0 \otimesdot 0$ of $Oc(G)$ after 
having performed $i$ essential steps 
(resp. $j$ essential steps) on the left (resp. right) chiral subgraphs.
These matrices have other uses and interpretations 
(in particular in terms of the cell calculus or in terms of
the ``chiral modular splitting'' (\cite{Ocneanu:MSRI})
but this will not be discussed here.

As written in the introduction, these toric matrices
were defined and obtained by Ocneanu (unpublished but advertised in
several conferences since 1995, for instance \cite{Ocneanu:Marseille}).
The article \cite{PetZub:Oc} gives closed formulae for the
determination of these objects, in the language of conformal field theory.
One of our purposes, in the present paper,
is to find them by another method, which consists in
a straightforward generalization of
the technique introduced in \cite{Coque:qtetra}. This method uses
explicitly our realization of the algebras ${\cal H}_{Oc(G)}$ in
terms of graph algebras.

Our first step is to compute the appropriate essential matrices
(those of the graph
associated with the graph algebra
involved in the previous step); generally, \ie not for $E_7$ or
$D_{odd}$, these are
the $r$ essential
matrices of the graph $G$ itself.  As discussed previously, they are
rectangular matrices $E_a$ of size
$(\kappa -1) \times r$.
We then construct ``reduced essential matrices'' $E_a^{red}$ by
keeping only those columns associated
with the subalgebra over which the tensor product is taken (\ie we
replace the other entries by $0$). These are again
rectangular matrices of
size $ (\kappa -1) \times r$.

We then define matrices $W[a,b]$ associated with elements $x =  a \otimes b$
of the algebra ${\cal H}_{Oc(G)}$ as square
matrices of size $(\kappa -1) \times (\kappa -1)$ by setting

$$
W[a,b] \doteq E_a . \widetilde{E_b}^{red} \equiv (E_a).
transpose(E_b^{red}) =
(E_a)^{red}. transpose(E_b^{red})
$$

The points $x$ of $Oc(G)$ are, in general, linear combinations of
elements of the type $a \otimesdot b$.
The toric matrices  $W_{x}$ associated with points $x = \sum a \otimesdot b$
of $Oc(G)$ are square matrices of size $(\kappa -1) \times (\kappa -1)$.
They are  obtained  by setting $$ W_{x}= \sum W[a,b] $$

In the case of $E_{7}$ and $D_{odd}$ the above construction should be
slightly twisted (see the relevant sections for details).

\smallskip

There are several ways to display the results: one possibility is to give
the collection of
all toric matrices
$W_{x} = W[a,b]$ with matrix elements $W[a,b](i,j)$, another one is to fix
$i$ and $j$ (with
$1\leq i,j\leq \kappa -1$)
and display the Ocneanu graph itself labeled by the entries
$W[a,b](i,j)$. For
physical reasons (at
least for traditional reasons)  we prefer to display the
corresponding (twisted)
partition functions: setting $\chi \doteq \{\chi_0, \chi_1, \chi_2, \ldots,
\chi_{\kappa -2}\}$, we
associate with $W_x = W[a,b]$
a partition function:
       $$Z_x \equiv Z[a,b] \doteq \overline{\chi} \, W[a,b] \, \chi .$$

       To ease the reading of the paper we  put all these partition functions in
       tables to be found at the end of the article. The matrix elements
       of all $W[a,b]$ are always positive integers. However, in order
       to display the results in tables, we had sometimes to group together
       several terms and introduce minus signs that will disappear if
       the sesquilinear forms are expanded.

These quantities can be interpreted in terms of twisted partition functions for
ADE boundary conformal field theories (see\footnote{This is also
discussed in a very recent preprint \cite{Otto}.}\cite{PetZub:Oc} and \cite{PetkovaZuber:bulk}).

The partition function $Z[0,0]$ associated with the origin $\sigma_0 \otimes
\sigma_0$ is the usual
modular invariant partition function of Itzykson, Capelli, Zuber. The
others are not modular invariant. We should remember, at that point, that
the representation of the modular group provided by the usual $S$ and $T$
matrices, in the representation of Verlinde-Hurwitz, is usually not effective:
for instance in the case of $E_{6}$, where $\kappa = 12$, on top of
relations  $S^4=(ST)^3=1$, one
gets $T^{4 \kappa = 48} = 1$ (and $T^s \neq 1$ for smaller powers of $T$). The
representation actually factorizes through a congruence subgroup of
$SL(2,\ZZ)$ and one obtains a representation of  $SL(2,\ZZ/48 \ZZ)$
(one can check that all the defining relations given in
\cite{CosteGannon:modular} are verified).

\eject

\subsubsection{Summary of notations}
\begin{itemize}

\item   $G$ is the chosen Dynkin diagram of type $ADE$. It has $r$
vertices. We also call $G$ the fusion algebra (graph algebra) of this
Dynkin diagram, when it exists.

\item  ${\cal G}$  is the adjacency matrix of $G$.

\item  ${\kappa }$ is the Coxeter number of $G$.

\item  $q$ is a primitive root of unity such that $q^{2\kappa}=1$.

\item  $A_{\kappa -1}$ is the graph of type $A$ with same Coxeter number
$\kappa$ as $G$.

\item  $N_{i}=(N_{i})^{j}_{k}$ are the fusion matrices for the graph algebra
(fusion algebra) of $A_{\kappa -1}$.

\item  $G_{a}=(G_{a})^{b}_{c}$ are the fusion matrices for the graph algebra
(fusion algebra) of $G$, when it exists.

\item $S_{G}$ is a $r \times r$ matrix that (in all cases but $E_{7}$
and $D_{odd}$) diagonalizes simultaneously the $r$ fusion matrices $G_{a}$
of the diagram $G$. When the diagram $G$ is of type $A$, we just call
it $S$.

\item  $E_{a} = (E_{a})^{i}_{b}$ are the essential matrices for the graph
$G$.

\item  $F_{i} = (F_{i})^{a}_{b} \doteq (E_{a})^{i}_{b}$ provide a
representation of the graph algebra of $A_{\kappa -1}$.
Matrices $F_i$ (or $E_a$) describe the couplings between vertices
$a,b$ of $G$ and the vertex $i$ of $A_{\kappa -1}$.

\item  $Oc(G)$ is the Ocneanu graph associated with $G$.

\item  $\Sigma_{x} = (\Sigma_{x})^{a}_{b}$ are matrices describing the dual
couplings between vertices $a,b$ of $G$  and the vertex $x$ of
the Ocneanu graph $Oc(G)$.

\item  $W_{x} = (W_{x})^{i}_{j}$ are the toric matrices (of size $(\kappa
-1)\times (\kappa -1)$) associated with the vertices of $Oc(G)$.

\item  $Z_{x}$ is the twisted partition function associated with $W_{x}$.

\end{itemize}


\section{The $A_n$ cases}

\subsection{$A_4$}
The $A_4$ Dynkin diagram and its adjacency matrix are displayed below,
where we use the following order for the basis: $\{ \tau_0, \tau_1,
\tau_2, \tau_3 \}$.

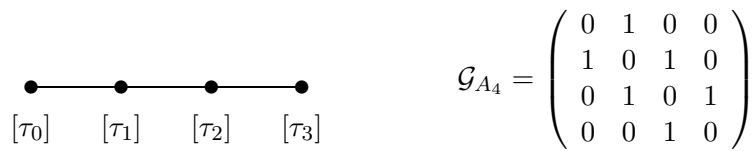
\begin{figure}[hhh]
\unitlength 0.8mm
\begin{center}
\begin{picture}(55,20)(0,10)
\put(5,10){\line(1,0){45}}
\multiput(5,10)(15,0){4}{\circle*{2}}
\put(5,3){\makebox(0,0){[$\tau_{0}$]}}
\put(20,3){\makebox(0,0){[$\tau_{1}$]}}
\put(35,3){\makebox(0,0){[$\tau_{2}$]}}
\put(50,3){\makebox(0,0){[$\tau_{3}$]}}
\end{picture}
\qquad \qquad
$
{\cal G}_{A_4} =
\left( \begin{array}{cccc}
     0 & 1 & 0 & 0 \\
     1 & 0 & 1 & 0  \\
     0 & 1 & 0 & 1   \\
     0 & 0 & 1 & 0   \\
\end{array}
\right)
$
\caption{The $A_4$ Dynkin diagram and its adjacency matrix}
\label{grA4}
\end{center}
\end{figure}

Here $\kappa = 5$ and the norm of the graph is the golden number
$\beta = 2 \cos
(\frac{\pi}{5}) = \frac{1+\sqrt 5}{2}$,
and the normalized Perron-Frobenius vector is
$D = \left( [1]_q, [2]_q, [2]_q, [1]_q \right) $. \\
The $A_4$ Dynkin diagram determines in a unique way the graph
algebra of $A_4$, whose multiplication table is displayed below.

\begin{table}[hhh]
$$
\begin{array}{|c||c|c|c|c|}
\hline
{} & 0 & 1 & 2 & 3 \\
\hline
\hline
0 & 0 & 1 & 2 & 3 \\
1 & 1 & 0+2 & 1+3 & 2 \\
2 & 2 & 1+3 & 0+2 & 1 \\
3 & 3 & 2 & 1 & 0 \\
\hline
\end{array}
$$
\caption{Multiplication table for the $A_4$ graph algebra}
\end{table}
The fusion matrices $N_{i}$ are given by the following polynomials:
$$
\begin{array}{ll}
N_0  = Id_4 \mathrm{\;(the \; identity \; matrix)} \qquad \qquad&
N_2  = N_1 . N_1 - N_0 \\
N_1  = {\cal G}_{A_4} &
N_3  = N_1 . N_1 . N_1 - 2 . N_1 \\
\end{array}
$$
They provide a faithful realization of the fusion algebra $A_4$.
In the chosen basis, they read:
\scriptsize
$$
N_0 = \left( \begin{array}{cccc}
     1 & 0 & 0 & 0 \\
     0 & 1 & 0 & 0 \\
     0 & 0 & 1 & 0 \\
     0 & 0 & 0 & 1 \\
\end{array}
\right)
\quad
N_1 = \left( \begin{array}{cccc}
     0 & 1 & 0 & 0 \\
     1 & 0 & 1 & 0 \\
     0 & 1 & 0 & 1 \\
     0 & 0 & 1 & 0 \\
\end{array}
\right)
\quad
N_2 = \left( \begin{array}{cccc}
     0 & 0 & 1 & 0 \\
     0 & 1 & 0 & 1 \\
     1 & 0 & 1 & 0 \\
     0 & 1 & 0 & 0 \\
\end{array}
\right)
\quad
N_3 = \left( \begin{array}{cccc}
     0 & 0 & 0 & 1 \\
     0 & 0 & 1 & 0 \\
     0 & 1 & 0 & 0 \\
     1 & 0 & 0 & 0 \\
\end{array}
\right)
$$

\normalsize

We form the tensor product $A_4 \otimes A_4$, whose dimension is 16, but
we take it over $A_4$. The Ocneanu algebra
of $A_4$  can be realized as the algebra of dimension 4 defined by:
$$
{\cal H}_{Oc(A_4)} = A_4 \otimesdot A_4 \doteq
\frac{A_4 \otimes A_4 }{A_4} = A_4 \otimes_{A_4} A_4.
$$
It is spanned by a basis with 4 elements:
$$
\ud0 = 0 \otimesdot 0, \qquad \qquad  \ud1 = 1 \otimesdot 0, \qquad \qquad
\ud2 = 2 \otimesdot 0, \qquad \qquad  \ud3 = 3 \otimesdot 0, \qquad \qquad
$$
and is isomorphic to the graph algebra $A_4$ itself.
For this reason, the Ocneanu graph $Oc(A_4)$ is the same as
the Dynkin diagram $A_4$.
Its elements are of the kind $m \otimesdot n = 0 \otimesdot mn = mn
\otimesdot 0$.
The dimensions $d_{n}$, with $n$ in $(0,1,2,3)$,
for the four blocks of the Racah-Wigner-Ocneanu bi-algebra $\cal A$
endowed with its first multiplicative law are respectively:
$(4,6,6,4).$
For its other multiplicative law (convolution), the dimensions $d_x$
of the four blocks, labeled with $x$ in the list ($0 \otimesdot 0, 1
\otimesdot 0, 2 \otimesdot 0, 3 \otimesdot 0$) are also respectively:
$(4,6,6,4)$.\\
We have of course $\sum d_{n} = \sum d_{x} = 20$ and $\sum d_{n}^2 = \sum
d_{n}^2 = 104$ but this observation is trivial in that case.

In the $A_4$ case (as in all $A_{n}$ cases) the essential matrices
$E_i$ happen to be the same as
the $N_i$ matrices.
The four toric matrices $W_{ab}$ of the $A_4$ model are also equal
to the $N_i$ matrices, $W_{00} = N_0$ being the  modular invariant.
We write them as sesquilinear forms (the twisted partition functions
given in the appendix).

\subsection{$A_n$}
We display below the Dynkin diagram of $A_n$, for $n > 4$.
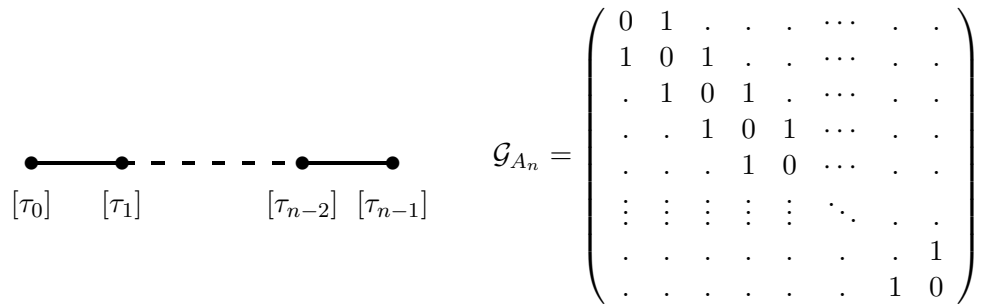
\begin{figure}[hhh]
\unitlength 0.8mm
\begin{center}
\begin{picture}(90,20)(0,10)
\thinlines
\multiput(15,10)(15,0){2}{\circle*{2}}
\multiput(60,10)(15,0){2}{\circle*{2}}
\thicklines
\multiput(30,10)(5,0){6}{\line(1,0){2}}
\put(15,10){\line(1,0){15}}
\put(60,10){\line(1,0){15}}
\put(15,3){\makebox(0,0){[$\tau_0$]}}
\put(30,3){\makebox(0,0){[$\tau_1$]}}
\put(60,3){\makebox(0,0){[$\tau_{n-2}$]}}
\put(75,3){\makebox(0,0){[$\tau_{n-1}$]}}
\end{picture}
$
{\cal G}_{A_n} =
\left( \begin{array}{cccccccc}
     0 & 1 & . & . & . & \cdots & . & .  \\
     1 & 0 & 1 & . & . & \cdots & . & .  \\
     . & 1 & 0 & 1 & . & \cdots & . & .  \\
     . & . & 1 & 0 & 1 & \cdots & . & .  \\
     . & . & . & 1 & 0 & \cdots & . & .  \\
     \vdots & \vdots & \vdots & \vdots & \vdots & \ddots & . & .  \\
     . & . & . & . & . & . & . & 1  \\
     . & . & . & . & . & . & 1 & 0  \\
\end{array}
\right)
$
\caption{The $A_n$ Dynkin diagram and its adjacency matrix}
\label{grAn}
\end{center}
\end{figure}

In all $A_n$ cases, the graph algebra is completely determined,
in a unique way, by the data of the corresponding Dynkin diagram.
The Ocneanu algebra of $A_n$ can be realized as:
$$
{\cal H}_{Oc(A_n)} = A_n \otimesdot A_n \doteq A_{n} \otimes_{A_n} \, A_n,
$$
and appears to be isomorphic to the graph algebra $A_n$ itself.
Due to this fact,
which occurs only in the $A_n$ cases, the Ocneanu graphs are also
equal to the corresponding Dynkin diagrams.
The fusion matrices $N_{i}$ are given by the following polynomials:
\begin{eqnarray*}
N_0 &=& Id_n      \\
N_1 &=& {\cal G}_{A_n}  \\
N_2 &=& N_1 . N_1 - N_0 \\
\vdots &=& \vdots \\
N_i &=& N_{i-1}.N_1 - N_{i-2}
\end{eqnarray*}
The essential matrices, as well as the $n$ toric matrices of the $A_n$
model are equal to these fusion matrices.
We just give the modular-invariant in sesquilinear form:
$$
A_n: \qquad Z_{\ud0} = \sum_{i=0}^{n} \xa{n} \qquad \forall n \geq 3
$$
It is easy to see that, for $A_{n}$, the dimensions $d_p$ of the blocks, for
$p$ from 0 to $n-1$ are given by $d_{p} = (p+1)(n-p)$.

\pagebreak

\section{The $E_{6}$ case}
The $E_6$ diagram and its adjacency matrix are displayed below. We
use the following order for the vertices: $\{\sigma_0, \sigma_1,
\sigma_2,
\sigma_5, \sigma_4, \sigma_3\}$.

\begin{figure}[hhh]
\unitlength 0.8mm
\begin{center}
\begin{picture}(80,30)(0,10)
\thinlines
\multiput(15,10)(15,0){5}{\circle*{2}}
\put(45,25){\circle*{2}}
\thicklines
\put(15,10){\line(1,0){60}}
\put(45,10){\line(0,1){15}}
\put(15,3){\makebox(0,0){$[\sigma_0]$}}
\put(30,3){\makebox(0,0){$[\sigma_1]$}}
\put(45,3){\makebox(0,0){$[\sigma_2]$}}
\put(60,3){\makebox(0,0){$[\sigma_5]$}}
\put(75,3){\makebox(0,0){$[\sigma_4]$}}
\put(51,25){\makebox(0,0){$[\sigma_3]$}}
\end{picture}
\qquad \qquad
$
{\cal G}_{E_6} =
\left( \begin{array}{cccccc}
      0 & 1 & 0 & 0 & 0 & 0  \\
      1 & 0 & 1 & 0 & 0 & 0  \\
      0 & 1 & 0 & 1 & 0 & 1  \\
      0 & 0 & 1 & 0 & 1 & 0  \\
      0 & 0 & 0 & 1 & 0 & 0  \\
      0 & 0 & 1 & 0 & 0 & 0  \\
\end{array}
\right)
$
\label{grE6}
\end{center}
\caption{The $E_6$ Dynkin diagram and its adjacency matrix}
\end{figure}
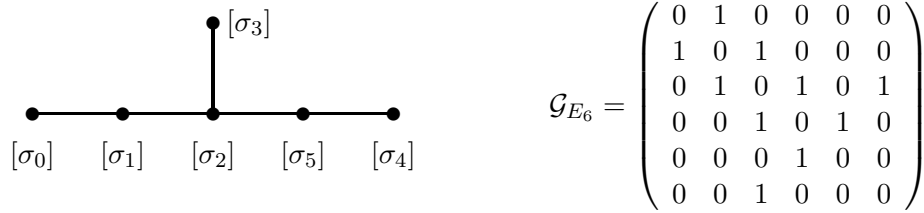

Here $\kappa = 12$, the norm of the graph is $\beta = 2 \cos
(\frac{\pi}{12}) = \frac{1 + \sqrt 3 }{\sqrt 2}$
and the normalized Perron-Frobenius vector is
$D = \left( [1]_q, [2]_q, [3]_q, [2]_q, [1]_q, \frac{[3]_q}{[2]_q}
\right)$.\\
The $E_6$ Dynkin diagram determines in a unique way the multiplication
table for the graph algebra of $E_6$, displayed below.

\begin{table}[hhh]
$$
\begin{array}{||c||c|c|c|c|c|c||}
\hline
{}& 0 & 1  & 2 & 5 & 4 & 3  \\
\hline
\hline
0 & 0  & 1      & 2       & 5     & 4 & 3    \\
1 & 1  & 0+2    & 1+3+5   & 2+4   & 5 & 2    \\
2 & 2  & 1+3+5  & 0+2+2+4 & 1+3+5 & 2 & 1+5  \\
5 & 5  & 2+4    & 1+3+5   & 0+2   & 1 & 2    \\
4 & 4  & 5      & 2       & 1     & 0 & 3    \\
3 & 3  & 2      & 1+5     & 2     & 3 & 0+4  \\
\hline
\end{array}
$$
\caption{Multiplication table for the graph algebra of $E_6$}
\end{table}

The fusion matrices $G_{i}$ are given by the following polynomials:
$$
\begin{array}{ll}
G_0 = Id_6  &
G_4 = G_1.G_1.G_1.G_1 - 4 G_1.G_1 + 2 G_0 \\
G_1 = {\cal G}_{E_6} &
G_5 = G_1.G_4 \\
G_2 = G_1.G_1 - G_0 \qquad \qquad \qquad &
G_3 =  - G_1.(G_4 - G_1.G_1 + 2 G_0)
\end{array}
$$

\begin{figure}[hhh]
\unitlength 0.7mm
\begin{center}
$
E_0 =
\left( \begin{array}{cccccc}
        1 & . & . & . & . & . \cr . &
        1 & . & . & . & . \cr . & . &
        1 & . & . & . \cr . & . & . &
        1 & . & 1 \cr . & . & 1 & . &
        1 & . \cr . & 1 & . & 1 & . &
        . \cr 1 & . & 1 & . & . & . \cr
        . & 1 & . & . & . & 1 \cr . &
        . & 1 & . & . & . \cr . & . &
        . & 1 & . & . \cr . & . & . &
        . & 1 & . \end{array} \right)
$
\begin{picture}(95,35)
\thinlines
\multiput(25,10)(15,0){5}{\circle*{2}}
\put(55,25){\circle*{2}}
\put(25,12){$\ast$}
\thicklines
\put(25,10){\line(1,0){60}}
\put(55,10){\line(0,1){15}}

\put(25,3){\makebox(0,0){$0$}}
\put(25,-2){\makebox(0,0){$6$}}

\put(40,3){\makebox(0,0){$1$}}
\put(40,-2){\makebox(0,0){$5$}}
\put(40,-7){\makebox(0,0){$7$}}

\put(55,3){\makebox(0,0){$2$}}
\put(55,-2){\makebox(0,0){$4$}}
\put(55,-7){\makebox(0,0){$6$}}
\put(55,-12){\makebox(0,0){$8$}}

\put(70,3){\makebox(0,0){$3$}}
\put(70,-2){\makebox(0,0){$5$}}
\put(70,-7){\makebox(0,0){$9$}}

\put(85,3){\makebox(0,0){$4$}}
\put(85,-2){\makebox(0,0){$10$}}

\put(63,27){\makebox(0,0){$3,7$}}
\end{picture}
\bigskip
\caption{Essential matrix $E_0$ and Essential Paths from the vertex 0
for the $E_6$-model}
\label{E6:E0}
\end{center}
\end{figure}
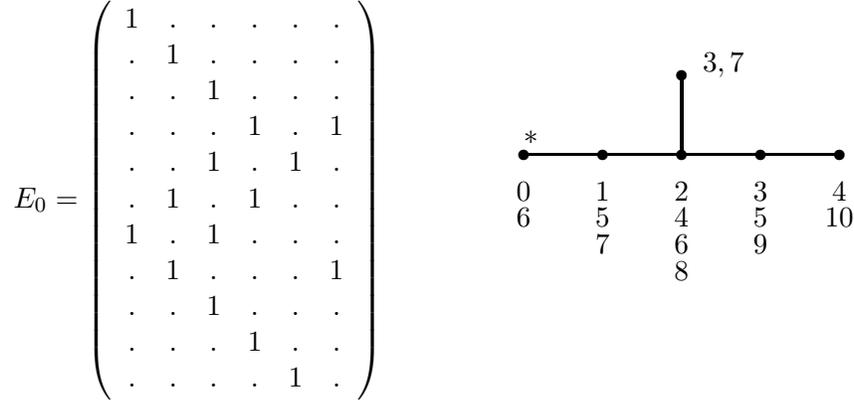

Essential matrices have $6$ columns and $11$ rows. They are labeled
by vertices of diagrams  $E_{6}$ and $A_{11}$.
They are calculated as explained in section 2.2.3.
With the order chosen for vertices $(012543)$ notice that  the first
row of matrix $E_5$, for example, is $E_5(0) = (000100)$.
The first essential matrix $E_{0}$ (essential paths leaving the
origin) is given in Fig \ref{E6:E0}, together with the corresponding
induction-restriction graph ($E_6$ diagram with vertices labeled by
$A_{11}$ vertices).

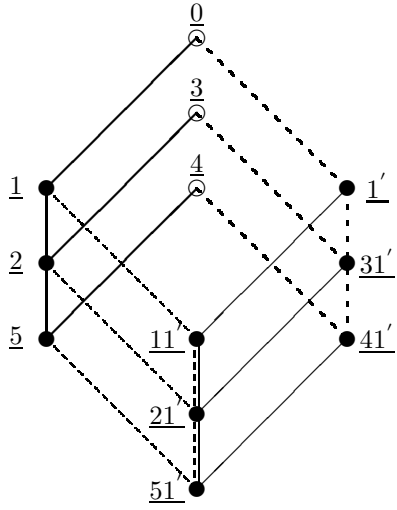
\begin{figure}[hhh]
\unitlength 1.0mm
\begin{center}
\begin{picture}(50,70)
\multiput(25,5)(0,10){3}{\circle*{2}}
\multiput(25,45)(0,10){3}{\circle{2}}
\multiput(5,25)(0,10){3}{\circle*{2}}
\multiput(45,25)(0,10){3}{\circle*{2}}

\thicklines
\put(5,45){\line(1,1){20}}
\put(5,35){\line(1,1){20}}
\put(5,25){\line(1,1){20}}
\put(5,25){\line(0,1){20}}

\thinlines
\put(45,45){\line(-1,-1){20}}
\put(45,35){\line(-1,-1){20}}
\put(45,25){\line(-1,-1){20}}
\put(25.3,5){\line(0,1){20}}

\thicklines
\dashline[50]{1}(45,45)(25,65)
\dashline[50]{1}(45,35)(25,55)
\dashline[50]{1}(45,25)(25,45)
\dashline[50]{1}(45,25)(45,45)

\thinlines
\dashline[50]{1}(5,45)(25,25)
\dashline[50]{1}(5,35)(25,15)
\dashline[50]{1}(5,25)(25,5)
\dashline[50]{1}(24.7,5)(24.7,25)

\small
\put(25,68){\makebox(0,0){\ud0}}
\put(25,58){\makebox(0,0){\ud3}}
\put(25,48){\makebox(0,0){\ud4}}
\put(1,45){\makebox(0,0){\ud1}}
\put(1,35){\makebox(0,0){\ud2}}
\put(1,25){\makebox(0,0){\ud5}}
\put(49,45){\makebox(0,0){$\ud{1^{'}}$}}
\put(49,35){\makebox(0,0){$\ud{31^{'}}$}}
\put(49,25){\makebox(0,0){$\ud{41^{'}}$}}
\put(21,25){\makebox(0,0){$\ud{11^{'}}$}}
\put(21,15){\makebox(0,0){$\ud{21^{'}}$}}
\put(21,5){\makebox(0,0){$\ud{51^{'}}$}}
\normalsize

\end{picture}
\caption{The $E_6$ Ocneanu graph}
\label{grocE6}
\end{center}
\end{figure}

The subspace $A_3$ generated by the elements $\{ 0, 3, 4 \}$ is a
subalgebra of the graph
algebra of $E_6$ and leaves invariant (by multiplication) the
complementary vector subspace generated by $\{ 1, 2, 5 \}$.
In other words the subalgebra $A_{3}$ of $E_{6}$ admits a two-sided
$A_{3}$-invariant complement.
We form the tensor product $E_6 \otimes E_6$, but we take it over
the subalgebra $A_3$ and define the following algebra:
$$
{\cal H}_{Oc(E_6)} = E_6 \otimesdot E_6 \doteq
\frac{E_6 \otimes E_6}{A_3} = E_6 \otimes_{A_3} E_6 .
$$
We have, for example, $3 \otimesdot 1 = 0 \otimesdot 31 = 0 \otimesdot
2$, and $4 \otimesdot 1 = 0 \otimesdot 41 = 0 \otimesdot 5$.\\
${\cal H}_{Oc(E_6)}$ is spanned by a basis with 12 elements:
$$
\begin{array}{lclcrcr}
\ud0 = 0 \otimesdot 0, & \qquad \qquad  & \ud3 = 3 \otimesdot 0, &
\qquad \qquad &
\ud{1^{'}} = 0 \otimesdot 1,  & \qquad \qquad & \ud{31^{'}}=  3
\otimesdot 1, \\
\ud1 = 1 \otimesdot 0, & \qquad & \ud4 = 4 \otimesdot 0, & \qquad &
\ud{11^{'}}=  1 \otimesdot 1, & \qquad & \ud{41^{'}}=  4 \otimesdot 1, \\
\ud2 = 2 \otimesdot 0, & \qquad & \ud5 = 5 \otimesdot 0, & \qquad &
\ud{21^{'}}=  2 \otimesdot 1, & \qquad & \ud{51^{'}}=  5 \otimesdot 1.
\end{array}
$$
The element $ 0 \otimesdot 0$ is the identity. The elements $1 \otimesdot
0$ and $0 \otimesdot 1$ are respectively the chiral left and right
generators; they span separately two subalgebras $E_{6}\otimes 0$ and
$0 \otimes E_{6}$, both isomorphic with the graph algebra itself.
    The ambichiral part is the linear span of $\{\ud0, \ud3, \ud4 \}$.
We can easily check that multiplication by generators of
${\cal H}_{Oc(E_6)}$ is indeed encoded by the Ocneanu graph
of $E_6$, represented in Fig \ref{grocE6}. The full lines encode
multiplication by the chiral left generator $\ud1$. For example:
$\ud1 . \ud2 = \ud1 + \ud3 + \ud5$ and in the $E_6$ Ocneanu graph the
vertices $\ud1, \ud3$ and $\ud5$ are joined to the vertex $\ud2$ by a
full line.
The dashed lines encode multiplication by the chiral right generator
$\ud{1^{'}}$. For example: $\ud{1^{'}} . \ud4 = \ud{41^{'}}$ and in
the $E_6$ Ocneanu graph the vertices $\ud4$ and $\ud{41^{'}}$ are
joined  by a dashed line.

The dimensions $d_{n}$, with $n$ in $(0,1,2,\ldots 10)$,
for the eleven blocks of the Racah-Wigner-Ocneanu
bi-algebra ${\cal A}$ endowed with its first multiplicative law are
respectively
$$
(6,10,14,18,20,20,20,18,14,10,6)
$$
For its other multiplicative law (convolution), the dimensions
$d_{x}$ of the twelve blocks, labeled with $x$ in the list
$(0\otimesdot 0, 3\otimesdot 0, 4\otimesdot 0,
1\otimesdot 0,2\otimesdot 0,5 \otimesdot 0,
0 \otimesdot1,0 \otimesdot 2,0 \otimesdot 5,
1 \otimesdot 1,2 \otimesdot 1,5 \otimesdot 1)$ are respectively
$$
(6,8,6,10,14,10,10,14,10,20,28,20)
$$
Notice that
$\sum {d_{n}} = \sum {d_{x}} = 156$ and $\sum {d_{n}^{2}} =
\sum {d_{x}^{2}} = 2512 .$

The twelve toric matrices $W_{ab}$ of the $E_{6}$ model are obtained
as explained in section 2.2.5;  for instance $W_{4\otimesdot 1}=
E_{4}.\widetilde{E_{1}}^{red}$.
We recall only the matrix expression
of $W_{00}$ (the modular invariant itself).  The eleven other
matrices\footnote{They were already given in \cite{Coque:qtetra}.}, are
written as sesquilinear forms  in
the appendix (they are the twisted partition functions).

$$
W_{00}=\left( \begin{array}{ccccccccccc}
1 & . & . & . & . & . & 1 & . & . & . & . \\
. & . & . & . & . & . & . & . & . & . & . \\
. & . & . & . & . & . & . & . & . & . & . \\
. & . & . & 1 & . & . & . & 1 & . & . & . \\
. & . & . & . & 1 & . & . & . & . & . & 1 \\
. & . & . & . & . & . & . & . & . & . & . \\
1 & . & . & . & . & . & 1 & . & . & . & . \\
. & . & . & 1 & . & . & . & 1 & . & . & . \\
. & . & . & . & . & . & . & . & . & . & . \\
. & . & . & . & . & . & . & . & . & . & . \\
. & . & . & . & 1 & . & . & . & . & . & 1 \\
\end{array}
\right)
$$



\section{The $E_{8}$ case}

The $E_8$ Dynkin diagram and its adjacency matrix are displayed
below. We use the following order for the vertices: $\{\sigma_0,
\sigma_1, \sigma_2,
\sigma_3, \sigma_4, \sigma_7, \sigma_6, \sigma_5 \}$.

\begin{figure}[hhh]
\unitlength 0.7mm
\begin{center}
\begin{picture}(120,30)(0,10)
\thinlines
\multiput(15,10)(15,0){7}{\circle*{2}}
\put(75,25){\circle*{2}}
\thicklines
\put(15,10){\line(1,0){90}}
\put(75,10){\line(0,1){15}}
\put(15,3){\makebox(0,0){[$\sigma_0$]}}
\put(30,3){\makebox(0,0){[$\sigma_1$]}}
\put(45,3){\makebox(0,0){[$\sigma_2$]}}
\put(60,3){\makebox(0,0){[$\sigma_3$]}}
\put(75,3){\makebox(0,0){[$\sigma_4$]}}
\put(90,3){\makebox(0,0){[$\sigma_7$]}}
\put(105,3){\makebox(0,0){[$\sigma_6$]}}
\put(81,25){\makebox(0,0){[$\sigma_5$]}}
\end{picture}
\footnotesize
$
{\cal G}_{E_8} =
\left( \begin{array}{cccccccc}
      0 & 1 & 0 & 0 & 0 & 0 & 0 & 0 \\
      1 & 0 & 1 & 0 & 0 & 0 & 0 & 0 \\
      0 & 1 & 0 & 1 & 0 & 0 & 0 & 0 \\
      0 & 0 & 1 & 0 & 1 & 0 & 0 & 0 \\
      0 & 0 & 0 & 1 & 0 & 1 & 0 & 1 \\
      0 & 0 & 0 & 0 & 1 & 0 & 1 & 0 \\
      0 & 0 & 0 & 0 & 0 & 1 & 0 & 0 \\
      0 & 0 & 0 & 0 & 1 & 0 & 0 & 0 \\
\end{array}
\right)
$
\normalsize
\caption{The $E_8$ Dynkin diagram and its adjacency matrix}
\label{grE8}
\end{center}
\end{figure}
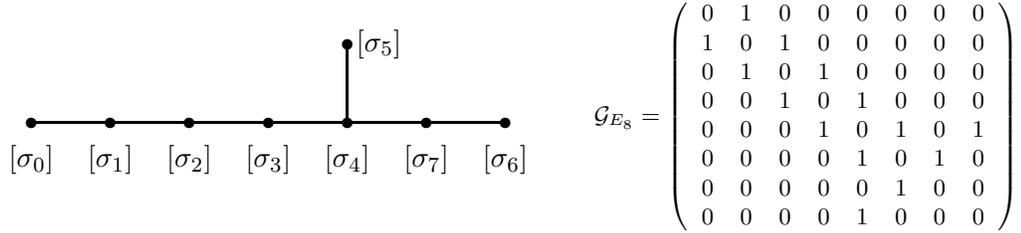
Here $\kappa = 30$, the norm of the graph is $\beta = 2 \cos
(\frac{\pi}{30})$ and the normalized Perron-Frobenius vector is
$D = \left( [1]_q, [2]_q, [3]_q, [4]_q, [5]_q, \frac{[7]_q}{[2]_q} ,
\frac{[5]_q}{[3]_q},
\frac{[5]_q}{[2]_q} \right)$.  \\
As for the $E_6$ case, the $E_8$ Dynkin diagram determines in a
unique way the multiplication table for the graph algebra of
$E_8$, displayed below.

\begin{table}[hhh]
\scriptsize
$$
\begin{array}{||c||c|c|c|c|c|c|c|c||}
\hline
{}& 0 & 1  & 2 & 3 & 4 & 7 & 6 & 5 \\
\hline
\hline
0 & 0 & 1     & 2       & 3           &   4         &  7        & 6
& 5       \\
1 & 1 & 0+2   & 1+3     & 2+4         & 3+5+7       & 4+6       & 7
& 4       \\
2 & 2 & 1+3   & 0+2+4   & 1+3+5+7     & 2+4_2+6     & 3+5+7     & 4
& 3+7     \\
3 & 3 & 2+4   & 1+3+5+7 & 0+2+4_2+6   & 1+3_2+5+7_2 & 2+4_2     & 3+5
& 2+4+6   \\
4 & 4 & 3+5+7 & 2+4_2+6 & 1+3_2+5+7_2 & 0+2_2+4_3+6 & 1+3_2+5+7 & 2+4
& 1+3+5+7 \\
7 & 7 & 4+6   & 3+5+7   & 2+4+4       & 1+3_2+5+7   & 0+2+4+6   & 1+7
& 2+4     \\
6 & 6 & 7     & 4       & 3+5         & 2+4         & 1+7       & 0+6
& 3       \\
5 & 5 & 4     & 3+7     & 2+4+6       & 1+3+5+7     & 2+4       & 3
& 0+4     \\
\hline
\end{array}
$$
\normalsize
\caption{Multiplication table for the graph algebra of $E_8$}
\end{table}

The fusion matrices $G_{i}$ are given by the following polynomials:
$$
\begin{array}{ll}
G_0 = Id_8   &
G_4 = G_1.G_1.G_1.G_1 - 3 G_1.G_1 + G_0 \\
G_1 = {\cal G}_{E_8} &
G_6 = G_2.G_4 - G_2 - 2 G_4 \\
G_2 = G_1.G_1 - G_0 &
G_7 = G_1.G_6 \\
G_3 = G_1.G_1.G_1 - 2 G_1 \qquad \qquad &
G_5 = G_2.G_7 - G_3 - G_7
\end{array}
$$

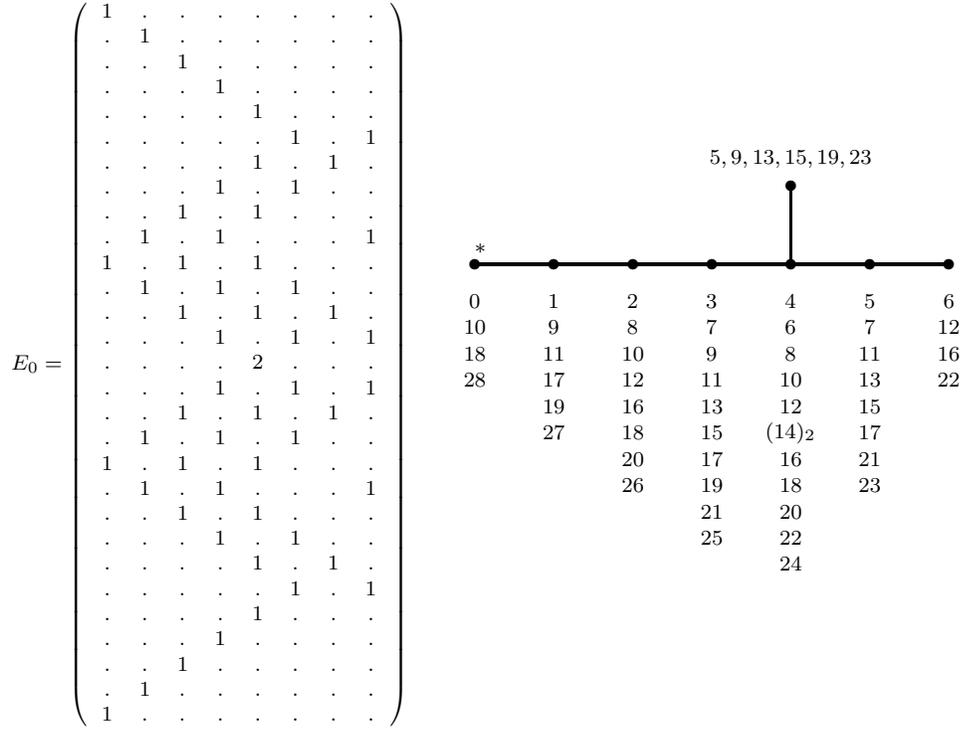
\begin{figure}[hhh]
\unitlength 0.7mm
\begin{center}
\scriptsize
$
E_0 =
\left(
\begin{array}{cccccccc}
1& .& .& .& .& .& .& .\\
.& 1& .& .& .& .& .& .\\
.& .& 1& .& .& .& .& .\\
.& .& .& 1& .& .& .& .\\
.& .& .& .& 1& .& .& .\\
.& .& .& .& .& 1& .& 1\\
.& .& .& .& 1& .& 1& .\\
.& .& .& 1& .& 1& .& .\\
.& .& 1& .& 1& .& .& .\\
.& 1& .& 1& .& .& .& 1\\
1& .& 1& .& 1& .& .& .\\
.& 1& .& 1& .& 1& .& .\\
.& .& 1& .& 1& .& 1& .\\
.& .& .& 1& .& 1& .& 1\\
.& .& .& .& 2& .& .& .\\
.& .& .& 1& .& 1& .& 1\\
.& .& 1& .& 1& .& 1& .\\
.& 1& .& 1& .& 1& .& .\\
1& .& 1& .& 1& .& .& .\\
.& 1& .& 1& .& .& .& 1\\
.& .& 1& .& 1& .& .& .\\
.& .& .& 1& .& 1& .& .\\
.& .& .& .& 1& .& 1& .\\
.& .& .& .& .& 1& .& 1\\
.& .& .& .& 1& .& .& .\\
.& .& .& 1& .& .& .& .\\
.& .& 1& .& .& .& .& .\\
.& 1& .& .& .& .& .& .\\
1& .& .& .& .& .& .& .
\end{array}
\right)
\normalsize
\qquad
$
\begin{picture}(90,40)
\thinlines
\multiput(0,20)(15,0){7}{\circle*{2}}
\put(60,35){\circle*{2}}
\put(0,22){$\ast$}
\thicklines
\put(0,20){\line(1,0){90}}
\put(60,20){\line(0,1){15}}

\put(0,13){\makebox(0,0){$0$}}
\put(0,8){\makebox(0,0){$10$}}
\put(0,3){\makebox(0,0){$18$}}
\put(0,-2){\makebox(0,0){$28$}}

\put(15,13){\makebox(0,0){$1$}}
\put(15,8){\makebox(0,0){$9$}}
\put(15,3){\makebox(0,0){$11$}}
\put(15,-2){\makebox(0,0){$17$}}
\put(15,-7){\makebox(0,0){$19$}}
\put(15,-12){\makebox(0,0){$27$}}

\put(30,13){\makebox(0,0){$2$}}
\put(30,8){\makebox(0,0){$8$}}
\put(30,3){\makebox(0,0){$10$}}
\put(30,-2){\makebox(0,0){$12$}}
\put(30,-7){\makebox(0,0){$16$}}
\put(30,-12){\makebox(0,0){$18$}}
\put(30,-17){\makebox(0,0){$20$}}
\put(30,-22){\makebox(0,0){$26$}}

\put(45,13){\makebox(0,0){$3$}}
\put(45,8){\makebox(0,0){$7$}}
\put(45,3){\makebox(0,0){$9$}}
\put(45,-2){\makebox(0,0){$11$}}
\put(45,-7){\makebox(0,0){$13$}}
\put(45,-12){\makebox(0,0){$15$}}
\put(45,-17){\makebox(0,0){$17$}}
\put(45,-22){\makebox(0,0){$19$}}
\put(45,-27){\makebox(0,0){$21$}}
\put(45,-32){\makebox(0,0){$25$}}

\put(60,13){\makebox(0,0){$4$}}
\put(60,8){\makebox(0,0){$6$}}
\put(60,3){\makebox(0,0){$8$}}
\put(60,-2){\makebox(0,0){$10$}}
\put(60,-7){\makebox(0,0){$12$}}
\put(60,-12){\makebox(0,0){$(14)_2$}}
\put(60,-17){\makebox(0,0){$16$}}
\put(60,-22){\makebox(0,0){$18$}}
\put(60,-27){\makebox(0,0){$20$}}
\put(60,-32){\makebox(0,0){$22$}}
\put(60,-37){\makebox(0,0){$24$}}

\put(75,13){\makebox(0,0){$5$}}
\put(75,8){\makebox(0,0){$7$}}
\put(75,3){\makebox(0,0){$11$}}
\put(75,-2){\makebox(0,0){$13$}}
\put(75,-7){\makebox(0,0){$15$}}
\put(75,-12){\makebox(0,0){$17$}}
\put(75,-17){\makebox(0,0){$21$}}
\put(75,-22){\makebox(0,0){$23$}}

\put(90,13){\makebox(0,0){$6$}}
\put(90,8){\makebox(0,0){$12$}}
\put(90,3){\makebox(0,0){$16$}}
\put(90,-2){\makebox(0,0){$22$}}

\put(60,40){\makebox(0,0){$5,9,13,15,19,23$}}

\end{picture}
\bigskip
\caption{Essential matrix $E_0$ and Essential Paths from the vertex 0
for the $E_8$-model}
\label{E8:E0}
\end{center}
\end{figure}

Essential matrices have $8$ columns and $29$ rows. They are labeled
by vertices of diagrams  $E_{8}$ and $A_{29}$.
The first essential matrix $E_{0}$ (essential paths leaving the
origin) is given in Fig \ref{E8:E0}, together with the corresponding
induction-restriction graph ($E_8$ diagram with vertices labeled by
$A_{29}$ vertices).

The subspace $A_2$ generated by the elements $\{ 0, 6 \}$ is a
subalgebra of the graph algebra of $E_8$ that admits a two-sided
$A_2$-invariant complement.
We form the tensor product $E_8 \otimes E_8$, but we take it over
the subalgebra $A_2$. The Ocneanu algebra of $E_8$ can be realized as:
$$
{\cal H}_{Oc(E_8)} = E_8 \otimesdot E_8 \doteq
\frac{E_8 \otimes E_8}{A_2} = E_8 \otimes_{A_2} E_8.
$$
For instance $6 \otimesdot 0 = 0 \otimesdot 6$, $6 \otimesdot 1 = 0 \otimesdot
7$, $6 \otimesdot 2 = 0 \otimesdot 4$, $6 \otimesdot 5 = 0 \otimesdot
3$.\\
${\cal H}_{Oc(E_8)}$ is spanned by a basis with 32 elements:
$$
\begin{array}{cccc}
\ud0 =0 \otimesdot 0, & \qquad\qquad \ud{1^{'}}  =0 \otimesdot 1, &
\qquad\qquad \ud{2^{'}}  =0 \otimesdot 2, & \qquad\qquad \ud{5^{'}} =0
\otimesdot 5,\\
\ud1 =1 \otimesdot 0, & \qquad\qquad \ud{11^{'}} =1 \otimesdot 1, &
\qquad\qquad \ud{12^{'}} =1 \otimesdot 2, & \qquad\qquad \ud{15^{'}}=1
\otimesdot 5,\\
\ud2 =2 \otimesdot 0, & \qquad\qquad \ud{21^{'}} =2 \otimesdot 1, &
\qquad\qquad \ud{22^{'}} =2 \otimesdot 2, & \qquad\qquad \ud{25^{'}}=2
\otimesdot 5,\\
\ud3 =3 \otimesdot 0, & \qquad\qquad \ud{31^{'}} =3 \otimesdot 1, &
\qquad\qquad \ud{32^{'}} =3 \otimesdot 2, & \qquad\qquad \ud{35^{'}}=3
\otimesdot 5,\\
\ud4 =4 \otimesdot 0, & \qquad\qquad \ud{41^{'}} =4 \otimesdot 1, &
\qquad\qquad \ud{42^{'}} =4 \otimesdot 2, & \qquad\qquad \ud{45^{'}}=4
\otimesdot 5,\\
\ud5 =5 \otimesdot 0, & \qquad\qquad \ud{51^{'}} =5 \otimesdot 1, &
\qquad\qquad \ud{52^{'}} =5 \otimesdot 2, & \qquad\qquad \ud{55^{'}}=5
\otimesdot 5,\\
\ud6 =6 \otimesdot 0, & \qquad\qquad \ud{61^{'}} =6 \otimesdot 1, &
\qquad\qquad \ud{62^{'}} =6 \otimesdot 2, & \qquad\qquad \ud{65^{'}}=6
\otimesdot 5,\\
\ud7 =7 \otimesdot 0, & \qquad\qquad \ud{71^{'}} =7 \otimesdot 1, &
\qquad\qquad \ud{72^{'}} =7 \otimesdot 2, & \qquad\qquad \ud{75^{'}}=7
\otimesdot 5.\\
\end{array}
$$

\begin{figure}[hhh]
\unitlength 1.0mm
\begin{center}
\begin{picture}(70,80)
\multiput(5,35)(0,10){2}{\circle*{2}}
\multiput(15,25)(0,10){4}{\circle*{2}}
\multiput(25,15)(0,10){6}{\circle*{2}}
\multiput(35,5)(0,10){6}{\circle*{2}}
\multiput(35,65)(0,10){2}{\circle{2}}
\multiput(45,15)(0,10){6}{\circle*{2}}
\multiput(55,25)(0,10){4}{\circle*{2}}
\multiput(65,35)(0,10){2}{\circle*{2}}

\thicklines
\put(35,75){\line(-1,-1){20}}
\put(5,35){\line(1,1){30}}
\put(5,35){\line(1,2){10}}
\put(5,45){\line(1,0){10}}

\thinlines
\put(45,65){\line(-1,-1){20}}
\put(15,25){\line(1,1){30}}
\put(15,25){\line(1,2){10}}
\put(15,35){\line(1,0){10}}

\put(55,55){\line(-1,-1){20}}
\put(25,15){\line(1,1){30}}
\put(25,15){\line(1,2){10}}
\put(25,25){\line(1,0){10}}

\put(65,45){\line(-1,-1){20}}
\put(35,5){\line(1,1){30}}
\put(35,5){\line(1,2){10}}
\put(35,15){\line(1,0){10}}

\thicklines
\dashline[50]{1}(35,75)(55,55)
\dashline[50]{1}(65,35)(35,65)
\dashline[50]{1}(65,35)(55,55)
\dashline[50]{1}(65,45)(55,45)

\thinlines
\dashline[50]{1}(25,65)(45,45)
\dashline[50]{1}(55,25)(25,55)
\dashline[50]{1}(55,25)(45,45)
\dashline[50]{1}(55,35)(45,35)

\dashline[50]{1}(15,55)(35,35)
\dashline[50]{1}(45,15)(15,45)
\dashline[50]{1}(45,15)(35,35)
\dashline[50]{1}(45,25)(35,25)

\dashline[50]{1}(5,45)(25,25)
\dashline[50]{1}(35,5)(5,35)
\dashline[50]{1}(35,5)(25,25)
\dashline[50]{1}(35,15)(25,15)

\scriptsize
\put(35,78){\makebox(0,0){0}}
\put(35,68){\makebox(0,0){6}}
\put(35,59){\makebox(0,0){$11^{'}$}}
\put(35,49){\makebox(0,0){$71^{'}$}}
\put(35,39){\makebox(0,0){$22^{'}$}}
\put(35,29){\makebox(0,0){$42^{'}$}}
\put(35,19){\makebox(0,0){$55^{'}$}}
\put(35.5,10){\makebox(0,0){$35^{'}$}}

\put(25,68){\makebox(0,0){1}}
\put(25,58){\makebox(0,0){7}}
\put(25,49){\makebox(0,0){$21^{'}$}}
\put(25,39){\makebox(0,0){$41^{'}$}}
\put(25,29){\makebox(0,0){$52^{'}$}}
\put(24.5,19){\makebox(0,0){$32^{'}$}}

\put(46,69){\makebox(0,0){$1^{'}$}}
\put(45,59){\makebox(0,0){$61^{'}$}}
\put(45,49){\makebox(0,0){$12^{'}$}}
\put(45,39){\makebox(0,0){$72^{'}$}}
\put(45,29){\makebox(0,0){$25^{'}$}}
\put(46,19){\makebox(0,0){$45^{'}$}}

\put(15,58){\makebox(0,0){2}}
\put(15,48){\makebox(0,0){4}}
\put(15,39){\makebox(0,0){$51^{'}$}}
\put(15,29){\makebox(0,0){$31^{'}$}}

\put(56,59){\makebox(0,0){$2^{'}$}}
\put(55,49){\makebox(0,0){$62^{'}$}}
\put(55,39){\makebox(0,0){$15^{'}$}}
\put(56,29){\makebox(0,0){$75^{'}$}}

\put(5,48){\makebox(0,0){5}}
\put(5,38){\makebox(0,0){3}}

\put(66,49){\makebox(0,0){$5^{'}$}}
\put(66.5,39){\makebox(0,0){$65^{'}$}}

\normalsize
\end{picture}
\caption{The $E_8$ Ocneanu graph}
\label{grocE8}
\end{center}
\end{figure}
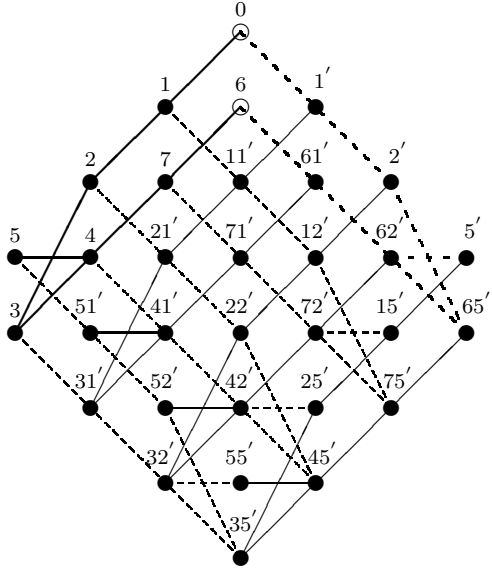

The element $ 0 \otimesdot 0$ is the identity. The elements $1 \otimesdot
0$ and $0 \otimesdot 1$ are respectively the chiral left and right
generators; they span independently the subalgebras $E_{8}\otimes 0$
and $0 \otimes E_{8}$.
One can easily check that
multiplication by these two generators is indeed encoded by the Ocneanu
graph of $E_8$, represented in Fig \ref{grocE8}. Full lines (resp.
dashed lines) encode multiplication by the chiral left (resp. chiral right)
generator.
The  ambichiral part is the linear span of $\{\ud0, \ud6 \}$.

The dimensions $d_{n}$, with $n$ in $(0,1,2,\ldots 28)$,
for the twenty nine blocks of the Racah-Wigner-Ocneanu
bi-algebra ${\cal A}$ endowed with its first multiplicative law are
respectively:
\small
$$
(8,14,20,26,32,38,44,48,52,56,60,62,64,64,64,64,64,62,60,56,52,48,44,38,32,26,20,14,8).
$$
\normalsize
For its other multiplicative law (convolution), the dimensions
$d_{x}$ of the thirty two blocks, labeled with $x$ in the list
$
    (0 \otimesdot 0 ,1 \otimesdot 0 ,2 \otimesdot 0 ,3 \otimesdot 0 ,4
\otimesdot 0 ,5 \otimesdot 0
		,6 \otimesdot 0 ,7 \otimesdot 0 ,0 \otimesdot 1 ,1
\otimesdot 1 ,2 \otimesdot 1 ,3 \otimesdot 1 ,4 \otimesdot 1 ,5
\otimesdot 1
		,6 \otimesdot 1 ,7 \otimesdot 1 ,0 \otimesdot 2 ,1
\otimesdot 2 ,2 \otimesdot 2 ,3 \otimesdot 2 ,4 \otimesdot 2 ,5
\otimesdot 2
		,6 \otimesdot 2 ,7 \otimesdot 2 ,0 \otimesdot 5 ,1
\otimesdot 5 ,2 \otimesdot 5 ,3 \otimesdot 5 ,4 \otimesdot 5 ,5
\otimesdot 5
		,6 \otimesdot 5 ,7 \otimesdot 5)
$
are respectively given by:
\footnotesize
$$
(8, 14, 20, 26, 32, 16, 12, 22, 14, 28, 40, 52, 64, 32, 22, 44, 20, 40,
         60, 78, 96, 48, 32, 64, 16, 32, 48, 64, 78, 40, 26, 52).
$$
\normalsize
Notice that
$\sum d_{n} = \sum d_{x} = 1240$ and $\sum d_{n}^{2} =
\sum d_{x}^{2} = 63136 .$

The thirty two toric matrices $W_{ab}$ of the $E_{8}$ model are obtained
as explained in section 2.2.5 ;  for instance $W_{\ud{52^{'}}} =
W_{5\otimesdot 2}=
E_{5}.\widetilde{E_{2}}^{red}$.
We recall only the matrix expression of
the modular invariant $W_{00}$.  The partition functions
corresponding to all the toric matrices are given as
sesquilinear forms in the appendix.


\scriptsize
$$
W_{00}=\left(
\begin{array}{ccccccccccccccccccccccccccccc}
1 &. &. &. &. &. &. &. &. &. &1 &. &. &. &. &. &. &. &1 &. &. &. &.
&. &. &. &. &. &1 \\
. &. &. &. &. &. &. &. &. &. &. &. &. &. &. &. &. &. &. &. &. &. &.
&. &. &. &. &. &. \\
. &. &. &. &. &. &. &. &. &. &. &. &. &. &. &. &. &. &. &. &. &. &.
&. &. &. &. &. &. \\
. &. &. &. &. &. &. &. &. &. &. &. &. &. &. &. &. &. &. &. &. &. &.
&. &. &. &. &. &. \\
. &. &. &. &. &. &. &. &. &. &. &. &. &. &. &. &. &. &. &. &. &. &.
&. &. &. &. &. &. \\
. &. &. &. &. &. &. &. &. &. &. &. &. &. &. &. &. &. &. &. &. &. &.
&. &. &. &. &. &. \\
. &. &. &. &. &. &1 &. &. &. &. &. &1 &. &. &. &1 &. &. &. &. &. &1
&. &. &. &. &. &. \\
. &. &. &. &. &. &. &. &. &. &. &. &. &. &. &. &. &. &. &. &. &. &.
&. &. &. &. &. &. \\
. &. &. &. &. &. &. &. &. &. &. &. &. &. &. &. &. &. &. &. &. &. &.
&. &. &. &. &. &. \\
. &. &. &. &. &. &. &. &. &. &. &. &. &. &. &. &. &. &. &. &. &. &.
&. &. &. &. &. &. \\
1 &. &. &. &. &. &. &. &. &. &1 &. &. &. &. &. &. &. &1 &. &. &. &.
&. &. &. &. &. &1 \\
. &. &. &. &. &. &. &. &. &. &. &. &. &. &. &. &. &. &. &. &. &. &.
&. &. &. &. &. &. \\
. &. &. &. &. &. &1 &. &. &. &. &. &1 &. &. &. &1 &. &. &. &. &. &1
&. &. &. &. &. &. \\
. &. &. &. &. &. &. &. &. &. &. &. &. &. &. &. &. &. &. &. &. &. &.
&. &. &. &. &. &. \\
. &. &. &. &. &. &. &. &. &. &. &. &. &. &. &. &. &. &. &. &. &. &.
&. &. &. &. &. &. \\
. &. &. &. &. &. &. &. &. &. &. &. &. &. &. &. &. &. &. &. &. &. &.
&. &. &. &. &. &. \\
. &. &. &. &. &. &1 &. &. &. &. &. &1 &. &. &. &1 &. &. &. &. &. &1
&. &. &. &. &. &. \\
. &. &. &. &. &. &. &. &. &. &. &. &. &. &. &. &. &. &. &. &. &. &.
&. &. &. &. &. &. \\
1 &. &. &. &. &. &. &. &. &. &1 &. &. &. &. &. &. &. &1 &. &. &. &.
&. &. &. &. &. &1 \\
. &. &. &. &. &. &. &. &. &. &. &. &. &. &. &. &. &. &. &. &. &. &.
&. &. &. &. &. &. \\
. &. &. &. &. &. &. &. &. &. &. &. &. &. &. &. &. &. &. &. &. &. &.
&. &. &. &. &. &. \\
. &. &. &. &. &. &. &. &. &. &. &. &. &. &. &. &. &. &. &. &. &. &.
&. &. &. &. &. &. \\
. &. &. &. &. &. &1 &. &. &. &. &. &1 &. &. &. &1 &. &. &. &. &. &1
&. &. &. &. &. &. \\
. &. &. &. &. &. &. &. &. &. &. &. &. &. &. &. &. &. &. &. &. &. &.
&. &. &. &. &. &. \\
. &. &. &. &. &. &. &. &. &. &. &. &. &. &. &. &. &. &. &. &. &. &.
&. &. &. &. &. &. \\
. &. &. &. &. &. &. &. &. &. &. &. &. &. &. &. &. &. &. &. &. &. &.
&. &. &. &. &. &. \\
. &. &. &. &. &. &. &. &. &. &. &. &. &. &. &. &. &. &. &. &. &. &.
&. &. &. &. &. &. \\
. &. &. &. &. &. &. &. &. &. &. &. &. &. &. &. &. &. &. &. &. &. &.
&. &. &. &. &. &. \\
1 &. &. &. &. &. &. &. &. &. &1 &. &. &. &. &. &. &. &1 &. &. &. &.
&. &. &. &. &. &1 \\
\end{array}
\right)
$$
\normalsize


\section{The $D_{even}$ case}

General formulae valid for all cases of this family are a bit heavy\ldots
We therefore only provide a detailed treatment of the cases $D_{4}$
and $D_{6}$ but
generalization is straightforward.

\subsection{The $D_4$ case}

The $D_4$ diagram and its adjacency matrix are displayed below. We
use the following order for the vertices: $\{\sigma_0, \sigma_1,
\sigma_2, \sigma_{2^{' }} \}$.

\begin{figure}[hhh]
\unitlength 0.8mm
\begin{center}
\begin{picture}(45,20)(0,10)
\thinlines
\multiput(5,10)(15,0){2}{\circle*{2}}
\put(35,17,5){\circle*{2}}
\put(35,2,5){\circle*{2}}
\thicklines
\put(5,10){\line(1,0){15}}
\put(20,10){\line(2,1){15}}
\put(20,10){\line(2,-1){15}}
\put(5,5){\makebox(0,0){[$\sigma_0$]}}
\put(20,5){\makebox(0,0){[$\sigma_1$]}}
\put(40,18){\makebox(0,0){[$\sigma_2$]}}
\put(41,3){\makebox(0,0){[$\sigma_{2^{'}}$]}}
\end{picture}
\qquad \qquad \qquad
$
{\cal G}_{D_4} =
\left( \begin{array}{cccc}
      0 & 1 & 0 & 0   \\
      1 & 0 & 1 & 1   \\
      0 & 1 & 0 & 0   \\
      0 & 1 & 0 & 0   \\
\end{array}
\right)
$
\caption{The $D_4$ Dynkin diagram and its adjacency matrix}
\label{grD4}
\end{center}
\end{figure}
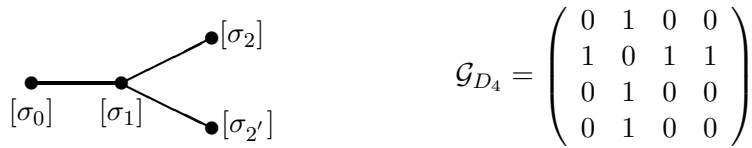
Here $\kappa = 6$, the norm of the graph is $\beta = 2 \cos
(\frac{\pi}{6}) = \sqrt 3$ and the normalized Perron-Frobenius vector is
$D = \left( [1]_q, [2]_q, \frac{[2]_q}{[2]_q} = 1,
\frac{[2]_q}{[2]_q}=1 \right)$.\\
For the $D_4$ case (as for all the $D_{2n}$ cases), we have to impose that
the structure constants of its graph algebra should be positive
integers, in order for the Dynkin diagram to determine in a unique way the
multiplication table of the graph algebra, displayed below.

\begin{table}[hhh]
$$
\begin{array}{||c||c|c|c|c||}
\hline
{}& 0 & 1  & 2 & 2^{'}   \\
\hline
\hline
0     & 0     & 1         & 2     & 2^{'}  \\
1     & 1     & 0+2+2^{'} & 1     & 1      \\
2     & 2     & 1         & 2^{'} & 0      \\
2^{'} & 2^{'} & 1         & 0     & 2      \\
\hline
\end{array}
$$
\caption{Multiplication table for the graph algebra of $D_4$ }
\end{table}
The fusion matrices $G_{i}$ are given by the following polynomials:
$$
G_0 = Id_4
\qquad \qquad \qquad
G_1 = {\cal G}_{D_4} \qquad \qquad \qquad
G_2 + G_{2^{'}} = G_1.G_1 - G_0
$$
Imposing that entries of $G_2$ and $G_{2^{'}}$ should be positive
integers leads to a unique solution (up to $G_2 \leftrightarrow
G_{2^{'}}$), namely:
$$
G_2 = \left( \begin{array}{cccc}
0 & 0 & 0 & 1 \\
0 & 1 & 0 & 0 \\
1 & 0 & 0 & 0 \\
0 & 0 & 1 & 0 \\
\end{array}
\right)
\qquad \qquad \qquad
G_2^{'} = \left( \begin{array}{cccc}
0 & 0 & 1 & 0 \\
0 & 1 & 0 & 0 \\
0 & 0 & 0 & 1 \\
1 & 0 & 0 & 0 \\
\end{array}
\right)
$$

\begin{figure}[hhh]
\unitlength 0.7mm
\begin{center}
$
E_0 =
\left(
\begin{array}{cccc}
1 & . & . & . \cr
. & 1 & . & . \cr
. & . & 1 & 1 \cr
. & 1 & . & . \cr
1 & . & . & . \cr
\end{array}
\right)
$
\qquad \qquad
\unitlength 1.0mm
\begin{picture}(45,20)(0,10)
\thinlines
\multiput(5,10)(15,0){2}{\circle*{1.5}}
\put(35,17,5){\circle*{1.5}}
\put(35,2,5){\circle*{1.5}}
\put(5,12){$\ast$}
\thinlines
\put(5,10){\line(1,0){15}}
\put(20,10){\line(2,1){15}}
\put(20,10){\line(2,-1){15}}
\put(5,5){\makebox(0,0){0}}
\put(5,0){\makebox(0,0){4}}
\put(20,5){\makebox(0,0){1}}
\put(20,0){\makebox(0,0){3}}
\put(35,13){\makebox(0,0){2}}
\put(35,-2){\makebox(0,0){2}}
\end{picture}
\bigskip
\caption{Essential matrix $E_0$ and Essential Paths from the vertex 0
for the $D_4$ model}
\label{D4:E0}
\end{center}
\end{figure}
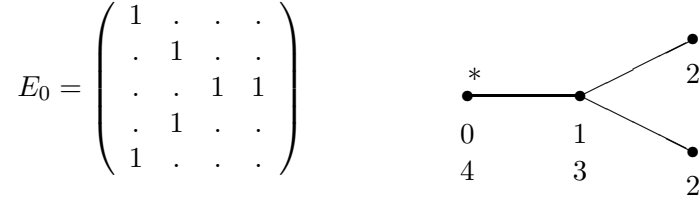

Essential matrices have 4 columns and 5 rows. They are labeled
by vertices of diagrams $D_4$ and $A_{5}$. The first essential
matrix $E_0$ is given in Fig \ref{D4:E0}, together with the corresponding
induction-restriction graph ($D_4$ diagram with vertices labeled
by $A_5$ vertices).

The subspace $J_3$ generated by the elements $\{ 0, 2, 2^{'} \}$, is a
subalgebra of the graph algebra of $D_4$ that admits a two-sided
$J_3$-invariant complement.
We first form the tensor product $D_4 \otimes D_4$, but we take it over
the subalgebra $J_3$.
We get the algebra $D_4^{\otimesdot} = D_4 \otimesdot D_4 \doteq
D_4 \otimes_{J_3} D_4$,
spanned by a basis with 6 elements:
$$
0 \otimesdot 0, \qquad \quad 1 \otimesdot 0, \qquad \quad  2 \otimesdot 0,
\qquad \quad  2^{'} \otimesdot 0, \qquad \quad 0 \otimesdot 1, \qquad \quad
1 \otimesdot 1.
$$
The Ocneanu algebra of $D_4$, ${\cal H}_{Oc(D_4)}$, can be realized as
a subalgebra of dimension 8 of the following non-commutative algebra:
$$
D_4^{\otimesdot} \; \oplus \; M(2,\mathbb{C})
$$
The 8 elements of the basis are given by:
$$
\begin{array}{cccclcc}
\underline{0} = 0     \otimesdot 0 &+&
\left( \begin{array}{cc} 1 & 0 \\ 0 & 1 \\ \end{array} \right)
& \qquad \qquad \qquad \qquad&
\; \underline{\epsilon} \; = \frac{1}{3} (1 \otimesdot 1) &+&
\theta \left( \begin{array}{cc} 0 & 1 \\ 1 & 0 \\ \end{array} \right)\\
\underline{1}         = 1     \otimesdot 0 &+&
\left( \begin{array}{cc} 0 & 0 \\ 0 & 0 \\ \end{array} \right)
& {} &
\underline{1\epsilon} = \quad \, 0     \otimesdot 1 &+&
\; \; \left( \begin{array}{cc} 0 & 0 \\ 0 & 0 \\ \end{array} \right) \\
\underline{2}         = 2     \otimesdot 0 &+&
\left( \begin{array}{cc} \alpha & 0 \\ 0 & \beta \\ \end{array} \right)
& {} &
\underline{2\epsilon} = \frac{1}{3} (1 \otimesdot 1) &+&
\theta \left( \begin{array}{cc} 0 & \alpha \\ \beta & 0 \\
\end{array} \right) \\
\underline{2^{'}}     = 2^{'} \otimesdot 0 &+&
\left( \begin{array}{cc} \beta & 0 \\ 0 & \alpha \\ \end{array} \right)
& {} &
\underline{2^{'}\epsilon} = \frac{1}{3}(1 \otimesdot 1) &+&
\theta \left( \begin{array}{cc} 0 & \beta \\ \alpha & 0 \\ \end{array} \right)
\end{array}
$$
where:
$$
\theta^2 = 0 \qquad \text{Grassmann parameter}, \qquad \qquad
\alpha   = \frac{-1+i\sqrt{3}}{2}, \qquad\qquad
\text{and} \qquad
\beta    = \frac{-1-i\sqrt{3}}{2}.
$$
The multiplication in this algebra is defined by:
$$
\left( (e1 \otimesdot f1) + A \right) . \left( (e2 \otimesdot f2) + B \right) =
      (e1.e2) \otimesdot (f1.f2)
      + A.B,
$$
where $ e_1, f_1, e_2, f_2 \in D_4^{\otimesdot}$ and $A, B \in M(2,
\mathbb{C})$.\\
The numbers $\alpha$ and $\beta$ are determined by the multiplication table
of ${\cal H}_{Oc(D_4)}$.
For example, the relations $\ud1 . \ud1 = \ud0 + \ud2 + \ud{2^{'}}$, $\ud2 .
\ud2 = \ud{2^{'}}$ and $\ud2 . \ud{2^{'}} = \ud0$ lead to the equations:
$\alpha + \beta = -1, \alpha . \beta = 1, \alpha^2=\beta$ and $\beta^2 =
\alpha$, that determines uniquely $\alpha$ and $\beta$.

\ud{Sketch of our construction}:
We first define $D_4^{\otimesdot}$ by quotienting the tensor square
of $D_{4}$ by the subalgebra $J_{3}$ that admits a two-sided
$J_3$-invariant complement.
   From the graph $Oc(D_{4})$ taken from \cite{Ocneanu:paths}, we see
that $\ud{1}$ and $\ud{1\epsilon}$ separately generate the left and
right subalgebras isomorphic with the graph algebra of $D_{4}$,
therefore we set $\ud{1} = 1 \otimesdot 0$ and $\ud{1\epsilon} = 0
\otimesdot 1$. We also see that $\ud{1} . \ud{\epsilon} =
\ud{1\epsilon}$; this equality implies that the $D_4^{\otimesdot}$ part of
$\ud{\epsilon}$ should be proportional to $1\otimesdot 1$ since
$(1 \otimesdot 0)(1 \otimesdot 1)=(0+2+2^{'})\otimesdot 1 = 3 (0 \otimesdot
1)$. The matrix part of ${\epsilon}$ and of the other generators (the
coefficients $\alpha$ and $\beta$) can then be determined
by imposing that the obtained multiplication table should
coincide with the multiplication table constructed from the Ocneanu
graph $Oc(D_{4})$. Such a construction can be generalized to all
$D_{even}$ cases.

\begin{figure}[hhh]
\unitlength 0.8mm
\begin{center}
\begin{picture}(120,70)
\multiput(25,5)(0,10){2}{\circle*{2}}
\multiput(25,25)(0,10){2}{\circle{2}}
\put(25,45){\circle*{2}}
\put(5,35){\circle*{2}}
\put(25,65){\circle{2}}
\put(45,35){\circle*{2}}

\thicklines
\put(5,35){\line(1,0){20}}
\put(5,35){\line(2,-1){20}}
\put(5,35){\line(2,3){20}}

\thinlines
\put(45,35){\line(-1,-1){20}}
\put(45,35){\line(-2,-3){20}}
\put(45,35){\line(-2,1){20}}

\thicklines
\dashline[50]{1}(45,35)(25,35)
\dashline[50]{1}(45,35)(25,65)
\dashline[50]{1}(45,35)(25,25)

\thinlines
\dashline[50]{1}(5,35)(25,45)
\dashline[50]{1}(5,35)(25,15)
\dashline[50]{1}(5,35)(25,5)

\scriptsize
\put(25,68){\makebox(0,0){$\ud{0}$}}
\put(25,48){\makebox(0,0){$\ud{\epsilon}$}}
\put(25,38){\makebox(0,0){$\ud{2}$}}
\put(25,29){\makebox(0,0){$\ud{2^{'}}$}}
\put(25,19){\makebox(0,0){$\ud{2\epsilon}$}}
\put(21,5){\makebox(0,0){$\ud{2^{'}\epsilon}$}}
\put(49,35){\makebox(0,0){$\ud{1\epsilon}$}}
\put(2,35){\makebox(0,0){$\ud{1}$}}
\normalsize
\put(100,35){\makebox(0,0){$
W_{00}=\left( \begin{array}{ccccc}
1 & . & . & . & 1 \\
. & . & . & . & . \\
. & . & 2 & . & . \\
. & . & . & . & . \\
1 & . & . & . & 1 \\
\end{array} \right)
$
}}
\end{picture}
\caption{The $D_4$ Ocneanu graph and the modular invariant matrix}
\label{grocD4}
\end{center}
\end{figure}
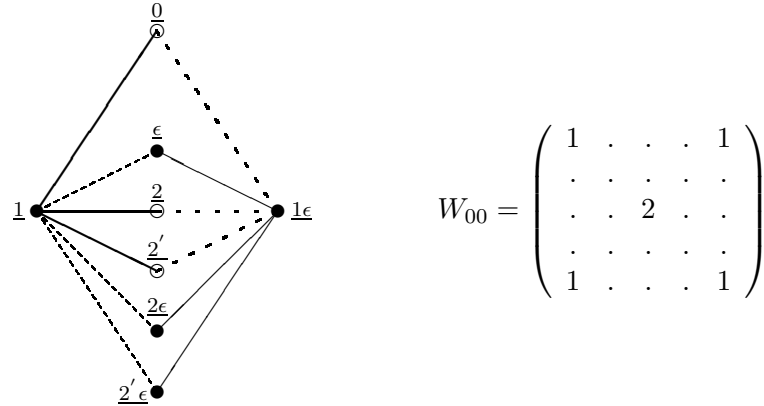

The element $\underline{0}$ is the identity. The elements
$\underline{1}$ and $\underline{1\epsilon}$ are respectively the
chiral left and right generators.
The multiplication table of this algebra is given in Table \ref{multOcd4},
and we can check that multiplication by the generators is indeed
encoded by the Ocneanu graph of $D_4$, represented in Fig \ref{grocD4}.
Warning: the table is not symmetric (the multiplication is not
commutative); for instance $\ud{2\epsilon} = \ud{2}.\ud{\epsilon}
\neq \ud{\epsilon}.\ud{2}$.
The  ambichiral part is the linear span of $\{\ud0, \ud{2}, \ud{2^{'}}  \}$.

\begin{table}[hhh]
$$
\begin{array}{|c||c|c|c|c|c|c|c|c|}
\hline
{}   & \ud0 & \ud1 & \ud{1\ep} & \ud{\ep} & \ud{2} & \ud{2^{'}} &
\ud{2\ep} & \ud{2^{'}\ep} \\
\hline
\hline
\ud0 & \ud0 & \ud1 & \ud{1\ep} & \ud{\ep} & \ud{2} & \ud{2^{'}} &
\ud{2\ep} & \ud{2^{'}\ep} \\
\ud1 & \ud1 & \ud0 + \ud2 + \ud{2^{'}} & \ud{\ep} + \ud{2\ep} +
\ud{2^{'}\ep} & \ud{1\ep} & \ud1 &
            \ud1 & \ud{1\ep} & \ud{1\ep} \\
\ud{1\ep} & \ud{1\ep} & \ud{\ep} + \ud{2\ep} + \ud{2^{'}\ep} & \ud0 +
\ud2 + \ud{2^{'}} & \ud1 &
            \ud{1\ep} & \ud{1\ep} & \ud1 & \ud1 \\
\ud{\ep} & \ud{\ep} & \ud{1\ep} & \ud1 & \ud{\eta} & \ud{2^{'}\ep} &
\ud{2\ep} & \ud{2^{'}\eta} &
            \ud{2\eta} \\
\ud2 & \ud2 & \ud1 & \ud{1\ep} & \ud{2\ep} & \ud{2^{'}} & \ud0 &
\ud{2^{'}\ep} & \ud{\ep} \\
\ud{2^{'}} & \ud{2^{'}} & \ud1 & \ud{1\ep} & \ud{2^{'}\ep} & \ud0 &
\ud2 & \ud{\ep} & \ud{2\ep} \\
\ud{2\ep} & \ud{2\ep} & \ud{1\ep} & \ud1 & \ud{2\eta} & \ud{\ep} &
\ud{2^{'}\ep} & \ud{\eta} &
            \ud{2^{'}\eta} \\
\ud{2^{'}\ep} & \ud{2^{'}\ep} & \ud{1\ep} & \ud1 & \ud{2^{'}\eta} &
\ud{2\ep} & \ud{\ep} &
            \ud{2\eta} & \ud{\eta} \\
\hline
\end{array}
$$
\qquad \qquad \qquad where $\eta = \frac{1}{3} (\ud0 + \ud2 + \ud{2^{'}})$
\caption{Multiplication table of the Ocneanu algebra of $D_4$}
\label{multOcd4}
\end{table}


The dimensions $d_{n}$, with $n = 0,1,2,3,4$
  are respectively  $(4,6,8,6,4)$.
We find that $\sum d_{n} = 28$ and $\sum d_{n}^{2} = 168.$

The eight toric matrices $W_{ab}$ of the $D_{4}$ model
and the corresponding partition functions
are obtained as usual. For instance $W_{\ud{\epsilon}} =
W_{\ud{2\epsilon}} = W_{\ud{2^{'}\epsilon}} = \frac{1}{3} E_{1} {\widetilde
E_{1}}^{red}.$
We recall the matrix expression of
the modular invariant $W_{00}$ and give the others toric matrices as
sesquilinear forms in the appendix\footnote{The toric matrices of $D_{4}$
were already published in \cite{PetZub:bcft}}.



\subsection{The $D_6$ case}

The $D_6$ Dynkin diagram and its adjacency matrix are displayed
below. We use the following order for the vertices: $\{\sigma_0,
\sigma_1, \sigma_2, \sigma_3, \sigma_4, \sigma_{4^{' }} \}$.

\begin{figure}[hhh]
\unitlength 0.8mm
\begin{center}
\begin{picture}(75,20)(0,10)
\thinlines
\multiput(5,10)(15,0){4}{\circle*{2}}
\put(65,17,5){\circle*{2}}
\put(65,2,5){\circle*{2}}
\thicklines
\put(5,10){\line(1,0){45}}
\put(50,10){\line(2,1){15}}
\put(50,10){\line(2,-1){15}}
\put(5,5){\makebox(0,0){[$\sigma_0$]}}
\put(20,5){\makebox(0,0){[$\sigma_1$]}}
\put(35,5){\makebox(0,0){[$\sigma_2$]}}
\put(50,5){\makebox(0,0){[$\sigma_3$]}}
\put(70,18){\makebox(0,0){[$\sigma_4$]}}
\put(71,3){\makebox(0,0){[$\sigma_4^{'}$]}}
\end{picture}
\qquad \qquad
$
{\cal G}_{D_6} =
\left( \begin{array}{cccccc}
      0 & 1 & 0 & 0 & 0 & 0   \\
      1 & 0 & 1 & 0 & 0 & 0   \\
      0 & 1 & 0 & 1 & 0 & 0   \\
      0 & 0 & 1 & 0 & 1 & 1   \\
      0 & 0 & 0 & 1 & 0 & 0   \\
      0 & 0 & 0 & 1 & 0 & 0   \\
\end{array}
\right)
$
\caption{The $D_6$ Dynkin diagram and its adjacency matrix}
\label{grD6}
\end{center}
\end{figure}
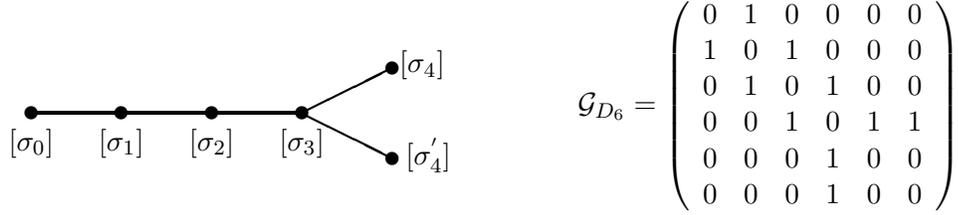
Here $\kappa = 10$, the norm of the graph is
$\beta = [2]_q = 2 \cos (\frac{\pi}{10})= \sqrt{\frac{5 + \sqrt
5}{2}}$ and the normalized Perron-Frobenius vector is
$D = \left( [1]_q, [2]_q, [3]_q, [4]_q, \frac{[4]_q}{[2]_q},
\frac{[4]_q}{[2]_q} \right)$.\\
Imposing positivity, the table of multiplication of the graph
algebra of $D_6$ is completely determined by its Dynkin diagram.

\begin{table}[hhh]
$$
\begin{array}{||c||c|c|c|c|c|c||}
\hline
{}& 0 & 1 & 2 & 3 & 4 & 4^{'}   \\
\hline
\hline
0     & 0     & 1         & 2           & 3             & 4       & 4^{'}    \\
1     & 1     & 0+2       & 1+3         & 2+4+4^{'}     & 3       & 3        \\
2     & 2     & 1+3       & 0+2+4+4^{'} & 1+3+3         & 2+4^{'} & 2+4      \\
3     & 3     & 2+4+4^{'} & 1+3+3       & 0+2+2+4+4^{'} & 1+3     & 1+3      \\
4     & 4     & 3         & 2+4^{'}     & 1+3           & 0+4     & 2        \\
4^{'} & 4^{'} & 3         & 2+4         & 1+3           & 2       & 0+4^{'}  \\
\hline
\end{array}
$$
\caption{Multiplication table for the graph algebra of $D_6$}
\end{table}
The fusion matrices $G_{i}$ are given by the following polynomials:
$$
\begin{array}{lll}
G_0 = Id_6 \qquad \qquad &   G_2 = G_1.G_1 - G_0 \qquad \qquad & G_4
+ G_{4^{'}} = G_1.G_3 - G_2 \\
G_1 = {\cal G}_{D_6}  & G_3 = G_2.G_1 - G_1 &  {}
\end{array}
$$
Imposing that entries of $G_4$ and $G_{4^{'}}$ should be positive
integers leads to a unique solution (up to $G_4 \leftrightarrow
G_{4^{'}}$), namely:
$$
G_4 = \left( \begin{array}{cccccc}
0 & 0 & 0 & 0 & 0 & 1 \\
0 & 0 & 0 & 1 & 0 & 0 \\
0 & 0 & 1 & 0 & 1 & 0 \\
0 & 1 & 0 & 1 & 0 & 0 \\
0 & 0 & 1 & 0 & 0 & 0 \\
1 & 0 & 0 & 0 & 0 & 1 \\
\end{array}
\right)
\qquad \qquad \qquad
G_4^{'} = \left( \begin{array}{cccccc}
0 & 0 & 0 & 0 & 1 & 0 \\
0 & 0 & 0 & 1 & 0 & 0 \\
0 & 0 & 1 & 0 & 0 & 1 \\
0 & 1 & 0 & 1 & 0 & 0 \\
1 & 0 & 0 & 0 & 1 & 0 \\
0 & 0 & 1 & 0 & 0 & 0 \\
\end{array}
\right)
$$

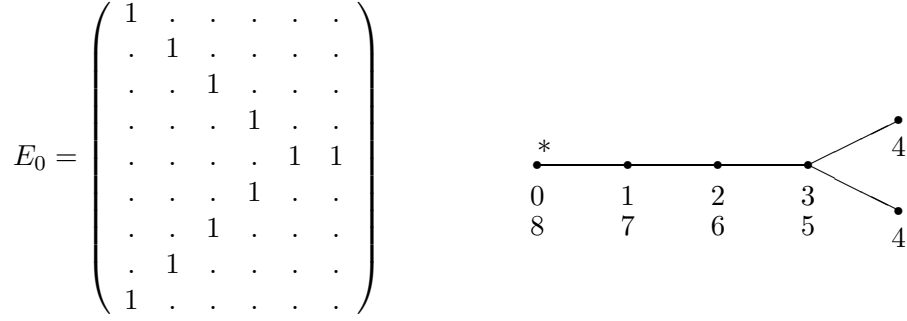
\begin{figure}[hhh]
\unitlength 0.7mm
\begin{center}
$
E_0 =
\left(
\begin{array}{cccccc}
1 & . & . & . & . & . \cr
. & 1 & . & . & . & . \cr
. & . & 1 & . & . & . \cr
. & . & . & 1 & . & . \cr
. & . & . & . & 1 & 1 \cr
. & . & . & 1 & . & . \cr
. & . & 1 & . & . & . \cr
. & 1 & . & . & . & . \cr
1 & . & . & . & . & . \cr
\end{array}
\right)
$
\qquad \qquad
\unitlength 0.8mm
\begin{picture}(75,20)(0,10)
\thinlines
\multiput(5,10)(15,0){4}{\circle*{1.5}}
\put(65,17,5){\circle*{1.5}}
\put(65,2,5){\circle*{1.5}}
\put(5,12){$\ast$}
\thinlines
\put(5,10){\line(1,0){45}}
\put(50,10){\line(2,1){15}}
\put(50,10){\line(2,-1){15}}
\put(5,5){\makebox(0,0){0}}
\put(5,0){\makebox(0,0){8}}
\put(20,5){\makebox(0,0){1}}
\put(20,0){\makebox(0,0){7}}
\put(35,5){\makebox(0,0){2}}
\put(35,0){\makebox(0,0){6}}
\put(50,5){\makebox(0,0){3}}
\put(50,0){\makebox(0,0){5}}
\put(65,13){\makebox(0,0){4}}
\put(65,-2){\makebox(0,0){4}}
\end{picture}
\bigskip
\caption{Essential matrix $E_0$ and Essential Paths from the vertex 0
for the $D_6$ model}
\label{D6:E0}
\end{center}
\end{figure}

Essential matrices have 6 columns and 9 rows. They are labeled
by vertices of diagrams $D_6$ and $A_9$. The first essential
matrix $E_0$ is given in Fig \ref{D6:E0}, together with the corresponding
induction-restriction graph ($D_6$ diagram with vertices labeled
by $A_9$ vertices).

\begin{figure}[hhh]
\unitlength 0.8mm
\begin{center}
\begin{picture}(160,110)
\multiput(25,5)(0,10){2}{\circle*{2}}
\multiput(25,25)(0,10){2}{\circle{2}}
\put(25,45){\circle*{2}}
\put(5,35){\circle*{2}}
\put(45,35){\circle*{2}}

\put(5,75){\circle*{2}}
\put(45,75){\circle*{2}}
\put(25,65){\circle{2}}
\put(25,85){\circle*{2}}
\put(25,105){\circle{2}}

\thicklines
\put(5,35){\line(1,0){20}}
\put(5,35){\line(2,-1){20}}
\put(5,35){\line(2,3){20}}
\put(5,75){\line(2,-1){20}}
\put(5,75){\line(2,3){20}}

\thinlines
\put(45,35){\line(-1,-1){20}}
\put(45,35){\line(-2,-3){20}}
\put(45,35){\line(-2,1){20}}
\put(45,75){\line(-2,1){20}}
\put(45,75){\line(-2,-3){20}}

\thicklines
\dashline[50]{1}(45,35)(25,35)
\dashline[50]{1}(45,35)(25,65)
\dashline[50]{1}(45,35)(25,25)
\dashline[50]{1}(45,75)(25,65)
\dashline[50]{1}(45,75)(25,105)

\thinlines
\dashline[50]{1}(5,35)(25,45)
\dashline[50]{1}(5,35)(25,15)
\dashline[50]{1}(5,35)(25,5)
\dashline[50]{1}(5,75)(25,85)
\dashline[50]{1}(5,75)(25,45)

\scriptsize
\put(25,109){\makebox(0,0){$\ud{0}$}}
\put(25,88){\makebox(0,0){$\ud{\epsilon}$}}
\put(2,75){\makebox(0,0){$\ud{1}$}}
\put(48,75){\makebox(0,0){$\ud{1\epsilon}$}}
\put(25,68){\makebox(0,0){$\ud{2}$}}
\put(25,49){\makebox(0,0){$\ud{2\epsilon}$}}
\put(25,38){\makebox(0,0){$\ud{4}$}}
\put(25,29){\makebox(0,0){$\ud{4^{'}}$}}
\put(25,19){\makebox(0,0){$\ud{4\epsilon}$}}
\put(21,5){\makebox(0,0){$\ud{4^{'}\epsilon}$}}
\put(48,35){\makebox(0,0){$\ud{3\epsilon}$}}
\put(2,35){\makebox(0,0){$\ud{3}$}}
\normalsize
\put(120,55){\makebox(0,0){
$
W_{00}=\left( \begin{array}{ccccccccc}
1 & . & . & . & . & . & . & . & 1 \\
. & . & . & . & . & . & . & . & . \\
. & . & 1 & . & . & . & 1 & . & . \\
. & . & . & . & . & . & . & . & . \\
. & . & . & . & 2 & . & . & . & . \\
. & . & . & . & . & . & . & . & . \\
. & . & 1 & . & . & . & 1 & . & . \\
. & . & . & . & . & . & . & . & . \\
1 & . & . & . & . & . & . & . & 1 \\
\end{array}
\right)
$}}
\end{picture}
\caption{The $D_6$ Ocneanu graph and the modular invariant matrix}
\label{grocD6}
\end{center}
\end{figure}
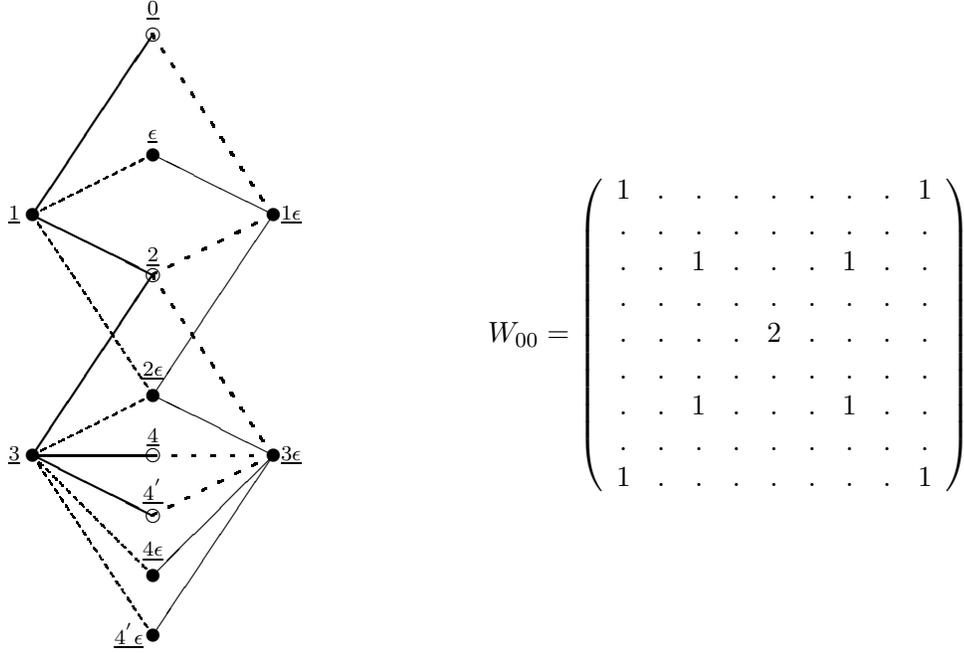

The subspace $J_4$ generated by the elements $\{ 0, 2, 4, 4^{'} \}$
is a subalgebra of the graph algebra of $D_6$
that admits a two-sided $J_4$-invariant complement.
We first form the tensor product $D_6 \otimes D_6$, but we take it over
the subalgebra $J_4$. We get the algebra $D_6^{\otimesdot} =
D_6 \otimesdot D_6 \doteq
D_6 \otimes_{J_4} D_6$, spanned by a basis with 10 elements:
\begin{eqnarray*}
&0 \otimesdot 0,& \qquad 1 \otimesdot 0, \qquad 2 \otimesdot 0, \qquad 3
\otimesdot 0, \qquad 4 \otimesdot 0, \\
&4^{'} \otimesdot 0,& \qquad 0 \otimesdot 1,\qquad 0 \otimesdot 3,\qquad 1
\otimesdot 1,\qquad 1 \otimesdot 3.
\end{eqnarray*}
The Ocneanu algebra of $D_6$, ${\cal H}_{Oc(D_6)}$, can be realized as
a subalgebra of dimension
12 of the following non-commutative algebra:
$$
D_6^{\otimesdot} \; \oplus \; M(2,\mathbb{C})
$$
The 12 elements of the basis are given by:
$$
\begin{array}{cccclcc}
\underline{0}         = 0     \otimesdot 0 &+&
\left( \begin{array}{cc} 1 & 0 \\ 0 & 1 \\ \end{array} \right)
& \qquad \qquad &
\underline{\epsilon}  = \frac{3}{5} (1 \otimesdot 1) - \frac{1}{5} (3
\otimesdot 1)                       &+&
\theta \left( \begin{array}{cc} 0 & 1 \\ 1 & 0 \\ \end{array} \right) \\
\underline{1}         = 1     \otimesdot 0 &+&
\left( \begin{array}{cc} 0 & 0 \\ 0 & 0 \\ \end{array} \right)
&{}&
\underline{1\epsilon} = 0     \otimesdot 1 &+&
\left( \begin{array}{cc} 0 & 0 \\ 0 & 0 \\ \end{array} \right) \\
\underline{2}         = 2     \otimesdot 0 &+&
\left( \begin{array}{cc} -1 & 0 \\ 0 & -1 \\ \end{array} \right)
&{}&
\underline{2\epsilon} = \frac{2}{5} (1 \otimesdot 1) + \frac{1}{5} (3
\otimesdot 1)                      &+&
\theta \left( \begin{array}{cc} 0 & -1 \\ -1 & 0 \\ \end{array} \right)\\
\underline{3}         = 3     \otimesdot 0 &+&
\left( \begin{array}{cc} 0 & 0 \\ 0 & 0 \\ \end{array} \right)
&{}&
\underline{3\epsilon} = 0     \otimesdot 3 &+&
\left( \begin{array}{cc} 0 & 0 \\ 0 & 0 \\ \end{array} \right)\\
\underline{4}         = 4     \otimesdot 0 &+&
\left( \begin{array}{cc} \alpha & 0 \\ 0 & \beta \\ \end{array} \right)
&{}&
\underline{4\epsilon} = -\frac{1}{5} (1 \otimesdot 1) + \frac{2}{5} (3
\otimesdot 1)                     &+&
\theta \left( \begin{array}{cc} 0 & \alpha \\ \beta & 0 \\
\end{array} \right) \\
\underline{4^{'}}     = 4^{'} \otimesdot 0 &+&
\left( \begin{array}{cc} \beta & 0 \\ 0 & \alpha \\ \end{array} \right)
&{}&
\underline{4^{'}\epsilon} = -\frac{1}{5} (1 \otimesdot 1) + \frac{2}{5}
(3 \otimesdot 1)                      &+&
\theta \left( \begin{array}{cc} 0 & \beta \\ \alpha & 0 \\ \end{array} \right)
\end{array}
$$
where:
$$
\theta^2 = 0, \qquad \qquad
\alpha   = \frac{-1+\sqrt{5}}{2}, \qquad \qquad
\text{and} \qquad
\beta    = \frac{-1-\sqrt{5}}{2}
$$
The element $\underline{0}$ is the identity. The elements
$\underline{1}$ and $\underline{1\epsilon}$ are respectively the
chiral left and right generators.
The multiplication by these generators is encoded by the Ocneanu
graph of $D_6$, represented in Fig \ref{grocD6}.
The ambichiral part is the linear span of $\{ \ud0, \ud{2}, \ud{4},
\ud{4^{'}} \}$

The dimensions $d_{n}$, with $n=0,1,2,\ldots,8$,
  are respectively $(6,10,14,16,18,16,14,10,6)$.
Therefore, $\sum {d_{n}} = 110$  and $\sum {d_{n}^{2}} = 1500.$

The twelve toric matrices $W_{ab}$ of the $D_{6}$ model
and the corresponding partition functions
are obtained as usual. For instance $W_{\ud{2 \epsilon}}=
\frac{2}{5} E_{1} {\widetilde E_{1}}^{red} + \frac{1}{5} E_{3}
{\widetilde E_{1}}^{red}.$
We recall the matrix expression of the modular invariant $W_{00}$ and
give the others as
sesquilinears forms in the appendix.


\subsection{The $D_{even}$ case}

In the case of $D_{2s}$, we first build $D_{2s}^{\otimesdot} =
D_{2s} \otimes D_{2s} / J_{s+1}$, of dimension $4s-2$
by dividing the tensor square of $D_{2s}$ by the two-sided ideal
generated by $u\otimes 0 - 0 \otimes u$, where $u$ belongs to the
subalgebra $J_{s+1}$ spanned by $\{0,2,4,6,\ldots, (2s-4), (2s-2),
(2s-2)^{'}\}$. This subalgebra admits a two-sided $J_{s+1}$-invariant
complement.
We then define ${\cal H}_{Oc(D_{2s})}$ as a subalgebra
of dimension $4s$ of  $D_{2s}^{\otimesdot} \oplus M(2,\CC)$. It is
enough to know $\ud{0}, \ud{1}$ and $\ud{\epsilon}$ to build explicitly
an algebra ${\cal H}_{Oc(D_{2s}})$ from the graph $Oc(D_{2s})$.
We fix:
$$\ud{0} = 0 \otimesdot 0 + \left(
\begin{array}{cc} 1 & 0 \\ 0 & 1 \end{array} \right)
\qquad
\ud{1} = 1 \otimesdot 0 + \left(
\begin{array}{cc} 0 & 0 \\ 0 & 0 \end{array} \right)
\qquad
\ud{1\epsilon} = 0 \otimesdot 1 + \left(
\begin{array}{cc} 0 & 0 \\ 0 & 0 \end{array} \right)
$$
and set:
$$\ud{\epsilon} = \sum a_{\alpha} \alpha \otimesdot 1 + \theta
\left( \begin{array}{cc} 0 & 1 \\ 1 & 0 \\ \end{array} \right)
$$
where $\alpha \in \{1,3,\ldots, 2s-3\}$ and where the $a_{\alpha}$ are
scalars uniquely determined by the (linear) equation $\ud1 .
\ud{\epsilon} = \ud{1\epsilon}$.
The $D_{2s}^{\otimesdot}$ parts of the other elements are then uniquely
fixed. For the elements $(\ud2, \ud3, \cdots, \ud{(2s-2)}, \ud{(2s-2)^{'}})$,
it is: $(2 \otimesdot 0, 3 \otimesdot 0, \cdots)$. \\
We write the matrix part of $\ud{(2s-2)}, \ud{(2s-2)^{'}})$ as:
$$
\ud{(2s-2)} = \cdots + \theta \left(
\begin{array}{cc} \alpha & 0 \\ 0 & \beta \end{array} \right)
\qquad \qquad
\ud{(2s-2)^{'}} = \cdots + \theta \left(
\begin{array}{cc} \beta & 0 \\ 0 & \alpha \end{array} \right)
$$
By imposing that the obtained multiplication table should coincide with the
multiplication table constructed from the Ocneanu graph $Oc(D_{2s})$,
we determine uniquely the expression of the others elements,
and find also the values of $\theta, \alpha$ and $\beta$.
In every case $\theta^2 = 0$. For $\alpha$ and $\beta$ we find:
\begin{itemize}
\item[-] s even : $\alpha$ and $\beta$ are complexes:
$$
\alpha = \frac{-1+i\sqrt{(2s-1)}}{2} \qquad \qquad
\beta = \frac{-1-i\sqrt{(2s-1)}}{2} = \overline{\alpha}.
$$
\item[-] s odd: $\alpha$ and $\beta$ are reals:
$$
\alpha = \frac{-1+ \sqrt{(2s-1)}}{2} \qquad \qquad
\beta = \frac{ -1- \sqrt{(2s-1)}}{2} .
$$
\end{itemize}
The tables of fusion for the cases s even and s odd have also a different
structure, as it is clear from the examples $D_4$ and $D_6$ given in the
previous sections.

\section{The $D_{odd}$ case}

General formulae valid for all cases of this family are a bit heavy\ldots
We therefore only provide a detailed treatment of the cases $D_{5}$ and
$D_{7}$ but  generalization is straightforward.

\subsection{The $D_5$ case}
The $D_5$ Dynkin diagram and its adjacency matrix are displayed
below. We use the following order for the vertices: $\{\sigma_0,
\sigma_1, \sigma_2, \sigma_3,
\sigma_{3^{'}} \}$.

\begin{figure}[hhh]
\unitlength 0.8mm
\begin{center}
\begin{picture}(60,20)(0,7.5)
\thinlines
\multiput(5,10)(15,0){3}{\circle*{2}}
\put(50,17,5){\circle*{2}}
\put(50,2,5){\circle*{2}}
\thicklines
\put(5,10){\line(1,0){30}}
\put(35,10){\line(2,1){15}}
\put(35,10){\line(2,-1){15}}
\put(5,5){\makebox(0,0){[$\sigma_0$]}}
\put(20,5){\makebox(0,0){[$\sigma_1$]}}
\put(35,5){\makebox(0,0){[$\sigma_2$]}}
\put(55,18){\makebox(0,0){[$\sigma_3$]}}
\put(56,3){\makebox(0,0){[$\sigma_3^{'}$]}}
\end{picture}
\qquad \qquad
$
{\cal G}_{D_5} =
\left( \begin{array}{ccccc}
      0 & 1 & 0 & 0 & 0   \\
      1 & 0 & 1 & 0 & 0   \\
      0 & 1 & 0 & 1 & 1   \\
      0 & 0 & 1 & 0 & 0   \\
      0 & 0 & 1 & 0 & 0   \\
\end{array}
\right)
$
\caption{The $D_5$ Dynkin diagram and its adjacency matrix}
\label{grD5}
\end{center}
\end{figure}
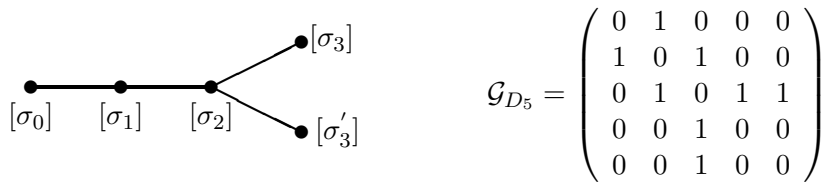

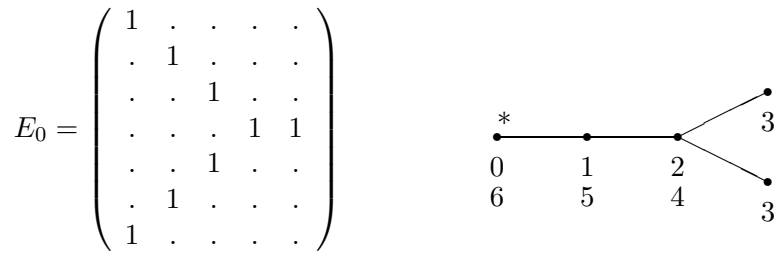
\begin{figure}[hhh]
\unitlength 0.7mm
\begin{center}
$
E_0 =
\left(
\begin{array}{ccccc}
1 & . & . & . & .  \cr
. & 1 & . & . & .  \cr
. & . & 1 & . & .  \cr
. & . & . & 1 & 1  \cr
. & . & 1 & . & .  \cr
. & 1 & . & . & .  \cr
1 & . & . & . & .  \cr
\end{array}
\right)
$
\qquad \qquad
\unitlength 0.8mm
\begin{picture}(60,20)(0,10)
\thinlines
\multiput(5,10)(15,0){3}{\circle*{1.5}}
\put(50,17,5){\circle*{1.5}}
\put(50,2,5){\circle*{1.5}}
\put(5,12){$\ast$}
\thinlines
\put(5,10){\line(1,0){30}}
\put(35,10){\line(2,1){15}}
\put(35,10){\line(2,-1){15}}
\put(5,5){\makebox(0,0){0}}
\put(5,0){\makebox(0,0){6}}
\put(20,5){\makebox(0,0){1}}
\put(20,0){\makebox(0,0){5}}
\put(35,5){\makebox(0,0){2}}
\put(35,0){\makebox(0,0){4}}
\put(50,12.5){\makebox(0,0){3}}
\put(50,-2.5){\makebox(0,0){3}}
\end{picture}
\bigskip
\caption{Essential matrix $E_0$ and Essential Paths from the vertex 0
for the $D_5$ model}
\label{D5:E0}
\end{center}
\end{figure}

Here $\kappa = 8$, the norm of the graph is $\beta = [2]_q = 2 \cos
(\frac{\pi}{8}) = \sqrt{2 + \sqrt 2}$ and the normalized
Perron-Frobenius vector is
$D = \left( [1]_q, [2]_q, [3]_q, \frac{[3]_q}{[2]_q}, \frac{[3]_q}{[2]_q}
     \right)$. \\
In the $D_{5}$ case, as in all $D_{odd}$ cases, it is not possible to define a
graph algebra at all.\\
Essential matrices of $D_{5}$ have 5 columns and 7 rows. They are labeled
by vertices of diagrams $D_5$ and $A_{7}$. The first essential
matrix $E_0$ is given in Fig \ref{D5:E0}, together with the corresponding
induction-restriction graph ($D_5$ diagram with vertices labeled
by $A_7$ vertices).

The Ocneanu algebra of $D_5$ can be realized by using the graph
algebra of $A_7$.
For $D_{2n+1}$, we have to use the graph algebra of $A_{4n-1}$.\\
We form the tensor product $A_7 \otimes A_7$, and define an application
$\rho: A_7 \rightarrow A_7$ such that:
$$
\rho(i) = i \qquad \text{for} i \in \{0,2,3,4,6,7 \} \qquad \text{and} \quad
\rho(1) = 5, \quad \rho(5) =1.
$$
We take the tensor product over $\rho$, and define the Ocneanu
algebra of $D_5$ as:
$$
{\cal{H}}_{Oc(D_5)}= A_7 \otimesdot A_7  \doteq \frac{A_7 \otimes
A_7}{\rho(A_7)}.
$$
For instance $2 \otimesdot 0 = 0 \otimesdot \rho (2) = 0 \otimesdot 2$, and
$1 \otimesdot 0 = 0 \otimesdot \rho (1) = 0 \otimesdot 5$.\\
${\cal{H}}_{Oc(D_5)}$ is spanned by a basis with 7 elements:
$$
\begin{array}{lcc}
\ud0 = 0 \otimesdot 0, & {} & \ud4 = 4 \otimesdot 0 = 0 \otimesdot 4, \\
\ud1 = 1 \otimesdot 0 = 0 \otimesdot 5, & \qquad \qquad \qquad & \ud5 = 5
\otimesdot 0 = 0 \otimesdot 1, \\
\ud2 = 2 \otimesdot 0 = 0 \otimesdot 2, & {} & \ud6 = 6 \otimesdot 0
= 0 \otimesdot 6. \\
\ud3 = 3 \otimesdot 0 = 0 \otimesdot 3, & {} & {} \\
\end{array}
$$

\begin{figure}[hhh]
\unitlength 0.8mm
\begin{center}
\begin{picture}(120,70)
\multiput(20,5)(0,15){5}{\circle{2}}
\put(5,35){\circle{2}}
\put(35,35){\circle{2}}

\thicklines
\put(5,35){\line(1,1){15}}
\put(5,35){\line(1,2){15}}
\put(35,35){\line(-1,-1){15}}
\put(35,35){\line(-1,-2){15}}
\put(19.5,50){\line(0,-1){15}}
\put(20.5,20){\line(0,1){15}}

\thicklines
\dashline[50]{1}(5,35)(20,20)
\dashline[50]{1}(5,35)(20,5)
\dashline[50]{1}(35,35)(20,50)
\dashline[50]{1}(35,35)(20,65)
\dashline[80]{1}(20.5,50)(20.5,35)
\dashline[80]{1}(19.5,20)(19.5,35)

\small
\put(20,69){\makebox(0,0){$\ud{0}$}}
\put(20,54){\makebox(0,0){$\ud{2}$}}
\put(23,35){\makebox(0,0){$\ud{3}$}}
\put(20,16){\makebox(0,0){$\ud{4}$}}
\put(1,35){\makebox(0,0){$\ud{1}$}}
\put(39,35){\makebox(0,0){$\ud{5}$}}
\put(20,1){\makebox(0,0){$\ud{6}$}}
\normalsize
\put(100,35){\makebox(0,0){$W_{00}=\left( \begin{array}{ccccccc}
1 & . & . & . & . & . & .  \\
. & . & . & . & . & 1 & .  \\
. & . & 1 & . & . & . & .  \\
. & . & . & 1 & . & . & .  \\
. & . & . & . & 1 & . & .  \\
. & 1 & . & . & . & . & .  \\
. & . & . & . & . & . & 1  \\
\end{array}
\right)
$}}

\end{picture}
\caption{The $D_5$ Ocneanu graph and the modular invariant matrix}
\label{grocD5}
\end{center}
\end{figure}
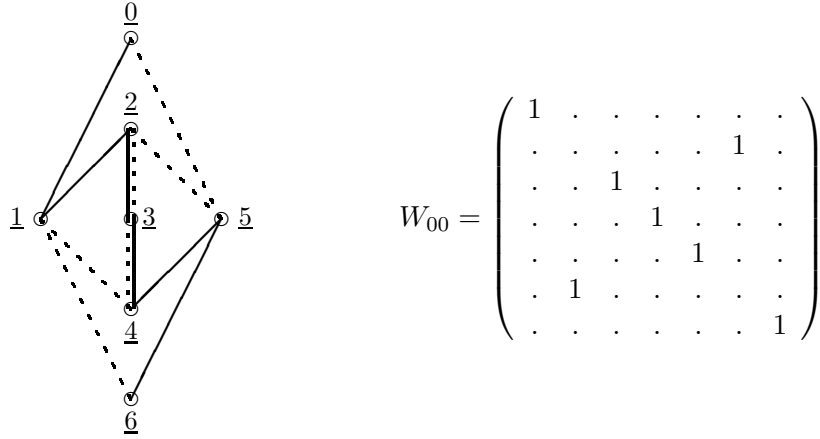

$1 \otimesdot 0$ and $0 \otimesdot 1$ are respectively the chiral
left and right
generators. The multiplication by these generators is encoded by the
Ocneanu graph
of $D_5$, represented in Fig \ref{grocD5}.
All the points are ambichiral.

To obtain the toric matrices of the $D_5$ model, we need the essential
matrices $E_i(A_7)$ of the $A_7$ case (we recall that in the
$A_n$ cases, the essential matrices are equal to the fusion matrices
$N_i$ ). We define new essential matrices $E_i^{\rho}(A_7)$ by
permuting the columns of $E_i(A_7)$ associated with the vertices 1 and 5.
The toric matrices of the $D_5$ model are then obtained by setting:
$$
W[a,b] \doteq E_a(A_7) . \widetilde{ (E_b^{\rho}(A_7))}
$$
We recall the matrix expression of the modular invariant $W_{00}$ and give
the others as sesquilinears forms in the appendix.

\subsection{The $D_7$ case}
The $D_7$ Dynkin diagram and its adjacency matrix are displayed
below. We use the following order for the vertices: $\{\sigma_0,
\sigma_1, \sigma_2, \sigma_3, \sigma_4, \sigma_5, \sigma_{5^{'}} \}$.

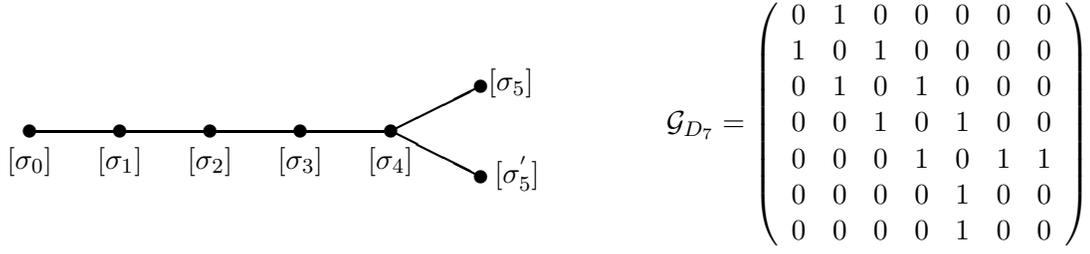
\begin{figure}[hhh]
\unitlength 0.8mm
\begin{center}
\begin{picture}(90,20)(0,10)
\thinlines
\multiput(5,10)(15,0){5}{\circle*{2}}
\put(80,17,5){\circle*{2}}
\put(80,2,5){\circle*{2}}
\thicklines
\put(5,10){\line(1,0){60}}
\put(65,10){\line(2,1){15}}
\put(65,10){\line(2,-1){15}}
\put(5,5){\makebox(0,0){[$\sigma_0$]}}
\put(20,5){\makebox(0,0){[$\sigma_1$]}}
\put(35,5){\makebox(0,0){[$\sigma_2$]}}
\put(50,5){\makebox(0,0){[$\sigma_3$]}}
\put(65,5){\makebox(0,0){[$\sigma_4$]}}
\put(85,18){\makebox(0,0){[$\sigma_5$]}}
\put(86,3){\makebox(0,0){[$\sigma_5^{'}$]}}
\end{picture}
\qquad \qquad
$
{\cal G}_{D_7} =
\left( \begin{array}{ccccccc}
      0 & 1 & 0 & 0 & 0 & 0 & 0  \\
      1 & 0 & 1 & 0 & 0 & 0 & 0  \\
      0 & 1 & 0 & 1 & 0 & 0 & 0  \\
      0 & 0 & 1 & 0 & 1 & 0 & 0  \\
      0 & 0 & 0 & 1 & 0 & 1 & 1  \\
      0 & 0 & 0 & 0 & 1 & 0 & 0  \\
      0 & 0 & 0 & 0 & 1 & 0 & 0  \\
      \end{array}
\right)
$
\caption{The $D_7$ Dynkin diagram and its adjacency matrix}
\label{grD7}
\end{center}
\end{figure}

\begin{figure}[hhh]
\unitlength 0.7mm
\begin{center}
$
E_0 =
\left(
\begin{array}{ccccccc}
1 & . & . & . & . & . & .  \cr
. & 1 & . & . & . & . & .  \cr
. & . & 1 & . & . & . & .  \cr
. & . & . & 1 & . & . & .  \cr
. & . & . & . & 1 & . & .  \cr
. & . & . & . & . & 1 & 1  \cr
. & . & . & . & 1 & . & .  \cr
. & . & . & 1 & . & . & .  \cr
. & . & 1 & . & . & . & .  \cr
. & 1 & . & . & . & . & .  \cr
1 & . & . & . & . & . & .  \cr
\end{array}
\right)
$
\qquad \qquad
\unitlength 0.8mm
\begin{picture}(90,20)(0,10)
\thinlines
\multiput(5,10)(15,0){5}{\circle*{1.5}}
\put(80,17,5){\circle*{1.5}}
\put(80,2,5){\circle*{1.5}}
\put(5,12){$\ast$}
\thinlines
\put(5,10){\line(1,0){60}}
\put(65,10){\line(2,1){15}}
\put(65,10){\line(2,-1){15}}
\put(5,5){\makebox(0,0){0}}
\put(5,0){\makebox(0,0){10}}
\put(20,5){\makebox(0,0){1}}
\put(20,0){\makebox(0,0){9}}
\put(35,5){\makebox(0,0){2}}
\put(35,0){\makebox(0,0){8}}
\put(50,5){\makebox(0,0){3}}
\put(50,0){\makebox(0,0){7}}
\put(65,5){\makebox(0,0){4}}
\put(65,0){\makebox(0,0){6}}
\put(80,12.5){\makebox(0,0){5}}
\put(80,-2.5){\makebox(0,0){5}}
\end{picture}
\bigskip
\caption{Essential matrix $E_0$ and Essential Paths from the vertex 0
for the $D_7$ model}
\label{D7:E0}
\end{center}
\end{figure}
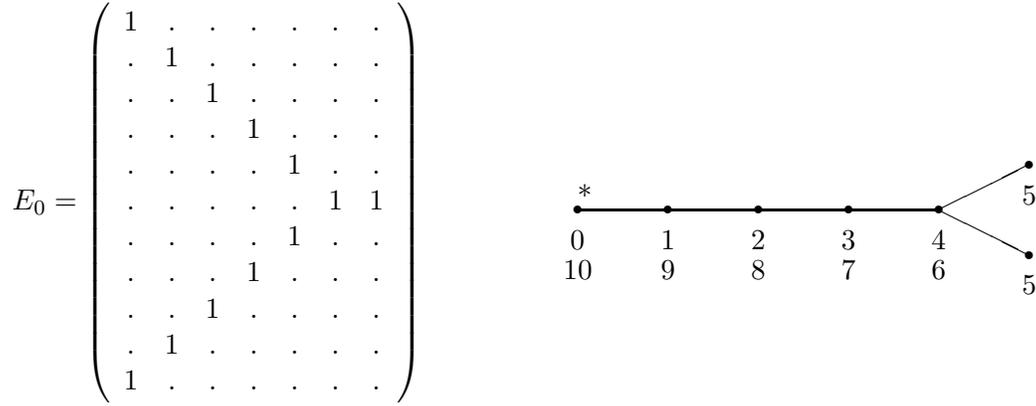

Here $\kappa = 12$, the norm of the graph is $\beta = [2]_q = 2 \cos
(\frac{\pi}{12}) = \frac{1+\sqrt 3}{\sqrt 2}$ and the normalized
Perron-Frobenius vector is
$D = \left( [1]_q, [2]_q, [3]_q, [4]_q, [5]_q, \frac{[5]_q}{[2]_q},
\frac{[5]_q}{[2]_q}
     \right)$.  \\
Essential matrices have 7 columns and 11 rows. They are labeled by
vertices of diagrams $D_7$ and $A_{11}$. The first essential matrix
$E_0$ is given in Fig \ref{D7:E0}, together with the corresponding
induction-restriction graph ($D_7$ diagram with vertices labeled
by $A_{11}$ vertices).

\begin{figure}[hhh]
\unitlength 0.7mm
\begin{center}
\begin{picture}(190,100)
\multiput(35,5)(0,15){7}{\circle{2}}
\multiput(5,50)(15,0){5}{\circle{2}}

\thicklines
\put(20,50){\line(1,1){15}}
\put(20,50){\line(1,2){15}}
\put(50,50){\line(-1,-1){15}}
\put(50,50){\line(-1,-2){15}}
\put(34.5,65){\line(0,-1){15}}
\put(35.5,35){\line(0,1){15}}

\put(5,50){\line(1,1){30}}
\put(5,50){\line(2,3){30}}
\put(65,50){\line(-1,-1){30}}
\put(65,50){\line(-2,-3){30}}

\thicklines
\dashline[50]{1}(20,50)(35,35)
\dashline[50]{1}(20,50)(35,20)
\dashline[50]{1}(50,50)(35,65)
\dashline[50]{1}(50,50)(35,80)
\dashline[80]{1}(35.5,65)(35.5,50)
\dashline[80]{1}(34.5,35)(34.5,50)

\dashline[50]{1}(65,50)(35,80)
\dashline[50]{1}(65,50)(35,95)
\dashline[50]{1}(5,50)(35,15)
\dashline[50]{1}(5,50)(35,5)

\small
\put(35,99){\makebox(0,0){$\ud{0}$}}
\put(35,84){\makebox(0,0){$\ud{2}$}}
\put(35,69){\makebox(0,0){$\ud{4}$}}
\put(38,50){\makebox(0,0){$\ud{5}$}}
\put(35,31){\makebox(0,0){$\ud{6}$}}
\put(35,16){\makebox(0,0){$\ud{8}$}}
\put(35,1){\makebox(0,0){$\ud{10}$}}
\put(1,50){\makebox(0,0){$\ud{1}$}}
\put(54,50){\makebox(0,0){$\ud{7}$}}
\put(69,50){\makebox(0,0){$\ud{9}$}}
\put(16,50){\makebox(0,0){$\ud{3}$}}
\normalsize
\put(140,50){\makebox(0,0){$W_{00}=\left( \begin{array}{ccccccccccc}
1 & . & . & . & . & . & . & . & . & . & . \\
. & . & . & . & . & . & . & . & . & 1 & . \\
. & . & 1 & . & . & . & . & . & . & . & . \\
. & . & . & . & . & . & . & 1 & . & . & . \\
. & . & . & . & 1 & . & . & . & . & . & . \\
. & . & . & . & . & 1 & . & . & . & . & . \\
. & . & . & . & . & . & 1 & . & . & . & . \\
. & . & . & 1 & . & . & . & . & . & . & . \\
. & . & . & . & . & . & . & . & 1 & . & . \\
. & 1 & . & . & . & . & . & . & . & . & . \\
. & . & . & . & . & . & . & . & . & . & 1 \\
\end{array}
\right)
$}}

\end{picture}
\caption{The $D_7$ Ocneanu graph and the modular invariant matrix}
\label{grocD7}
\end{center}
\end{figure}
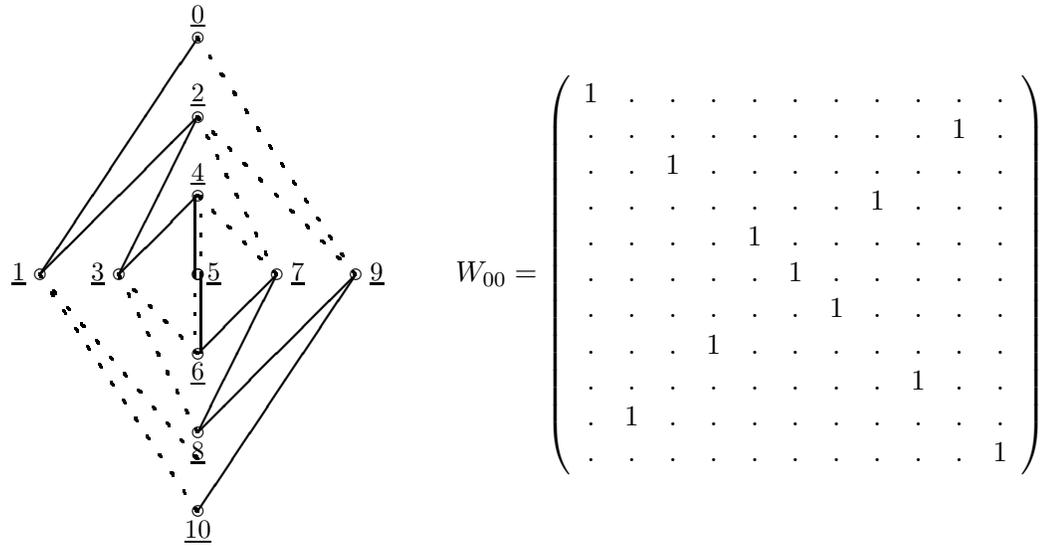

The Ocneanu algebra of $D_7$ can be realized by using the graph algebra
of $A_{11}$. We form the tensor product $A_{11}\otimes A_{11}$,
and define an application
$\rho: A_{11} \rightarrow A_{11}$ such that:
$$
\rho(i) = i \quad \text{for} \quad i \in \{0,2,4,5,6,8,10\}  \quad
\text{and} \quad
\rho(1) = 9, \quad \rho(3) =7, \quad \rho(7) = 3, \quad \rho(9) =1.
$$
We take the tensor product over $\rho$, and define the Ocneanu
algebra of $D_7$ as:
$$
{\cal{H}}_{Oc(D_7)} = A_{11} \otimesdot A_{11} \doteq \frac{A_{11}
\otimes A_{11}}{\rho(A_{11})}.
$$
It is spanned by a basis with 11 elements:
$$
\begin{array}{lll}
\ud0 = 0 \otimesdot 0               & \qquad \qquad \ud4 = 4 \otimesdot 0 =
0 \otimesdot 4 &
\qquad \qquad \; \; \ud8 = \; 8 \, \otimesdot 0 = 0 \otimesdot 8 \\
\ud1 = 0 \otimesdot 0 = 0 \otimesdot 9 & \qquad \qquad \ud5 = 5 \otimesdot 0 =
0 \otimesdot 5 &
\qquad \qquad \; \; \ud9 = \; 9 \, \otimesdot 0 = 0 \otimesdot 1 \\
\ud2 = 2 \otimesdot 0 = 0 \otimesdot 2 & \qquad \qquad \ud6 = 6 \otimesdot 0 =
0 \otimesdot 6 &
\qquad \qquad \ud10 = 10 \otimesdot 0 = 0 \otimesdot 10 \\
\ud3 = 3 \otimesdot 0 = 0 \otimesdot 7 & \qquad \qquad \ud7 = 7 \otimesdot 0 =
0 \otimesdot 3 &
\qquad \qquad
\end{array}
$$
$1 \otimesdot 0$ and $0 \otimesdot 1$ are respectively the chiral
left and right
generators. The multiplication by these generators is encoded by the
Ocneanu graph
of $D_{7}$, represented in Fig \ref{grocD7}.
All the points are ambichiral.

To obtain the toric matrices of the $D_7$ model, we need the
essential matrices $E_i(A_{11})$ of the $A_{11}$ case. We define new
essential matrices $E_i^{\rho}(A_{11})$ defined by
permuting the columns of $E_i(A_{11})$ associated to the vertices 1
and 9, 3 and 7.
The toric matrices of the $D_7$ model are then obtained by setting:
$$
W[a,b] \doteq E_a(A_{11}) . \widetilde{  (E_b^{\rho}(A_{11}))}
$$
We recall the matrix expression of the modular invariant $W_{00}$ and
give the others as sesquilinears forms in the appendix.


\section{The $E_{7}$ case}

The $E_7$ Dynkin diagram and its adjacency matrix are displayed
below. We use the following order for the vertices: $\{\sigma_0,
\sigma_1, \sigma_2, \sigma_3,
\sigma_6, \sigma_5, \sigma_4 \}$.

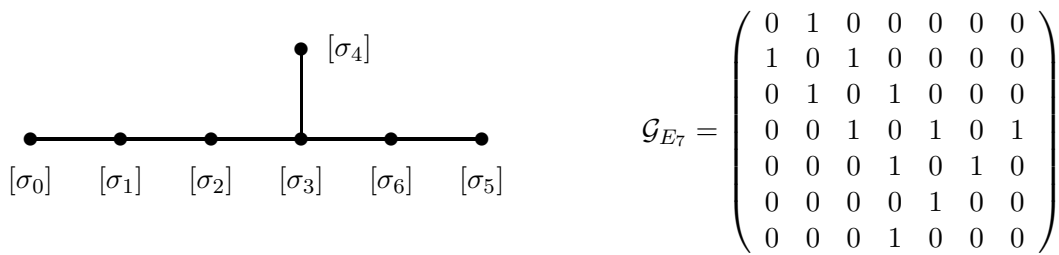
\begin{figure}[hhh]
\unitlength 0.8mm
\begin{center}
\begin{picture}(115,30)(0,10)
\thinlines
\multiput(15,10)(15,0){6}{\circle*{2}}
\put(60,25){\circle*{2}}
\thicklines
\put(15,10){\line(1,0){75}}
\put(60,10){\line(0,1){15}}
\put(15,3){\makebox(0,0){[$\sigma_0$]}}
\put(30,3){\makebox(0,0){[$\sigma_1$]}}
\put(45,3){\makebox(0,0){[$\sigma_2$]}}
\put(60,3){\makebox(0,0){[$\sigma_3$]}}
\put(75,3){\makebox(0,0){[$\sigma_6$]}}
\put(90,3){\makebox(0,0){[$\sigma_5$]}}
\put(68,25){\makebox(0,0){[$\sigma_4$]}}
\end{picture}
$
{\cal G}_{E_7} =
\left( \begin{array}{ccccccc}
      0 & 1 & 0 & 0 & 0 & 0 & 0 \\
      1 & 0 & 1 & 0 & 0 & 0 & 0 \\
      0 & 1 & 0 & 1 & 0 & 0 & 0 \\
      0 & 0 & 1 & 0 & 1 & 0 & 1 \\
      0 & 0 & 0 & 1 & 0 & 1 & 0 \\
      0 & 0 & 0 & 0 & 1 & 0 & 0 \\
      0 & 0 & 0 & 1 & 0 & 0 & 0 \\
\end{array}
\right)
$
\caption{The $E_7$ Dynkin diagram and its adjacency matrix}
\label{grE7}
\end{center}
\end{figure}
Here $\kappa = 18$, the norm of the graph is
$\beta = [2]_q = 2 \cos (\frac{\pi}{18})$ and the normalized
Perron-Frobenius vector is
$D = \left( [1]_q, [2]_q, [3]_q, [4]_q, \frac{[6]_q}{[2]_q},
\frac{[4]_q}{[3]_q}, \frac{[4]_q}{[2]_q} \right) $.  \\
The graph algebra of the Dynkin diagram $E_7$ is not a positive integral
graph algebra. We give it for illustration but it will not be used in the
sequel.

\begin{table}[hhh]
$$
\begin{array}{||c||c|c|c|c|c|c|c||}
\hline
E_7& 0 & 1  & 2 & 3 & 6 & 5 & 4  \\
\hline
\hline
0 & 0 & 1     & 2       & 3           & 6     & 5     & 4     \\
1 & 1 & 0+2   & 1+3     & 2+4+6       & 3+5   & 6     & 3     \\
2 & 2 & 1+3   & 0+2+4+6 & 1+3_2+5     & 2+4+6 & 3     & 2+6   \\
3 & 3 & 2+4+6 & 1+3_2+5 & 0+2_2+4+6_2 & 1+3_2 & 2+4   & 1+3+5 \\
6 & 6 & 3+5   & 2+4+6   & 1+3_2       & 0+2+6 & 1+5   & 2+4   \\
5 & 5 & 6     & 3       & 2+4         & 1+5   & 0-4+6 & 3-5   \\
4 & 4 & 3     & 2+6     & 1+3+5       & 2+4   & 3-5   & 0+6   \\
\hline
\end{array}
$$
\caption{Multiplication table of the graph algebra $E_7$}
\end{table}
The fusion matrices $G_{i}$ are given by the following polynomials:
$$
\begin{array}{ll}
G_0 = Id_7  \qquad \qquad \qquad \qquad \qquad \qquad &
G_5 = G_3.G_2 - G_1 - 2.G_3 \\
G_1 = \cal{G} &
G_4 = G_5.G_2 \\
G_2 = G_1.G_1 - G_0 &
G_6 = G_5.G_4 - G_2 \\
G_3 = G_1.G_2 - G_1 & {}
\end{array}
$$

Essential matrices of $E_{7}$ have 7 columns and 17 rows. They are labeled
by vertices of diagrams $E_7$ and $A_{17}$. The first essential
matrix $E_0$ is given below, together with the corresponding
induction-restriction graph. To obtain the toric matrices, we also
need to know the essential matrices for the $D_{10}$ case. They are
obtained as usual (we also display
the essential matrix $E_0$ of the $D_{10}$ case below).

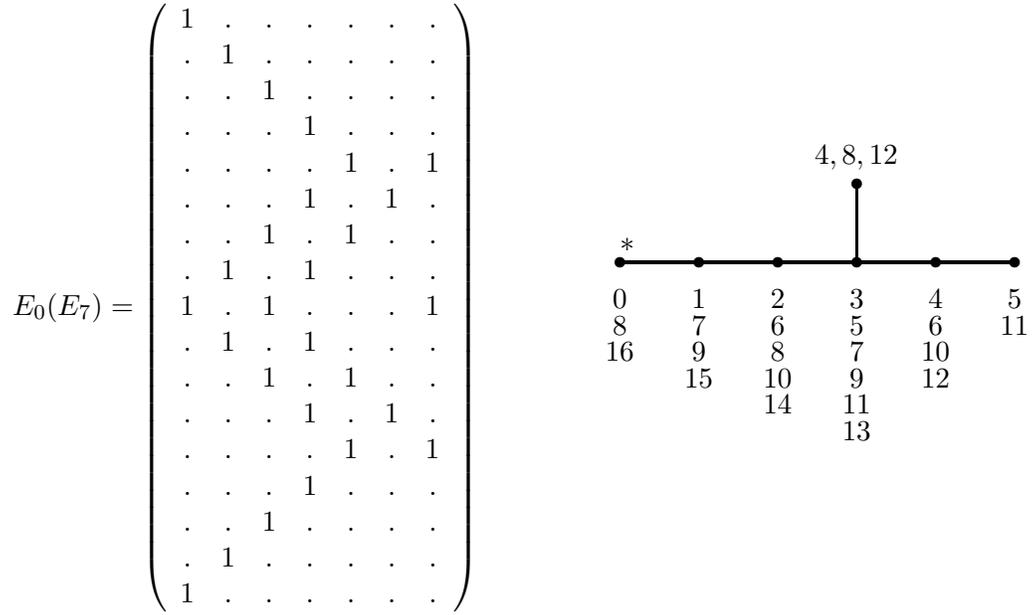
\begin{figure}[hhh]
\unitlength 0.7mm
\begin{center}
$
E_0 (E_7)=
\left(
\begin{array}{ccccccc}
1 & . & . & . & . & . & . \cr
. & 1 & . & . & . & . & . \cr
. & . & 1 & . & . & . & . \cr
. & . & . & 1 & . & . & . \cr
. & . & . & . & 1 & . & 1 \cr
. & . & . & 1 & . & 1 & . \cr
. & . & 1 & . & 1 & . & . \cr
. & 1 & . & 1 & . & . & . \cr
1 & . & 1 & . & . & . & 1 \cr
. & 1 & . & 1 & . & . & . \cr
. & . & 1 & . & 1 & . & . \cr
. & . & . & 1 & . & 1 & . \cr
. & . & . & . & 1 & . & 1 \cr
. & . & . & 1 & . & . & . \cr
. & . & 1 & . & . & . & . \cr
. & 1 & . & . & . & . & . \cr
1 & . & . & . & . & . & . \cr
\end{array}
\right)
$
\begin{picture}(110,35)
\thinlines
\multiput(25,10)(15,0){6}{\circle*{2}}
\put(70,25){\circle*{2}}
\put(25,12){$\ast$}
\thicklines
\put(25,10){\line(1,0){75}}
\put(70,10){\line(0,1){15}}

\put(25,3){\makebox(0,0){$0$}}
\put(25,-2){\makebox(0,0){$8$}}
\put(25,-7){\makebox(0,0){$16$}}

\put(40,3){\makebox(0,0){$1$}}
\put(40,-2){\makebox(0,0){$7$}}
\put(40,-7){\makebox(0,0){$9$}}
\put(40,-12){\makebox(0,0){$15$}}

\put(55,3){\makebox(0,0){$2$}}
\put(55,-2){\makebox(0,0){$6$}}
\put(55,-7){\makebox(0,0){$8$}}
\put(55,-12){\makebox(0,0){$10$}}
\put(55,-17){\makebox(0,0){$14$}}

\put(70,3){\makebox(0,0){$3$}}
\put(70,-2){\makebox(0,0){$5$}}
\put(70,-7){\makebox(0,0){$7$}}
\put(70,-12){\makebox(0,0){$9$}}
\put(70,-17){\makebox(0,0){$11$}}
\put(70,-22){\makebox(0,0){$13$}}

\put(85,3){\makebox(0,0){$4$}}
\put(85,-2){\makebox(0,0){$6$}}
\put(85,-7){\makebox(0,0){$10$}}
\put(85,-12){\makebox(0,0){$12$}}

\put(100,3){\makebox(0,0){$5$}}
\put(100,-2){\makebox(0,0){$11$}}

\put(70,30){\makebox(0,0){$4,8,12$}}
\end{picture}
\bigskip
\caption{Essential matrix $E_0$ and Essential Paths from the vertex 0
for the $E_7$-model}
\label{E7:E0}
\end{center}
\end{figure}

\begin{figure}
\begin{center}
$$
E_0 (D_{10})=
\left(
\begin{array}{cccccccccc}

1 & . & . & . & . & . & . & . & . & . \cr

. & 1 & . & . & . & . & . & . & . & . \cr

. & . & 1 & . & . & . & . & . & . & . \cr

. & . & . & 1 & . & . & . & . & . & . \cr

. & . & . & . & 1 & . & . & . & . & . \cr

. & . & . & . & . & 1 & . & . & . & . \cr

. & . & . & . & . & . & 1 & . & . & . \cr

. & . & . & . & . & . & . & 1 & . & . \cr

. & . & . & . & . & . & . & . & 1 & 1 \cr

. & . & . & . & . & . & . & 1 & . & . \cr

. & . & . & . & . & . & 1 & . & . & . \cr

. & . & . & . & . & 1 & . & . & . & . \cr

. & . & . & . & 1 & . & . & . & . & . \cr

. & . & . & 1 & . & . & . & . & . & . \cr

. & . & 1 & . & . & . & . & . & . & . \cr

. & 1 & . & . & . & . & . & . & . & . \cr

1 & . & . & . & . & . & . & . & . & . \cr

\end{array}
\right)
$$
\end{center}
\caption{Essential matrix $E_0$ for $D_{10}$}
\end{figure}

We form the tensor product $D_{10} \otimes D_{10}$, and
identify $a u \otimes b$ with $a \otimes \rho(u) b$ where

$$
\begin{array}{ccc}
\rho(0)=0, & \qquad \qquad \rho(4)=4, & \qquad \qquad \rho(8)=2,  \\
\rho(2)=8, & \qquad \qquad \rho(6)=6, & \qquad \qquad \rho(8^{'})=8^{'},
\end{array}
$$

The Ocneanu algebra of $E_7$ can be realized as:
$$
{\cal{H}}_{Oc(E_7)}= D_{10} \otimesdot D_{10} \doteq
\frac{D_{10} \otimes D_{10}}{\rho}.
$$
It is spanned by a basis with 17 elements:
$$
\begin{array}{cclcccl}
\ud0 &=& 0 \otimesdot 0  & \qquad \qquad & \ud{(0)} &=& 0 \otimesdot 1   \\
\ud1 &=& 1 \otimesdot 0  & \qquad \qquad & \ud{(1)} &=& 1 \otimesdot 1   \\
\ud2 &=& 2 \otimesdot 0 = 0 \otimesdot 8 & \qquad \qquad & \ud{(2)} &=& 2
\otimesdot 1 = 0 \otimesdot 7  \\
\ud3 &=& 3 \otimesdot 0  & \qquad \qquad & \ud{(3)} &=& 3 \otimesdot 1 = 1
\otimesdot 3  \\
\ud4 &=& 4 \otimesdot 0 = 0 \otimesdot 4 & \qquad \qquad & \ud{(4)} &=& 0
\otimesdot 3   \\
\ud5 &=& 5 \otimesdot 0  & \qquad \qquad & \ud{(5)} &=& 5 \otimesdot 1 - 3
\otimesdot 1  \\
\ud6 &=& 6 \otimesdot 0 = 0 \otimesdot 6 & \qquad \qquad & &=&1 \otimesdot 5 -
1 \otimesdot 3   \\
\ud7 &=& 7 \otimesdot 0  & \qquad \qquad & \ud{(6)} &=& 0 \otimesdot 5   \\
\ud8 &=& 8 \otimesdot 0 = 0 \otimesdot 2 & \qquad \qquad & & &   \\
\ud{8^{'}} &=& 8^{'} \otimesdot 0 = 0 \otimesdot 8^{'} & \qquad
\qquad & & &   \\
\end{array}
$$
$\ud1$ and $\ud{(0)}$ are respectively the left and right generators.
The ambichiral part is the linear span of
$\{\ud{0},\ud{2},\ud{4},\ud{6},\ud{8},\ud{8^{'}}\}$.
The multiplication of the elements of this algebra by the generators
is shown in the following table.
We can observe on the Ocneanu graph $Oc(E_{7})$ that $E_7$
does not appear as a subalgebra of ${\cal H}_{Oc(E_{7}})$
but as a quotient (there are two such quotients).

The seventeen toric matrices $W_{ab}$ of the $E_{7}$ model are obtained
as explained in section 2.2.5, but with a twist: We use the essential
matrices $E_{a}(D_{10})$, and replace the matrix elements of the
columns associated with
vertices $1,3,5,7$ of the graph $D_{10}$ by $0$; this being done, we
permute the columns associated with vertices $2$ and $8$ of $D_{10}$
(with our ordering, these are columns $3$ and $9$). The reduced and twisted
matrix so obtained is called $E_{a}^{\rho}(D_{10})$.
The seventeen toric matrices of the $E_{7}$ model are then obtained by setting:
$$
W[a,b] \doteq E_a(D_{10}) . \widetilde{ (E_b^{\rho}(D_{10}))}
$$
For instance $W_{\ud{(5)}}=  W[5,1] - W[3,1] = E_5(D_{10}) .
\widetilde{ (E_1^{\rho}(D_{10}))}
- E_3(D_{10}) . \widetilde{ (E_1^{\rho}(D_{10}))}$.
We recall the matrix expression of the modular invariant $W_{00}$ and give
the others as sesquilinears forms in the appendix.

\begin{table}[hhh]
$$
\begin{array}{|c|c|}
\hline
"D_{10}" & \ud1 \\
\hline
\ud0 & \ud1 \\
\ud1 & \ud0 + \ud2 \\
\ud2 & \ud1 + \ud3 \\
\ud3 & \ud2 + \ud4 \\
\ud4 & \ud3 + \ud5 \\
\ud5 & \ud4 + \ud6 \\
\ud6 & \ud5 + \ud7 \\
\ud7 & \ud6 + \ud8 + \ud{8^{'}}\\
\ud8 & \ud7 \\
\ud{8^{'}} & \ud7 \\
\hline
\hline
"E_7" & \ud1 \\
\hline
\ud{(0)} & \ud{(1)} \\
\ud{(1)} & \ud{(0)} + \ud{(2)} \\
\ud{(2)} & \ud{(1)} + \ud{(3)} \\
\ud{(3)} & \ud{(2)} + \ud{(4)} + \ud{(6)} \\
\ud{(4)} & \ud{(3)} \\
\ud{(6)} & \ud{(3)} + \ud{(5)} \\
\ud{(5)} & \ud{(6)} \\
\hline
\end{array}
\qquad \qquad \qquad \qquad
\begin{array}{|c|c|}
\hline
"D_{10}" & \ud{(0)} \\
\hline
\ud0 & \ud{(0)} \\
\ud{(0)} & \ud0 + \ud8 \\
\ud8 & \ud{(0)} + \ud{(4)} \\
\ud{(4)} & \ud8 + \ud4 \\
\ud4 & \ud{(4)} + \ud{(6)} \\
\ud{(6)} & \ud4 + \ud6 \\
\ud6 & \ud{(6)} + \ud{(2)} \\
\ud{(2)} & \ud2 + \ud6 + \ud{8^{'}}\\
\ud2 & \ud{(2)} \\
\ud8^{'} & \ud{(2)} \\
\hline
\hline
"E_7" & \ud{(0)} \\
\hline
\ud1 & \ud{(1)} \\
\ud{(1)} & \ud1 + \ud7 \\
\ud7 & \ud{(1)} + \ud{(3)} \\
\ud{(3)} & \ud3+\ud5+\ud7 \\
\ud3 & \ud{(3)} \\
\ud5 & \ud{(3)} + \ud{(5)} \\
\ud{(5)} & \ud5 \\
\hline
\end{array}
$$
\caption{Multiplication of the elements of the Ocneanu algebra of
$E_7$ by the generators}
\end{table}

\begin{figure}[hhh]
\unitlength 0.8mm
\begin{center}
\begin{picture}(50,90)
\put(25,5){\circle*{2}}
\multiput(25,15)(0,10){2}{\circle{2}}
\put(25,35){\circle*{2}}
\put(25,45){\circle{2}}
\put(25,75){\circle*{2}}
\put(25,85){\circle{2}}
\multiput(5,10)(0,20){2}{\circle*{2}}
\multiput(5,60)(0,20){2}{\circle*{2}}
\multiput(45,10)(0,20){2}{\circle*{2}}
\multiput(45,60)(0,20){2}{\circle*{2}}
\put(20,60){\circle{2}}
\put(30,60){\circle{2}}

\thicklines
\put(5,80){\line(4,1){20}}
\put(5,80){\line(5,-4){25}}
\put(5,60){\line(1,0){15}}
\put(5,60){\line(4,-3){20}}
\dashline[100]{8}(5,60)(25,25)
\put(5,30){\line(5,6){25}}
\put(5,30){\line(4,-3){20}}
\put(5,10){\line(4,3){20}}
\put(5,10){\line(4,1){20}}

\thinlines
\put(45,80){\line(-4,-1){20}}
\put(45,60){\line(-4,3){20}}
\put(45,60){\line(-4,-5){20}}
\put(45,30){\line(-4,1){20}}
\put(45,10){\line(-4,5){20}}
\put(45,10){\line(-4,-1){20}}

\thicklines
\dashline[50]{1}(45,80)(25,85)
\dashline[50]{1}(45,80)(20,60)
\dashline[50]{1}(45,60)(30,60)
\dashline[50]{1}(45,60)(25,45)
\dashline[50]{1}(45,60)(25,25)
\dashline[50]{1}(45,30)(20,60)
\dashline[50]{1}(45,30)(25,15)
\dashline[50]{1}(45,10)(25,25)
\dashline[50]{1}(45,10)(25,15)

\thinlines
\dashline[50]{1}(5,80)(25,75)
\dashline[50]{1}(5,60)(25,75)
\dashline[50]{1}(5,60)(25,35)
\dashline[50]{1}(5,30)(25,35)
\dashline[50]{1}(5,10)(25,35)
\dashline[50]{1}(5,10)(25,5)

\scriptsize
\put(25,88){\makebox(0,0){$\ud{0}$}}
\put(25,78){\makebox(0,0){$(\ud{1})$}}
\put(25,49.5){\makebox(0,0){$\ud{8^{'}}$}}
\put(25,39.5){\makebox(0,0){$(\ud{3})$}}
\put(25,28){\makebox(0,0){$\ud{6}$}}
\put(25,18){\makebox(0,0){$\ud{4}$}}
\put(25,8){\makebox(0,0){$(\ud{5})$}}

\put(1,10){\makebox(0,0){$\ud{5}$}}
\put(1,30){\makebox(0,0){$\ud{3}$}}
\put(1,60){\makebox(0,0){$\ud{7}$}}
\put(1,80){\makebox(0,0){$\ud{1}$}}

\put(49,10){\makebox(0,0){$(\ud{6})$}}
\put(49,30){\makebox(0,0){$(\ud{4})$}}
\put(49,60){\makebox(0,0){$(\ud{2})$}}
\put(49,80){\makebox(0,0){$(\ud{0})$}}

\put(31,63){\makebox(0,0){$\ud{2}$}}
\put(19,63){\makebox(0,0){$\ud{8}$}}

\normalsize
\end{picture}
\caption{Ocneanu graph $E_7 $}
\label{grocE7}
\end{center}
\end{figure}

$$
W_{00}=\left( \begin{array}{ccccccccccccccccc}
1 & . & . & . & . & . & . & . & . & . & . & . & . & . & . & . & 1 \\
. & . & . & . & . & . & . & . & . & . & . & . & . & . & . & . & . \\
. & . & . & . & . & . & . & . & 1 & . & . & . & . & . & . & . & . \\
. & . & . & . & . & . & . & . & . & . & . & . & . & . & . & . & . \\
. & . & . & . & 1 & . & . & . & . & . & . & . & 1 & . & . & . & . \\
. & . & . & . & . & . & . & . & . & . & . & . & . & . & . & . & . \\
. & . & . & . & . & . & 1 & . & . & . & 1 & . & . & . & . & . & . \\
. & . & . & . & . & . & . & . & . & . & . & . & . & . & . & . & . \\
. & . & 1 & . & . & . & . & . & 1 & . & . & . & . & . & 1 & . & . \\
. & . & . & . & . & . & . & . & . & . & . & . & . & . & . & . & . \\
. & . & . & . & . & . & 1 & . & . & . & 1 & . & . & . & . & . & . \\
. & . & . & . & . & . & . & . & . & . & . & . & . & . & . & . & . \\
. & . & . & . & 1 & . & . & . & . & . & . & . & 1 & . & . & . & . \\
. & . & . & . & . & . & . & . & . & . & . & . & . & . & . & . & . \\
. & . & . & . & . & . & . & . & 1 & . & . & . & . & . & . & . & . \\
. & . & . & . & . & . & . & . & . & . & . & . & . & . & . & . & . \\
1 & . & . & . & . & . & . & . & . & . & . & . & . & . & . & . & 1
\end{array}
\right)
$$

We conclude this section with the determination of the integers $d_{n}$ and
$d_{x}$.\\
The integers $d_{n}$ associated with endomorphisms of essential paths
on the graph $E_{7}$ are determined easily by the usual method.
For $n=1,2,\ldots 17$ one finds
$$
d_{n} = 7,12,17,22,27,30,33,34,35,34,33,30,27,22,17,12,7.
$$
In order to determine the integers $d_x$, we need to know the
multiplication table defined by the Ocneanu graph $Oc(E_7)$.
   The full table of ${\cal H}_{Oc(E_7)}$ has the
following structure: $"D_{10}" \times "D_{10}" \rightarrow "D_{10}"$,
$"D_{10}" \times "E_7" \rightarrow "E_7"$, and
$"E_7" \times "E_7" \rightarrow "D_{10}"$, where $"D_{10}"$ is the
subalgebra linearly spanned by $\ud{0}, \ud{1}, \ldots, \ud{8}, \ud{8^{'}}$ and
$"E_7"$ is the linear subspace  linearly spanned by $\ud{(0)},
\ud{(1)}, \ldots,
\ud{(6)}$. Actually, it is enough, four our purpose, to know the smaller table
obtained by restriction to the $E_7$ quotients, \ie the
$"E_7" \times "E_7" \rightarrow "D_{10}"$ table.

We encode this multiplication table by a set of $10$ matrices $s_p$
(labeled by
$D_{10}$, of size $7\times 7$ (labeled by $E_7$)).
The result of a given multiplication, such as
$\ud{(4)} \times \ud{(5)} = \ud{3} + \ud{7}$
is indicated by the presence of the integer $1$ in position $(4,5)$ in both
matrices $s_3$ and $s_7$.

Since we have an explicit realization of ${\cal H}_{Oc(E_7)}$, it is not too
difficult to find

{\small
$$
s_0 = \left( \begin{array}{ccccccc}
   1 & 0 & 0 & 0 & 0 & 0 & 0 \\ 0 & 1 & 0 & 0 & 0 & 0 & 0 \\ 0 & 0 & 1 & 0 \
& 0 & 0 & 0 \\ 0 & 0 & 0 & 1 & 0 & 0 & 0 \\ 0 & 0 & 0 & 0 & 1 & 0 & 0 \
\\ 0 & 0 & 0 & 0 & 0 & 1 & 0 \\ 0 & 0 & 0 & 0 & 0 & 0 & 1 \\
\end{array}
\right)
\qquad \qquad \qquad
s_1 = \left( \begin{array}{ccccccc}
   0 & 1 & 0 & 0 & 0 & 0 & 0 \\ 1 & 0 & 1 & 0 & 0 & 0 & 0 \\ 0 & 1 & 0 & 1 \
& 0 & 0 & 0 \\ 0 & 0 & 1 & 0 & 1 & 0 & 1 \\ 0 & 0 & 0 & 1 & 0 & 0 & 0 \
\\ 0 & 0 & 0 & 0 & 0 & 0 & 1 \\ 0 & 0 & 0 & 1 & 0 & 1 & 0 \\
\end{array}
\right)
$$

$$
s_2 = \left( \begin{array}{ccccccc}
   0 & 0 & 1 & 0 & 0 & 0 & 0 \\ 0 & 1 & 0 & 1 & 0 & 0 & 0 \\ 1 & 0 & 1 & 0 \
& 1 & 0 & 1 \\ 0 & 1 & 0 & 2 & 0 & 1 & 0 \\ 0 & 0 & 1 & 0 & 0 & 0 & 1 \
\\ 0 & 0 & 0 & 1 & 0 & 0 & 0 \\ 0 & 0 & 1 & 0 & 1 & 0 & 1 \\
\end{array}
\right)
\qquad \qquad \qquad
s_3 = \left( \begin{array}{ccccccc}
   0 & 0 & 0 & 1 & 0 & 0 & 0 \\ 0 & 0 & 1 & 0 & 1 & 0 & 1 \\ 0 & 1 & 0 & 2 \
& 0 & 1 & 0 \\ 1 & 0 & 2 & 0 & 1 & 0 & 2 \\ 0 & 1 & 0 & 1 & 0 & 1 & 0 \
\\ 0 & 0 & 1 & 0 & 1 & 0 & 0 \\ 0 & 1 & 0 & 2 & 0 & 0 & 0 \\
\end{array}
\right)
$$

$$
s_4 = \left( \begin{array}{ccccccc}
   0 & 0 & 0 & 0 & 1 & 0 & 1 \\ 0 & 0 & 0 & 2 & 0 & 1 & 0 \\ 0 & 0 & 2 & 0 \
& 1 & 0 & 2 \\ 0 & 2 & 0 & 3 & 0 & 1 & 0 \\ 1 & 0 & 1 & 0 & 1 & 0 & 1 \
\\ 0 & 1 & 0 & 1 & 0 & 0 & 0 \\ 1 & 0 & 2 & 0 & 1 & 0 & 1 \\
\end{array}
\right)
\qquad \qquad \qquad
s_5= \left( \begin{array}{ccccccc}
   0 & 0 & 0 & 1 & 0 & 1 & 0 \\ 0 & 0 & 1 & 0 & 1 & 0 & 2 \\ 0 & 1 & 0 & 3 \
& 0 & 1 & 0 \\ 1 & 0 & 3 & 0 & 2 & 0 & 2 \\ 0 & 1 & 0 & 2 & 0 & 0 & 0 \
\\ 1 & 0 & 1 & 0 & 0 & 0 & 1 \\ 0 & 2 & 0 & 2 & 0 & 1 & 0 \\
\end{array}
\right)
$$

$$
s_6 = \left( \begin{array}{ccccccc}
   0 & 0 & 1 & 0 & 0 & 0 & 1 \\ 0 & 1 & 0 & 2 & 0 & 1 & 0 \\ 1 & 0 & 2 & 0 \
& 2 & 0 & 2 \\ 0 & 2 & 0 & 4 & 0 & 1 & 0 \\ 0 & 0 & 2 & 0 & 1 & 0 & 1 \
\\ 0 & 1 & 0 & 1 & 0 & 1 & 0 \\ 1 & 0 & 2 & 0 & 1 & 0 & 2 \\
\end{array}
\right)
\qquad \qquad \qquad
s_7 = \left( \begin{array}{ccccccc}
   0 & 1 & 0 & 1 & 0 & 0 & 0 \\ 1 & 0 & 2 & 0 & 1 & 0 & 1 \\ 0 & 2 & 0 & 3 \
& 0 & 1 & 0 \\ 1 & 0 & 3 & 0 & 2 & 0 & 3 \\ 0 & 1 & 0 & 2 & 0 & 1 & 0 \
\\ 0 & 0 & 1 & 0 & 1 & 0 & 1 \\ 0 & 1 & 0 & 3 & 0 & 1 & 0 \\
\end{array}
\right)
$$

$$
s_8 = \left( \begin{array}{ccccccc}
   1 & 0 & 0 & 0 & 1 & 0 & 0 \\ 0 & 1 & 0 & 1 & 0 & 0 & 0 \\ 0 & 0 & 2 & 0 \
& 0 & 0 & 1 \\ 0 & 1 & 0 & 2 & 0 & 1 & 0 \\ 1 & 0 & 0 & 0 & 1 & 0 & 1 \
\\ 0 & 0 & 0 & 1 & 0 & 0 & 0 \\ 0 & 0 & 1 & 0 & 1 & 0 & 1 \\
\end{array}
\right)
\qquad \qquad \qquad
s_{8^{'}} = \left( \begin{array}{ccccccc}
   0 & 0 & 1 & 0 & 0 & 0 & 0 \\ 0 & 1 & 0 & 1 & 0 & 0 & 0 \\ 1 & 0 & 1 & 0 \
& 1 & 0 & 1 \\ 0 & 1 & 0 & 2 & 0 & 1 & 0 \\ 0 & 0 & 1 & 0 & 0 & 0 & 1 \
\\ 0 & 0 & 0 & 1 & 0 & 0 & 0 \\ 0 & 0 & 1 & 0 & 1 & 0 & 1 \\
\end{array}
\right)
$$
}

The sum of matrix elements of the $10$ matrices $s_{p}$, for
$p = 0,1,2,\ldots 8, 8^{'}$ is \\
$7,12,17,22,27,30,33,34,18,17$.

To each linear generator $x = \sum a \otimes b$ (for instance $\ud{(5)} =
5\otimesdot 1 - 3 \otimesdot 1$) of the Ocneanu algebra of $E_{7}$ (the
basis with $17$ elements was given previously) we associate
a matrix $\Sigma_{x} = \sum s_{a}s_{b}$ (for instance
$\Sigma_{\ud{(5)}} = s_{5}s_{1} - s_{3}s_{1})$.  The integer $d_{x}$
is the sum of matrix elements of the matrix $\Sigma_{x}$ (for
instance $d_{\ud{(5)}} = 16$). In particular, the $d_{x}$ associated
with the $"D_{10}"$ part of the graph are just given by sum of
matrix elements of matrices $s_{p}$.

The final list of integers $d_{x}$, associated with blocks
$\ud{0},\ud{1},\ldots, \ud{8},\ud{8^{'}}; \ud{(0)}, \ud{(1)},\ldots \ud{(6)}$
is
$$
d_x = 7,12,17,22,27,30,33,34,18,17,12,24,34,44,22,16,30.
$$
Note that $\sum d_{n} = \sum d_{x} = 399$ and that  $\sum d_{n}^2 = \sum
d_{x}^2 = 10905$.

The above results agree\footnote{The preprint version of \cite{PetZub:Oc},
available on
the web, contains a typing misprint: the last values of $d_{x}$ should be
read $44,30,16,22$ and not $442, 30, 16, 22$.}
with those obtained by \cite{PetZub:Oc}.


\section{Acknowledgements}

G. Schieber would like to thank the Conselho Nacional de Desenvolvimento
Cient\'{\i}fico e Tecnol\'ogico, CNPq (Brazilian Research Agency) for financial
support.

We thank C. Mercat, O. Ogievetsky and J.B. Zuber for their comments about the first preprint version
of this manuscript.


\appendix

\section{The general notion of essential paths on a graph $G$}

The general definitions given here are adapted from (\cite{Ocneanu:paths}).

Call $\beta$ the norm of the graph $G$ (the biggest eigenvalue of  the
adjacency matrix ${\cal G}$)
and  $D_{i}$ the components of the (normalized) Perron-Frobenius eigenvector.
Call $\sigma_{i}$ the vertices of $G$ and, if
$\sigma_{j}$ is a neighbour of
$\sigma_{i}$, call $\xi_{ij}$ the oriented edge
from $\sigma_{i}$ to $\sigma_{j}$. If $G$ is unoriented (the case for $ADE$
and affine $ADE$ diagrams), each edge should be considered  as carrying
both orientations.

An elementary path can be written either as a finite
sequence of consecutive (neighbours on the graph) vertices,
$[\sigma_{a_1} \sigma_{a_2} \sigma_{a_3} \ldots ]$,
or as a sequence $(\xi(1)\xi(2)\ldots)$ of consecutive edges, with
$\xi(1) = \xi_{a_{1}a_{2}}= \sigma_{a_1} \sigma_{a_2} $,
$\xi(2) = \xi_{a_{2}a_{3}} = \sigma_{a_2}  \sigma_{a_3} $, \etc
Vertices are considered as paths of length $0$.

The length of the (possibly backtracking) path $( \xi(1)\xi(2)\ldots
\xi(p) )$ is $p$.
We call $r(\xi_{ij})=\sigma_{j}$, the range of $\xi_{ij}$
and $s(\xi_{ij})=\sigma_{i}$, the source of $\xi_{ij}$.

For all edges $\xi(n+1) = \xi_{ij}$ that appear in an elementary path,
we set  ${\xi(n+1)}^{-1} \doteq \xi{ji}$.

For every integer $n >0$, the annihilation operator $C_{n}$,
acting on elementary paths of length $p$ is defined
as follows:  if $p \leq n$, $C_{n}$ vanishes and if $ p \geq  n+1$ then
$$
C_{n} (\xi(1)\xi(2)\ldots\xi(n)\xi(n+1)\ldots) =
\sqrt\frac{D_{r(\xi(n))}}{D_{s(\xi(n))}}
\delta_{\xi(n),{\xi(n+1)}^{-1}}
      (\xi(1)\xi(2)\ldots{\hat\xi(n)}{\hat\xi(n+1)}\ldots)
$$
Here, the symbol ``hat'' ( like  in $\hat \xi$) denotes omission.
The result is therefore either $0$ or a linear combination of paths
of length $p-2$.
Intuitively, $C_{n}$ chops the round trip that possibly appears
at positions $n,n+1$.

A path is called essential if it belongs to
the intersection of the kernels
of the annihilators $C_{n}$'s.

For instance, in the case of the diagram $E_{6}$,
\begin{eqnarray*}
C_{3}(\xi_{01}\xi_{12}\xi_{23}\xi_{32}) &=&  \sqrt \frac {1}{[2]}
(\xi_{01}\xi_{12}) \\
C_{3}(\xi_{01}\xi_{12}\xi_{25}\xi_{52}) &=&  \sqrt \frac {[2]}{[3]}
(\xi_{01}\xi_{12})
      \end{eqnarray*}

The following difference of non essential paths of length $4$
starting at $\sigma_{0}$ and ending at $\sigma_{2}$
is an essential path of length $4$  on $E_{6}$:
$$\sqrt{[2]} (\xi_{01}\xi_{12}\xi_{23}\xi_{32})  - \sqrt
\frac{[3]}{[2]} (\xi_{01}\xi_{12}\xi_{25}\xi_{52})
= \sqrt {[2]}  [0,1,2,3,2]- \sqrt \frac{[3]}{[2]}[0,1,2,5,2]$$
Here the values of q-numbers are $[2] = \frac{\sqrt 2}{\sqrt 3 -1}$ and $[3] =
\frac{2}{\sqrt 3 -1}$.

Acting on an elementary path of length $p$, the creating operators
$C^{\dag}_{n}$ are defined as follows:
if $n > p+1$, $C^{\dag}_{n}$ vanishes and, if $n \leq p+1$ then,
setting $j = r(\xi(n-1))$,
$$
C^{\dag}_{n} (\xi(1)\ldots\xi(n-1)\ldots) = \sum_{d(j,k)=1}
\sqrt(\frac{D_{k}}{D_{j}})  (\xi(1)\ldots\xi(n-1)\xi_{jk}\xi_{kj}\ldots)
$$
The above sum is taken over the neighbours $\sigma_{k}$ of
$\sigma_{j}$ on the graph.
Intuitively, this operator adds one  (or several) small round trip(s)
at position $n$.
The result is therefore either $0$ or a linear combination of paths of
length $p+2$.

For instance, on paths of length zero (\ie vertices),
$$
C^{\dag}_{1} (\sigma_{j}) = \sum_{d(j,k)=1}
\sqrt(\frac{D_{k}}{D_{j}}) \xi_{jk}\xi_{kj} = \sum_{d(j,k)=1}
\sqrt(\frac{D_{k}}{D_{j}}) \, [\sigma_{j}\sigma_{k}\sigma_{j}]
$$

Jones' projectors $e_{k}$ can be defined (as endomorphisms of
$Path^p$) by
$$
e_{k} \doteq \frac{1}{\beta} C^{\dag}_{k} C_{k}
$$

The reader can check that all Jones-Temperley-Lieb relations
between the $e_i$ are satisfied.
Essential paths can also be defined as elements of the intersection of the
kernels of the Jones projectors $e_{i}$'s.



\section{Twisted partition functions for the $ADE$ models}


\begin{table}[hhh]
$$
\begin{array}{|c||l|}
\hline
Point & Z \\
\hline
{}   & {} \\
\ud0 & \xa{0} + \xa{1} + \xa{2} + \xa{3} \\
{}   & {} \\
\hline
{}   & {} \\
\ud1 & [(\xx{0}{1} + \xx{1}{2} + \xx{2}{3}) + h.c.] \\
{}   & {} \\
\hline
{}   & {} \\
\ud2 & \xa{1} + \xa{2} + [(\xx{0}{2} + \xx{1}{3}) + h.c.]  \\
{}   & {} \\
\hline
{}   & {} \\
\ud3 & [(\xx{0}{3} + \xx{1}{2}) + h.c.] \\
{}   & {} \\
\hline
\end{array}
$$
\normalsize
\caption{Twisted partition functions for the $A_4$ model}
\end{table}


\begin{table}
\scriptsize
$$
\begin{array}{|c||l|}
\hline
Point & \mathcal{Z} \\
\hline
\hline
{}  &  {}  \\
\ud0  & \xaa{0}{6} + \xaa{3}{7} + \xaa{4}{10}   \\
{}  &  {}  \\
\hline
{}  &  {}  \\
\ud3  & (\ch{0} + \ch{4} + \ch{6} + \ch{10}).(\och{3} + \och{7}) + h.c. \\
{}  &  {}  \\
\hline
{}  &  {}  \\
\ud4  & \xaa{3}{7} + [ (\ch{0} + \ch{6}).(\och{4} + \och{10})+ h.c.] \\
{}  &  {}  \\
\hline
\hline
{}  &  {}  \\
\ud{11^{'}}  & \xaaa{1}{5}{7} + \xaaaa{2}{4}{6}{8} + \xaaa{3}{5}{9}\\
{}  &  {}  \\
\hline
{}  &  {}  \\
\ud{21^{'}}  & (\ch{1} + \ch{3} + 2\, (\ch{5}) + \ch{7} + \ch{9}).
         (\och{2} + \och{4} + \och{6} + \och{8}) + h.c.  \\
{}  &  {}  \\
\hline
{}  &  {}  \\
\ud{51^{'}}  &  \xaaaa{2}{4}{6}{8} + 2\, \xa{5} + \left( [(\ch{1} + \ch{7}).
         (\och{3} + \och{5} + \och{9}) + \xx{3}{5} + \xx{5}{9}]
         + h.c. \right) \\
{}  &  {}  \\
\hline
\hline
{}  &  {}  \\
\ud1  &   (\ch{0} + \ch{6}).(\och{1} + \och{5} + \och{7}) +
    (\ch{3} + \ch{7}).(\och{2} + \och{4} + \och{6} + \och{8})
+ (\ch{4} + \ch{10}).(\och{3} + \och{5} + \och{9})  \\
{}  &  {}  \\
\hline
{}  &  {}  \\
\ud{1^{'}} & h.c. (Z_{\ud1}) \\
{}  &  {}  \\
\hline
{}  &  {}  \\
\ud2  & \xaa{3}{7} + \xaa{4}{6} +
(\ch{0} + \ch{10}).(\och{2} + \och{4} + \och{6} + \och{8})
+ (\ch{3} + \ch{7}).(\och{1} + 2\,(\och{5}) + \och{9})
+ (\ch{4} + \ch{6}).(\och{2} + \och{8}) \\
{}  &  {}  \\
\hline
{}  &  {}  \\
\ud{31^{'}} & h.c. (Z_{\ud2}) \\
{}  &  {}  \\
\hline
{}  &  {}  \\
\ud5  & (\ch{0} + \ch{6}).(\och{3} + \och{5} + \och{9})
+  (\ch{3} + \ch{7}).(\och{2} + \och{4} + \och{6} + \och{8})
+  (\ch{4} + \ch{10}).(\och{1} + \och{5} + \och{7})  \\
{}  &  {}  \\
\hline
{}  &  {}  \\
\ud{41^{'}} & h.c. (Z_{\ud5}) \\
{}  &  {}  \\
\hline
\end{array}
$$
\normalsize
\caption{Twisted partition functions for the $E_6$ model}
\end{table}


\begin{table}
\scriptsize
$$
\begin{array}{|c||l|}
\hline
Point & \mathcal{Z} \\
\hline
\hline
{}  &  {}  \\
\ud0  & \xaaaa{0}{10}{18}{28} + \xaaaa{6}{12}{16}{22}  \\
{}  &  {}  \\
\hline
{}  &  {}  \\
\ud6  & \xaaaaaa{0}{6}{12}{16}{22}{28} - \xaa{0}{28} + [(\ch{6} + \ch{12} +
    \ch{16} + \ch{22}).(\och{10} + \och{18}) + h.c.] \\
{}  &  {}  \\
\hline
\hline
{}  & {} \\
\ud{11^{'}}  &  \xaaaaaa{1}{9}{11}{17}{19}{27} +
\xaaaaaaaa{5}{7}{11}{13}{15}{17}{21}{23} \\
{}  &  {} \\
\hline
{}  & {} \\
\ud{71^{'}}  & |\ch{1} + \ch{5} + \ch{7} + \ch{11} + \ch{13} +
\ch{15} + \ch{17} + \ch{21} + \ch{23} + \ch{27}|^{2} +
\xaaaaaa{9}{11}{13}{15}{17}{19} - \\
{}  & \xaa{1}{27} + \xaa{11}{17}  - \xaa{13}{15} - \xaa{9}{19} +
[(\ch{5} + \ch{7} + \ch{21} + \ch{23}).(\och{9} + \och{11} + \och{17}
+ \och{19}) + h.c.]  \\
{}  &  {} \\
\hline
{}  & {} \\
\ud{22^{'}}  & \xaa{2}{26} + | \sum_{i=2}^{12}(\ch{2i})|^{2} +
|\sum_{i=4}^{10}(\ch{2i})|^{2}
         + 2|\ch{14}|^{2} + \\
{}  & [\left( (\ch{2} + \ch{26}).(\och{8} + \och{10} + \och{12} +
\och{16} + \och{18} + \och{20}) + (\ch{4} + \ch{6} + \ch{22} +
\ch{24}).(\och{14}) \right) + h.c.] \\
{}  & {} \\
\hline
{}  & {}  \\
\ud{42^{'}}  & |\sum_{i=1}^{13}(\ch{2i})|^{2} - \xaa{2}{24} +
|\sum_{i=2}^{12}(\ch{2i})|^{2}
        - \xaaaa{4}{6}{22}{24} + \\
{}  & |\sum_{i=4}^{10}(\ch{2i})|^{2} + [(\ch{2}+\sum_{i=4}^{10}(\ch{2i})
+ \ch{26}).\och{14} + h.c.] \\
{}  & {} \\
\hline
{}  & {} \\
\ud{55^{'}}  & \xaaaaaa{5}{9}{13}{15}{19}{23} + |\ch{3} +
\sum_{i=3}^{10}(\ch{2i+1}) + \ch{25}|^{2} \\
{}  & {}  \\
\hline
{}  & {}  \\
\ud{35^{'}}  & |\sum_{i=1}^{12}(\ch{2i+1})|^{2} +
|\sum_{i=3}^{10}(\ch{2i+1})|^{2} +
          \xaaaa{9}{13}{15}{19} - \xaaaa{7}{11}{17}{21} - \xaa{5}{23} + \\
{}  & [(\ch{3} + \ch{25}).(\och{9} + \och{13} + \och{15} + \och{19}) + h.c.] \\
{}  & {} \\
\hline
\hline
{}  &  {}  \\
\ud1  &  \quad (\ch{0} + \ch{10} + \ch{18} + \ch{28}).
(\och{1} + \och{9} + \och{11} + \och{17} + \och{19} + \och{27})
           \\
{}  & + \,(\ch{6} + \ch{12} + \ch{16} + \ch{22}).
(\och{5} + \och{7} + \och{11} + \och{13} + \och{15} + \och{17}
+ \och{21} + \och{23})\\
{}  &  {}  \\
\hline
{}  &  {}  \\
\ud{1^{'}}  &  h.c. (Z_{\ud1}) \\
{}  &  {}  \\
\hline
{}  &  {} \\
\ud2  & \quad \xaaaa{6}{12}{16}{22} + (\ch{0} +\ch{10} + \ch{18} +
\ch{28}).(\och{2} + \och{8} + \och{10} + \och{12} + \och{16} +
\och{18} + \och{20} + \och{26}) \\
{}  & + \, (\ch{6} +\ch{12} + \ch{16} + \ch{22}).(\och{4} + \och{8} +
\och{10} + 2\,(\och{14}) + \och{18} + \och{20} + \och{24}) \\
{}  & {} \\
\hline
{}  &  {} \\
\ud{2^{'}}  &  h.c. (Z_{\ud2}) \\
{}  &  {}  \\
\hline
{}  &  {} \\
\ud3  &  \quad (\ch{0} +\ch{10} + \ch{18} + \ch{28}) .\left( \och{3}
+ \sum_{i=3}^{10}(\och{2i+1}) + \och{25} \right)  \\
{}  & + \, (\ch{6} +\ch{12} + \ch{16} + \ch{22}). \left(
\sum_{i=1}^{12}(\och{2i+1}) + \och{9} + \och{13} + \och{15}
+ \och{19} \right)\\
{}  & {} \\
\hline
{}  & {} \\
\ud{65^{'}}  &  h.c. (Z_{\ud3}) \\
{}  &  {}  \\
\hline
{}  &  {}  \\
\ud4  & \xaaaaaa{6}{10}{12}{16}{18}{22} + \xaa{12}{16} + (\ch{6} +
\ch{22}).(\och{12} + \och{16}) + \\
{}  & (\ch{0} + \ch{10} + \ch{18} + \ch{28}).(\och{4} + \och{8} +
2\,(\och{14}) + \och{20} + \och{24})
+ (\ch{0} + \ch{28}).(\och{6} + \och{10} + \och{12} + \och{16} +
\och{18} + \och{22}) + \\
{}  & (\ch{6} + \ch{12} + \ch{16} + \ch{22}).(\och{2} + \och{4} +
2\,(\och{8}) +  \och{10} + 2\,(\och{14}) + \och{18} + 2\,(\och{20}) +
\och{24} + \och{26}) \\
{}  & {} \\
\hline
{}  & {} \\
\ud{62^{'}}  & h.c. (Z_{\ud4}) \\
{}  &  {}  \\
\hline
\end{array}
$$
\normalsize
\caption{Twisted partition functions for the $E_8$ model (part 1.)}
\end{table}

\begin{table}
\scriptsize
$$
\begin{array}{|c||l|}
\hline
Point & Z \\
\hline
{}  & {} \\
\ud5  & (\ch{0} + \ch{10} + \ch{18} + \ch{28}).(\och{5} + \och{9} +
\och{13} + \och{15} + \och{19} + \och{23}) + (\ch{6} + \ch{12} +
\ch{16} + \ch{22}).(\och{3} + \sum_{i=3}^{10}(\och{2i+1}) + \och{25})
\\
{}  & {} \\
\hline
{}  & {} \\
\ud{5^{'}}  & h.c. (Z_{\ud5}) \\
{}  & {} \\
\hline
{}  & {} \\
\ud7  & (\ch{0} + \ch{10} + \ch{18} + \ch{28}).(\och{5} + \och{7} +
\och{11} + \och{13} + \och{15} + \och{17} + \och{21} + \och{23}) +  \\
{}  & (\ch{6} + \ch{12} + \ch{16} + \ch{22}).(\och{1} +
\sum_{i=2}^{11}(\och{2i+1}) + \och{11} + \och{17} + \och{27}) \\
{}  & {} \\
\hline
{}  & {} \\
\ud{61^{'}}  & h.c. (Z_{\ud7}) \\
{}  & {} \\
\hline
\hline
{}   & {} \\
\ud{21^{'}} & (\ch{1} + \ch{9} + \ch{11} + \ch{17} + \ch{19} +
\ch{27}).(\och{2} + \och{8} + \och{10} + \och{12} + \och{16} +
\och{18} + \och{20}
+ \och{26})\\
{} & + (\ch{5} + \ch{7} + \ch{11} + \ch{13} + \ch{15} + \ch{17} +
\ch{21} + \ch{23}).(\sum_{i=2}^{12}(\och{2i}) + \och{14}) \\
{}  & {} \\
\hline
{}  & {} \\
\ud{12^{'}}  & h.c.(Z_{\ud{21^{'}}}) \\
{}  & {} \\
\hline
{}  & {} \\
\ud{41^{'}}  & (\ch{1} + \ch{9} + \ch{11} + \ch{17} + \ch{19} +
\ch{27}).(\sum_{i=2}^{12}(\och{2i}) + \och{14}) +  \\
{}  & (\ch{5} + \ch{7} + \ch{11} + \ch{13} + \ch{15} + \ch{17} +
\ch{21} + \ch{23}).(\sum_{i=1}^{13}(\och{2i}) +
\sum_{i=4}^{10}(\och{2i})) \\
{}  & {} \\
\hline
{}  & {} \\
\ud{72^{'}}  & h.c.(Z_{\ud{41^{'}}}) \\
{}  & {} \\
\hline
{}  & {} \\
\ud{52^{'}}  & (\ch{2} + \ch{8} + \ch{10} + \ch{12} + \ch{16} +
\ch{18} + \ch{20}+ \ch{26}).(\och{5} + \och{9} + \och{13} + \och{15}
+ \och{19} + \och{23}) + \\
{}  & (\sum_{i=2}^{12}(\ch{2i}) + \ch{14}).(\och{3} +
\sum_{i=3}^{10}(\och{2i+1}) + \och{25}) \\
{}  & {} \\
\hline
{}  & {} \\
\ud{25^{'}}  & h.c.(Z_{\ud{52^{'}}}) \\
{}  & {} \\
\hline
{}  & {} \\
\ud{32^{'}}  & (\ch{2} + \ch{8} + \ch{10} + \ch{12} + \ch{16} +
\ch{18} + \ch{20}+ \ch{26}).(\och{3} + \sum_{i=3}^{10}(\och{2i+1}) +
\och{25})+  \\
{}  & (\sum_{i=2}^{12}(\ch{2i}) +
\ch{14}).(\sum_{i=1}^{12}(\och{2i+1}) + \och{9} + \och{13} + \och{15}
+ \och{19}) \\
{}  & {} \\
\hline
{}  & {} \\
\ud{45^{'}}  & h.c.(Z_{\ud{32^{'}}}) \\
{}  & {} \\
\hline
{}  & {} \\
\ud{51^{'}}  & (\ch{1} + \ch{9} + \ch{11} + \ch{17} + \ch{19} +
\ch{27}).(\och{5} + \och{9} + \och{13} + \och{15} + \och{19} +
\och{23}) + \\
{}  & (\ch{5} + \ch{7} + \ch{11} + \ch{13} + \ch{15} + \ch{17} +
\ch{21}+ \ch{23}).(\och{3} + \sum_{i=3}^{10}(\och{2i+1}) + \och{25})
\\
{}  & {} \\
\hline
{}  & {} \\
\ud{15^{'}}  & h.c.(Z_{\ud{51^{'}}}) \\
{}  & {} \\
\hline
{}  & {} \\
\ud{31^{'}}  & |\sum_{i=2}^{11}(\ch{2i+1})|^2 + (\ch{1} + \ch{11} +
\ch{17} + \ch{27}).(\och{3} + \sum_{i=3}^{10}(\och{2i+1}) + \och{25})
+ (\ch{9} + \ch{19}) .(\och{3} - \och{5} - \och{23} + \och{25})\\
{}  & (\ch{5} + \ch{7} + \ch{11} + \ch{13} + \ch{15} + \ch{17} +
\ch{21}+ \ch{23}).(\och{3} + \och{9} + \och{13} + \och{15} + \och{19}
+ \och{25}) \\
{}  & {} \\
\hline
{}  & {} \\
\ud{75^{'}}  & h.c.(Z_{\ud{31^{'}}}) \\
{}  & {} \\
\hline
\end{array}
$$
\normalsize
\caption{Twisted partition functions for the $E_8$ model (part 2.)}
\end{table}


\begin{table}
\scriptsize
$$
\begin{array}{|c||l|}
\hline
Point & Z \\
\hline
{}   & {} \\
\ud0 & \xaa{0}{4} + 2\xa{2} \\
{}   & {} \\
\hline
{}   & {} \\
\ud2, \; \ud{2^{'}} & \xa{2} + [(\ch{0} + \ch{4}).\och{2} + h.c.] \\
{}   & {} \\
\hline
\hline
{}   & {} \\
\ud{\ep}, \; \ud{2\ep}, \; \ud{2^{'}\ep} & \xaa{1}{3} \\
{}   & {} \\
\hline
\hline
{}   & {} \\
\ud1 & (\ch{0} + 2\, (\ch{2}) + \ch{4}).(\och{1} + \och{3}) \\
{}   & {} \\
\hline
{}   & {} \\
\ud{1\ep} & h.c. (Z_{\ud1}) \\
{}   & {} \\
\hline
\end{array}
$$
\normalsize
\caption{Twisted partition functions for the $D_4$ model}
\end{table}


\begin{table}
\scriptsize
$$
\begin{array}{|c||l|}
\hline
Point & Z \\
\hline
{}   & {} \\
\ud0 & \xaa{0}{8} + \xaa{2}{6} + 2\xa{4} \\
{}   & {} \\
\hline
{}   & {} \\
{}   & {} \\
\ud2 & \xaa{2}{6} + 2\xa{4} + [(\ch{0} + 2\, (\ch{4}) +
\ch{8}).(\och{2} + \och{6}) +  h.c.] \\
{}   & {} \\
\hline
{}   & {} \\
\ud4, \; \ud{4^{'}} & \xaaa{2}{4}{6} +[(\ch{0} + \ch{8}).\och{4})+ h.c.] \\
{}   & {} \\
\hline
\hline
{}   & {} \\
\ud{\ep} & \xaa{1}{7} + \xaa{3}{5} \\
{}   & {} \\
\hline
{}   & {} \\
\ud{2\ep} & \xaaaa{1}{3}{5}{7} + \xaa{3}{5} \\
{}   & {} \\
\hline
{}   & {} \\
\ud{4\ep}, \; \ud{4^{'}\ep} & \xaaaa{1}{3}{5}{7} - \xaa{1}{7} \\
{}   & {} \\
\hline
\hline
{}   & {} \\
\ud1 & (\ch{0} + \ch{2} + \ch{6} + \ch{8}).(\och{1} + \och{7}) +
(\ch{2} + 2\, (\ch{4})+ \ch{6}).(\och{3} + \och{5})  \\
{}   & {} \\
\hline
{}   & {} \\
\ud{1\ep} & h.c. (Z_{\ud1}) \\
{}   & {} \\
\hline
{}   & {} \\
\ud3 & (\ch{0} + \ch{2} + \ch{6} + \ch{8}).(\och{3} + \och{5}) +
(\ch{2} + 2\, (\ch{4}) + \ch{6}).(\och{1} + \och{3} + \och{5} +
\och{7}) \\
{}   & {} \\
\hline
{}   & {} \\
\ud{3\ep} & h.c. (Z_{\ud3}) \\
{}   & {} \\
\hline
\end{array}
$$
\normalsize
\caption{Twisted partition functions for the $D_6$ model}
\end{table}


\begin{table}
\scriptsize
$$
\begin{array}{|c||l|}
\hline
Point & \mathcal{Z} \\
\hline
\hline
{}  &  {}  \\
\ud0  & \xa{0} + \xa{2} + \xa{3} + \xa{4} + \xa{6} + (\xx{1}{5} + h.c.) \\
{}  &  {}  \\
\hline
{}  &  {}  \\
\ud2  & \xaa{2}{4} + \xa{3} + [( \xx{0}{2} + \xx{1}{3} + \xx{1}{5} +
\xx{3}{5} + \xx{4}{6}) + h.c.] \\
{}  &  {}  \\
\hline
{}  &  {}  \\
\ud3  & [(\ch{0} + \ch{2} + \ch{4} + \ch{6}).\och{3} + (\ch{1} + \ch{5}).
(\och{2} + \ch{4})  + h.c.]  \\
{}  &  {}  \\
\hline
{}  &  {}  \\
\ud4  & \xa{1} + \xa{3} + \xa{5} + \xaa{2}{4} + [ (\xx{0}{4} + \xx{1}{3}
       + \xx{2}{6} + \xx{3}{5} ) + h.c. ] \\
{}  &  {}  \\
\hline
{}  &  {}  \\
\ud6  &  \xa{1} + \xa{3} + \xa{5} + [(\xx{0}{6} + \xx{2}{4}) + h.c.] \\
{}  &  {}  \\
\hline
\hline
{}  &  {}  \\
\ud1  & (\ch{0} + \ch{2}).\och{1} + \ch{5}.(\och{0} + \och{2}) +
\ch{1} .(\och{4} + \och{6})+ (\ch{4} + \ch{6}).\och{5} + [(\xx{2}{3}
+ \xx{3}{4}) + h.c.]\\
{}  &  {}  \\
\hline
{}  &  {}  \\
\ud5  &  h.c.(Z_{\ud1}) \\
{}  &  {}  \\
\hline
\end{array}
$$
\normalsize
\caption{Twisted partition functions for the $D_5$ model}
\end{table}


\begin{table}
\scriptsize
$$
\begin{array}{|c||l|}
\hline
Point & \mathcal{Z} \\
\hline
\hline
{}  &  {}  \\
\ud0  & \xa{0} + \xa{2} + \xa{4} + \xa{5} + \xa{6} + \xa{8} + \xa{10} +
       [(\xx{1}{9} + \xx{3}{7}) + h.c.] \\
{}  &  {}  \\
\hline
{}  &  {}  \\
\ud2  & \xaa{2}{4} + \xa{5} + \xaa{6}{8} + [( \xx{0}{2} + (\ch{1} + \ch{3}).
(\och{7} + \och{9}) + \xx{3}{5} + \xx{4}{6}+  \xx{5}{7}  +
\xx{8}{10}) + h.c.] \\
{}  &  {}  \\
\hline
{}  &  {}  \\
\ud4  & \xaaaa{2}{4}{6}{8} + \xaaa{3}{5}{7} + [\xx{0}{4} +
\ch{1}.(\och{5} + \och{7}) - \xx{2}{8} + (\ch{3} + \ch{5}).\och{9} +
\xx{6}{10}) + h.c.] \\
{}  &  {}  \\
\hline
{}  &  {}  \\
\ud5  &  [(\ch{0} + \ch{2} + \ch{4} + \ch{6} + \ch{8} + \ch{10}).\och{5} +
(\ch{1} + \ch{3} + \ch{7} + \ch{9}).(\och{4} + \och{6}) +
(\ch{2} + \ch{8}).(\och{3} + \och{7}) + h.c.] \\
{}  &  {}  \\
\hline
{}  &  {}  \\
\ud6  &  \xaaaa{2}{4}{6}{8} + \xaaa{3}{5}{7} - \xa{2} - \xa{8} + [(\xx{0}{6}
+ \ch{1}.(\och{3} + \och{5}) + \xx{4}{10} + (\ch{5} +
\ch{7}).\och{9}) + h.c.] \\
{}  &  {}  \\
\hline
{}  &  {}  \\
\ud8  &   \xaa{1}{3} + \xaa{4}{6}+ \xaa{7}{9} + \xa{5} + [(\xx{0}{8} +
\ch{2}.(\och{6} + \och{8} + \och{10}) + \xx{3}{5} + \xx{4}{8} + \xx{5}{7})
+ h.c.] \\
{}  &  {}  \\
\hline
{}  &  {}  \\
\ud{10} & \xa{1} + \xa{3} + \xa{5} + \xa{7} + \xa{9} + [(\xx{0}{10} +
\xx{2}{8} + \xx{4}{6}) + h.c.]   \\
{}  &  {}  \\
\hline
\hline
{}  & {} \\
\ud1  & (\ch{0} + \ch{2}).\och{1} + \ch{9}.(\och{0} + \och{2}) +
         (\ch{2} + \ch{4}).\och{3} + \ch{7}.(\och{2} + \och{4}) +
         (\ch{6} + \ch{8}).\och{7} + \ch{3}.(\och{6} + \och{8})  + \\
{}  & (\ch{8} + \ch{10}).\och{9} + \ch{1}.(\och{8} + \och{10})+
      [(\ch{4} + \ch{6}).\och{5} + h.c.] \\
{}  &  {}  \\
\hline
{}  &  {}  \\
\ud9  &  h.c.(Z_{\ud1})  \\
{}  &  {}  \\
\hline
{}  &  {}  \\
\ud3  &  \ch{7}.(\och{0} + \och{2}) + (\ch{2} + \ch{4}).\och{1} +
          \ch{9}.(\och{2} + \och{4})+ (\ch{0} + \ch{2}).\och{3} +
          \ch{1}.(\och{6} + \och{8}) + (\ch{8} + \ch{10}).\och{7}+ \\
{}  &  \ch{3}.(\och{8} + \och{10}) + (\ch{6} + \ch{8}).\och{9} +
          [(\ch{2} + \ch{4} + \ch{6} + \ch{8}).\och{5} +
           (\ch{3} + \ch{7}).(\ch{4} + \ch{6}) + h.c.] \\
{}  &  {}  \\
\hline
{}  &  {}  \\
\ud7  &  h.c.(Z_{\ud3})  \\
{}  &  {}  \\
\hline
\end{array}
$$
\normalsize
\caption{Twisted partition functions for the $D_7$ model}
\end{table}


\begin{table}
\scriptsize
$$
\begin{array}{|c||l|}
\hline
Point & \mathcal{Z} \\
\hline
\hline
{}  &  {}  \\
\ud0 & \xaa{0}{16}+ \xaa{4}{12} + \xaa{6}{10} + \xa{8} + [(\ch{2} +
\ch{14}).\och{8} + h.c.] \\
{}  &  {}  \\
\hline
{}  &  {}  \\
\ud2  & \xaaaaa{4}{6}{8}{10}{12} - (\ch{4} + \ch{12}).\och{8} +
\ch{8}.(\och{0} + \och{16}) + \\
{} & (\ch{0} + \ch{4} + \ch{8} + \ch{12} + \ch{16}).(\och{2} +
\och{14}) + (\ch{2} + \ch{14}).(\och{6} + \och{8} + \och{10}) +
(\ch{6} + \ch{10}).\och{8} \\
{}  &  {}  \\
\hline
{}  &  {}  \\
\ud4  & \xaaaaaaa{2}{4}{6}{8}{10}{12}{14} - \xaa{2}{14} +
\xaaa{6}{8}{10} - \xa{8} + \\
{}  & [(\ch{0} + \ch{8} + \ch{16}).(\och{4} + \och{12}) + h.c.] \\
{}  &  {}  \\
\hline
{}  &  {}  \\
\ud6  & \xaaaaaaa{2}{4}{6}{8}{10}{12}{14} + \xaaaaa{4}{6}{8}{10}{12}
- \xaa{4}{12} + \xa{8} + \\
{}  & [(\ch{0} + \ch{16}).(\och{6} + \och{12}) + h.c.] \\
{}  &  {}  \\
\hline
{}  &  {}  \\
\ud8  & \xaaaaa{4}{6}{8}{10}{12} - \ch{8}.(\och{4} + \och{12}) +
(\ch{2} + \ch{14}).(\och{0} + \och{4} + \och{8} + \och{12} +
\och{16}) + \\
{}  & (\ch{6} + \ch{8} + \ch{10}).(\och{2} + \och{14}) + (\ch{0} +
\ch{16}).\och{8} + \ch{8}.(\och{6} + \och{10}) \\
{}  &  {}  \\
\hline
{}  &  {}  \\
\ud{8^{'}} & \xaaaa{2}{6}{10}{14} - \xaa{6}{10} +
\xaaaaa{4}{6}{8}{10}{12} + \xa{8} + [(\ch{0} + \ch{16}).\och{8}+
h.c.] \\
{}  &  {}  \\
\hline
\hline
{}  &  {}  \\
(\ud1)  &  \xaaaa{1}{7}{9}{15} + \xaaaaaa{3}{5}{7}{9}{11}{13} + \xaa{5}{11} \\
{}  &  {}  \\
\hline
{}  &  {}  \\
(\ud3)  & |\sum_{i=0}^{7}(\ch{2i+1})|^{2} + |\sum_{i=1}^{6}(\ch{2i+1})|^{2} +
|\sum_{i=2}^{5}(\ch{2i+1})|^{2} - \xaa{1}{15} - \xaa{3}{13} - \xaa{5}{11} \\
{}  &  {}  \\
\hline
{}  &  {}  \\
(\ud5)   & \xaaaa{3}{7}{9}{13} + \xaaaa{5}{7}{9}{11}- \xaa{7}{9} +
[(\ch{1}+  \ch{15}).(\och{5} + \och{11}) + h.c.] \\
{}  &  {}  \\
\hline
\hline
{}  &  {} \\
\ud1  & (\ch{0} + \ch{8} + \ch{16}).(\och{1} + \och{15}) + (\ch{4} +
\ch{8} + \ch{12}).(\och{3} + \och{13}) + (\ch{4} + \ch{6} + \ch{10} +
\ch{12}).(\och{5} + \och{11}) + \\
{} &  (\ch{2} + \ch{6} + \ch{8} +\ch{10} + \ch{14}).(\och{7} + \och{9}) \\
{}  &  {}  \\
\hline
{}  &  {}  \\
(\ud0)  &  h.c.(Z_{\ud1}) \\
{}  &  {}  \\
\hline
{}  &  {}  \\
\ud7  &  (\ch{2} + \ch{6} + \ch{8} + \ch{10} +
\ch{14}).(\sum_{i=0}^{7}(\och{2i+1}))+ (\ch{4} + \ch{6} + \ch{10} +
\ch{12}).(\sum_{i=1}^{6}(\och{2i+1})) + \\
{}  & (\ch{4} + \ch{8} + \ch{12}).(\och{5} + \och{7} + \och{9} + \och{11})
+ (\ch{0} + \ch{8} + \ch{16}).(\och{7} + \och{9}) \\
{}  &  {}  \\
\hline
{}  &  {}  \\
(\ud2)  &  h.c.(Z_{\ud7}) \\
{}  &  {}  \\
\hline
{}  &  {}  \\
\ud3  & (\ch{4} + \ch{8} + \ch{12}).(\sum_{i=0}^{7}(\och{2i+1})) +
(\ch{0} + \ch{6} + \ch{10} + \ch{16}).(\och{3} + \och{12}) + \\
{}  & (\ch{2} + \ch{6} + \ch{10} + \ch{14}).(\och{5} + \och{7} +
\och{9} + \och{11}) + (\ch{6} + \ch{10}).(\och{7} + \och{9}) +
\ch{8}.(\och{5} + \och{11}) \\
{}  &  {}  \\
\hline
{}  &  {}  \\
(\ud4)  &  h.c.(Z_{\ud3}) \\
{}  &  {}  \\
\hline
{}  &  {}  \\
\ud5  & (\ch{4} + \ch{6} + \ch{10} + \ch{12}).(\sum_{i=0}^{7}(\och{2i+1})) +
(\ch{2} + 2\, (\ch{8}) + \ch{14}).(\sum_{i=1}^{6}(\och{2i+1})) + \\
{}  & (\ch{0} + \ch{6} + \ch{10} + \ch{16}).(\och{5} + \och{11}) +
(\ch{4} + \ch{6} + \ch{10} + \ch{12}).(\och{7} + \och{9}) \\
{}  &  {}  \\
\hline
{}  &  {}  \\
(\ud6)  &  h.c.(Z_{\ud5}) \\
{}  &  {}  \\
\hline
\end{array}
$$
\caption{Twisted partition functions for the $E_7$ model}
\end{table}


\end{document}